\documentclass[a4paper]{article}
\pdfoutput=1
\usepackage{jheppub}
\usepackage{graphicx}
\usepackage{amsmath}
\usepackage{amssymb}
\usepackage{amstext}
\usepackage{amsfonts}
\usepackage{hyperref}
\usepackage{physics}
\usepackage[utf8]{inputenc}
\usepackage{hhline}
\usepackage{bm}
\usepackage{pifont}
\usepackage{ragged2e}

\usepackage{xcolor}
\usepackage{fontawesome5}
\definecolor{orcidlogocol}{rgb}{0.65, 0.807, 0.223}
\newcommand{\orcid}[1]{\,\href{https://orcid.org/#1}{\textcolor{orcidlogocol}{\footnotesize\faOrcid}}\,}

\newcommand{\ZenodoComment}{Animation(s) of the dynamics for these parameters are available at \cite{ZenodoVideos}.}

\newcommand{\cmark}{\ding{51}}
\newcommand{\xmark}{\ding{55}}

\begin{document}

\title{Periodic Cosmic String Formation and Dynamics}

\author[a]{Michael A.~Fedderke\orcid{0000-0002-1319-1622},}
\author[a]{Junwu Huang\orcid{0000-0001-6007-7315},}
\author[b,c]{and Nils Siemonsen\orcid{0000-0001-5664-3521}}

\affiliation[a]{Perimeter Institute for Theoretical Physics, 31 Caroline St.~N., Waterloo, Ontario N2L 2Y5, Canada}
\affiliation[b]{Princeton Gravity Initiative, Princeton University, Princeton, NJ 08544, USA}
\affiliation[c]{Department of Physics, Princeton University, Princeton, NJ 08544, USA}

\emailAdd{mfedderke@perimeterinstitute.ca}
\emailAdd{jhuang@perimeterinstitute.ca}
\emailAdd{nils.siemonsen@princeton.edu}

\arxivnumber{2503.03116}

\abstract{%
We study string formation and dynamics in a scalar field theory with a global $U(1)$ symmetry. 
If a scalar field $\Phi$ subject to a wine-bottle potential is initially displaced from the potential minimum, and even if this is done uniformly and coherently over large spatial patches, we show that small spatial perturbations to $\Phi$ grow through parametric resonance as $\Phi$ oscillates; this observation holds over a wide range of initial $U(1)$ charge densities. 
We show that the growth of these perturbations leads to the formation of spatially coherent, temporally stable \emph{counter-rotating regions}; i.e., spatially connected regions that exhibit $\Phi$ evolution with large and opposite-sign rotation speeds in field space and that persist over long durations. 
These counter-rotating regions are separated by domain boundaries characterized by a large field gradient and zero rotational speed in field space.
We find that string or vortex topological defects form, are confined to, and then annihilate periodically on these domain boundaries.
We demonstrate these periodic dynamics with numerical simulations in both $2+1$ and $3+1$ dimensions, in both Minkowski spacetime and in a radiation-dominated Friedmann--Lema{\^i}tre--Robertson--Walker (FLRW) universe, and we explain some features of the evolution (semi-)analytically. 
At late times in an expanding universe, when $\Phi$ approaches the minimum of the potential, we find counter-rotating regions and vortices to dissipate into scalar radiation.
Phenomenologically, periodic bursts of string formation and annihilation are expected to lead to periodic bursts of gravitational-wave production. 
For small initial $U(1)$ charge density, these gravitational-wave bursts can be synchronized across the whole Universe.
Owing to their periodic nature, it is possible that they could give rise to a gravitational-wave frequency spectrum consisting of a forest of fully or partially resolved peaks. 
We find that these periodic scalar field dynamics also occur with large (but not fine-tuned) initial $U(1)$ charge density; they may thus have implications for models that depend on a coherent field rotation, such as kination and the axion kinetic-misalignment mechanism.
}

\maketitle

\section{Introduction and Summary}\label{sec:intro}

The formation of topological defects in theories with global or gauged $U(1)$ symmetries has been extensively studied in both high-energy and condensed-matter contexts~\cite{Kibble:1976sj,Kibble:1980mv,Zurek:1996sj}. 
The Kibble--Zurek mechanism~\cite{Kibble:1976sj,Zurek:1993ek,Zurek:1985qw} is a well-known example in thermal systems, in which topological defects form as the early Universe cools below the critical temperature for spontaneous symmetry breaking. 
Such defects can have important phenomenological consequences (for instance for axion dark matter, where details are still under active investigation; see, e.g., \cite{Gorghetto:2020qws,Buschmann:2021sdq,Benabou:2023ghl,Benabou:2023npn,Gorghetto:2023vqu,Benabou:2024msj,Gorghetto:2024vnp} for a selection of recent work).
Formation of cosmic strings can also occur non-thermally as a result of quantum fluctuations during inflation~\cite{Linde:1990yj,PhysRevD.44.340,Vishniac:1986sk,Lyth:1992tw}.
Defect formation during preheating scenarios has also been studied extensively~\cite{Kasuya:1997ha,Kasuya:1998td,Kasuya:1999hy}.
After inflation, nonlinearities may drive the formation of topological defects in a large background field~\cite{Abrikosov:1956sx,osti_4551407,Galaiko1966FormationOV,East:2022ppo,East:2022rsi,Kofman:1997yn,Kawasaki:2013iha}. 
In the gauged-$U(1)$ case (i.e., the Abelian Higgs model), in a system where the gauge field strength grows~\cite{East:2022ppo,East:2022rsi} (either in the early Universe or via black-hole superradiance) and surpasses a critical threshold, a superheated phase transition occurs and a dense network of strings is produced.%
\footnote{%
    We will use the terms `vortices' and `strings' when discussing the dynamics presented in this paper. 
    Although vortices are generally associated with $(2+1)$-dimensional systems and strings with $(3+1)$-dimensional systems, we use them interchangeably, depending on the context. 
    For all practical purposes of this paper, there is no difference between these two terms.} %
This string network absorbs energy from the background gauge fields, grows in size, and persists in the system. 
These dynamics resemble the vortex-forming phase transition in a superconductor, which is described in the framework of the Ginzburg--Landau model~\cite{Ginzburg:1950sr,Abrikosov:1956sx} (see App.~\ref{app:cmtvortex} for a short review of the relevant literature).

A theory with a global $U(1)$ symmetry, resembling a superfluid, can also potentially support phase transitions leading to the non-thermal formation of cosmic strings.
In the superfluid context, Feynman~\cite{feynman1955chapter,ANDERSON:1966bps,PhysRevLett.19.822} demonstrated the existence of a critical minimum fluid velocity above which vortices may form.
The subsequent evolution of these superfluid vortices was studied in a series of seminal papers~\cite{Ao:1993zz,1985PhRvL..55.2887H,Thouless:1996mt,1993PhyA..200...42T,Sonin_1997}, in which it was demonstrated that their dynamics are governed by the so-called `Magnus force'. 
As we will show in more detail, these properties of superfluid vortices can be employed as useful heuristic tools to develop an understanding of non-thermal formation and subsequent dynamics of global $U(1)$ strings in the cosmological context.

In this paper, we consider non-thermal vortex formation in the early Universe in a complex scalar theory with a $U(1)$ global symmetry:%
\footnote{\label{ftnt:lagNormalization}%
    The normalization of the Lagrangian \eqref{eq:lagrangianintro} is non-standard by an overall factor of 2, but results in a canonically normalized real radial mode $\rho(t,\bm{x})$ given our definition for the latter in terms of $\Phi$.
    The more conventional choices would be to define $\mathcal{L} = \mathcal{L}'/2$, $\rho = \rho'/\sqrt{2}$, and $\lambda=2\lambda'$.
    In that case, $\mathcal{L'}$ takes the more standard form, the $\rho'$ field is canonically normalized, and the potential minimum of $V=0$ is at $\rho'=v'=\mu/\sqrt{\lambda'} \Leftrightarrow |\Phi| = v'/\sqrt{2}$.
}\textsuperscript{,}%
\footnote{%
    Throughout this paper, we set $\hbar=c=1$. 
    Greek tensor indices are spatiotemporal and run $\mu= 0,1,\ldots,D$ in $D$ spatial dimensions; Latin indices are spatial and run $i = 1,\ldots,D$.
    Repeated indices are summed over.
} %
\begin{align}\label{eq:lagrangianintro}
    \mathcal{L} &= \frac{1}{2}| \partial_\mu \Phi |^2 + \frac{1}{2}\mu^2 |\Phi|^2 - \frac{1}{4} \lambda |\Phi|^4 - \frac{1}{4} \frac{\mu^4}{\lambda} \:; & \Phi(t,\bm{x}) &\equiv \rho(t,\bm{x}) \, e^{i\theta(t,\bm{x})}\:. 
\end{align}
Here, $\Phi(t,\bm{x})$ is a complex scalar field, while $\rho(t,\bm{x})$ and $\theta(t,\bm{x})$ are both real scalar fields, and $\mu$ and $\lambda>0$ are real parameters.
The wine-bottle potential for $\Phi$,
\begin{align}
    V \equiv -\frac{1}{2}\mu^2 |\Phi|^2 + \frac{1}{4} \lambda |\Phi|^4 + \frac{1}{4} \frac{\mu^4}{\lambda}\:, \label{eq:V}
\end{align}
has a minimum at $ |\Phi| = v \equiv \mu/\sqrt{\lambda}$ (with zero vacuum energy in this minimum by definition), while the (canonically normalized) radial mode $\rho$ has a mass of $\sqrt{2}\mu$ (as measured around the potential minimum).
The conserved global-$U(1)$ Noether current is
\begin{align}
    j^\mu \equiv \frac{i}{2} \left[ \Phi \partial^{\mu} \Phi^* - \Phi^* \partial^\mu \Phi \right] = \rho^2 \partial^\mu \theta \: ,
    \label{eq:noethercurrent}
\end{align}
where ${}^*$ denotes complex conjugation.

Global strings can form non-thermally in such a model, similar to the gauged examples~\cite{East:2022ppo,East:2022rsi}, in both compact objects~\cite{Siemonsen:2023hko} and as a result of cosmological dynamics~\cite{Tkachev:1998dc}. 
In this paper, we consider initial conditions motivated by early Universe cosmological dynamics, where the radial mode is displaced initially to  $\rho_i = |\Phi_i| > v$, together with a non-vanishing initial angular velocity, $\dot{\theta}_i \neq 0$ (which may however be very small).%
\footnote{%
    Vortex formation also occurs for $\dot{\theta}_i = 0$; cf.~\cite{Tkachev:1998dc}.
    } %
A displaced radial mode can occur naturally at the end of inflation due to the misalignment mechanism, or modifications to the radial-mode potential during inflation due to couplings to curvature, the inflaton, and other particles~\cite{Tkachev:1998dc,Bezrukov:2007ep}.
On the other hand, a small (or even potentially large) non-zero initial angular velocity could stem from higher dimensional operators that explicitly break the global $U(1)$ symmetry~\cite{Co:2017mop,Co:2020dya}. 
Kinematically, a nonzero initial angular velocity ensures that the field does not initially oscillate through the field-space origin (even were the initial displacement large enough for it to be energetically possible for it to do so). 
However, as we demonstrate, a small and uniform initial charge density $\mathcal{J}_0 \equiv \rho_i^2 \dot{\theta}_i$ does not prevent the formation of vortices in the system, even though the field does not approach the origin initially. 
In fact, we find that vortex formation occurs even in the presence of large (but not fine-tuned) $\mathcal{J}_0$, which is motivated by models that rely on a coherent field rotation, such as kination and the axion kinetic-misalignment mechanism~\cite{Affleck:1984fy,Co:2017mop,Co:2020dya}.

In this work, we explore in detail the following dynamics that lead to global vortex or string formation in this model under such conditions (see \cite{ZenodoVideos} for relevant animations showing our results):
(1) the initial growth of field perturbations through parametric resonance, leading to
(2) the saturation of the instability when the system fragments into {\it regions of counter rotation} characterized by field evolution with large and opposite sign $\dot{\theta}$. Eventually, these regions facilitate the (3) {\it periodic} vortex pair production%
\footnote{%
    Throughout this paper, we use the terms ``vortex pair production'' and ``vortex--anti-vortex pair production'' interchangeably to refer a single production event that creates exactly one vortex and exactly one anti-vortex.
    Additionally, we will often state that vortices have some or other property; unless the context makes clear that such a reading would be nonsensical, such statements should be read inclusively to apply also to anti-vortices.
} %
at the {\it domain boundaries} between regions with counter rotation, once the field gradient $\nabla \theta$ reaches a critical value that is analogous to Feynman's critical velocity in a superfluid (we also identify an alternative, global criterion for vortex formation, as discussed below).
This then leads to (4) a long-lived quasi-steady state, during which strings repeatedly (i.e., periodically) form and annihilate, releasing energy in scalar radiation (and possibly as gravitational waves).
And finally, (5) at very late times in an expanding universe, the eventual disappearance of the vortices from the system when it {\it thermalizes} owing to the transfer of energy to radiation.%
\footnote{The eventual disappearance of all vortices may also happen for a flat background spacetime, but the timescale for this process to complete can, depending on parameters, be longer than the total duration of our simulations.} %

Phenomenologically, periodic gravitational-wave bursts in the early Universe give a intriguing experimental target for current and future gravitational-wave detectors. 
Moreover, our findings may have implications for any mechanism that generates or relies on a stable, coherently rotating field in the early Universe, highlighting the importance of dedicated numerical simulations to assess the viability of mechanisms that appear to be operational at the level of background field (i.e., zero-mode) evolution. 

This work provides another example (see also, e.g.~\cite{East:2022ppo,East:2022rsi}) showing that formation of topological defects can occur when too much energy is added into a nonlinear system, which can significantly alter the evolution of the system, leading to striking observable consequences.

The paper is organized as follows.
In Sec.~\ref{sec:growth}, we begin with a purely analytical perturbative analysis valid for small $U(1)$ charge density that demonstrates the existence of a parametric resonance instability~\cite{Tkachev:1998dc} for spatial field perturbations $\delta \Phi$ around a spatially homogeneous background field evolution $\Phi_0$ [Secs.~\ref{sec:backgroundevo} and \ref{sec:pert}]. 
We extend this perturbative analysis to larger $U(1)$ charge densities via a numerical study of the linearized system of equations governing the evolution of the perturbations $\delta \Phi$~\cite{Co:2020dya}, demonstrating that this parametric resonance remains present in this case [Sec.~\ref{sec:linearizednumerics}].
We then analyze the fully non-linear evolution of the $\Phi$ field from the epoch of the initial parametric resonance through to the onset of the formation of the counter-rotating regions~(CRR) that emerge when the perturbations grow beyond the linear regime phase-coherently over a range of momentum modes [Sec.~\ref{sec:pertfullnonlinear}]; we do this both for small and, for the first time, large $U(1)$ charge densities.
Finally, we summarize the main points of this section [Sec.~\ref{sec:Sec2Coda}].

In Sec.~\ref{sec:formation}, we then move to a numerical examination of the vortex-formation dynamics that occur during the quasi-steady state once counter-rotating regions have developed.
After a brief primer on global-$U(1)$ vortices generally [Sec.~\ref{sec:globalstring}], we show [Sec.~\ref{sec:vortexformationdynamics}] the dynamics of a representative vortex--anti-vortex pair creation event occurring on the domain boundary between two counter-rotating regions at a location of large spatial phase gradient $|\nabla \theta|$, and we discuss some of the associated vortex kinematics (see also~\cite{Tkachev:1998dc}).
We then analyze the conditions for vortex formation more generally in two spatial dimensions [Sec.~\ref{sec:formation_conditions}], showing that, with rather generic initial conditions $\rho_i$ and $\dot{\theta}_i$, vortex--anti-vortex pairs will form on the boundaries between the counter-rotating regions provided that certain simple global conditions are met. 
These conditions are weaker requirements as compared to earlier results~\cite{Tkachev:1998dc}.
For small $\dot{\theta}_i$, the global criterion that we identify is $V(\Phi_i) > c_D V(0)$, where $c_{D=2} \approx 0.29< 1$ is a coefficient that might depend on the spatial dimensionality $D$ of the system. 
We also identify a similar condition for larger $\dot{\theta}_i$ that is applicable as long as $\dot{\theta}_i$ is not precisely tuned to match the size of the initial displacement $\rho_i$ (i.e., as long as $\ddot{\rho} \not\approx 0$ initially).
We then turn to the case of three spatial dimensions [Sec.~\ref{sec:VortexFormation3D}], and show that global $U(1)$-string loop formation occurs similarly to the case of vortex--anti-vortex formation in two spatial dimensions.
Finally, we summarize our findings in this section [Sec.~\ref{sec:Sec3Summary}].

In Sec.~\ref{sec:dynamics}, we explore the subsequent evolution of vortex pairs or string loops after their formation, and find, using numerical simulations, several qualitatively new surprising features.
In particular, we elaborate on [Sec.~\ref{sec:ConfinmentAndAnnihilation}] how vortices, which are produced in synchronized bursts across large scales, are confined to move on codimension-1 domain boundaries separating long-lived counter-rotating regions, and how this confinement allows for the \emph{complete} annihilation of all vortices formed in each burst of defect production. 
We find this in both two and three spatial dimensions (in contrast to the gauged U(1) case~\cite{East:2022rsi}).
Furthermore, we find that the synchronized production and subsequent annihilation of a large number of vortices recurs \emph{periodically}, potentially over as many as hundreds of bursts. 
We also show that the periodicity found in our numerical simulations is well-captured by a simple (semi-)analytical expression.
Concluding this section, we explore the late-time evolution of the system for both small~[Sec.~\ref{sec:lateTimeStringEvol}] and large~[Sec.~\ref{sec:endstatelargec0}] initial $U(1)$ charge densities.
For small $\mathcal{J}_0$, we find that, toward late times, the system is dominated by (relativistic) angular-mode and semi-relativistic radial-mode radiation (having energy and momentum both comparable to the mass of the radial mode $\mu$), with few to no vortices present.
For large $\mathcal{J}_0$, the quasi-steady state of periodic vortex production likely evolves only on the timescale of the merger of counter-rotating regions, and persists longer than the total duration of our simulations.

Up to the end of Sec.~\ref{sec:dynamics}, our analysis proceeds in a non-expanding, Minkowski spacetime; in Sec.~\ref{sec:FLRW}, we extend this analysis to an expanding, radiation-dominated FLRW background spacetime.%
\footnote{%
    In this analysis, we take the $\Phi$ field to be a subdominant component of the total energy density of the Universe, which does not affect its overall expansion.
} %
After some preliminaries to establish the context for this computation, we derive some basic scaling results [Sec.~\ref{sec:FLRWanalysis}], and then discuss [Sec.~\ref{sec:FLRWnumerics}] whether and how our earlier conclusions are modified in this context, for both small~[Sec.~\ref{sec:FLRWsmallC0}] and large~[Sec.~\ref{sec:FLRWlargeC0}] initial $U(1)$ charge densities.
Most importantly, we find that the periodic bursts of string formation and annihilation (and therefore the associated gravitational-wave bursts) persist in an expanding universe, with a periodicity in the scale factor that can be predicted semi-analytically.
We discuss some phenomenological implications of our findings in Sec.~\ref{sec:stringburst}, focusing on gravitational-wave observables.
We conclude in Sec.~\ref{sec:conclusion}.

A number of appendices provide extra information.
In App.~\ref{app:cmtvortex}, we provide a concise review of vortices in condensed matter systems, which provides a useful heuristic framework to understand some of our observations.
In App.~\ref{app:simdetails}, we provide details of our numerical implementation.
In App.~\ref{app:mathieuAnalysis}, we provide further details of the analytical perturbative analysis presented in Sec.~\ref{sec:pert}. 
In App.~\ref{app:scalings}, we provide the derivations of some scaling results used in Sec.~\ref{sec:FLRW}.
Finally, in App.~\ref{app:burstp}, we give a derivation of a result for the spacing between sequential bursts of string production in FLRW spacetime that is used in Sec.~\ref{sec:FLRW}.

\section{Perturbation Growth to the Onset of Vortex Formation}
\label{sec:growth}

In this section we show that a spatially homogeneous scalar field subject to a wine-bottle potential, initially either with or without field-space angular momentum, is not a stable field configuration: rather, it exhibits (classical) parametric instabilities that lead to growth of spatial inhomogeneities. 
Some of the results presented in this section were previously discussed in, e.g.,~\cite{Dolgov:1989us,Traschen:1990sw,Kofman:1994rk,Kofman:1997yn,Co:2020dya}. 

For now, we ignore Hubble expansion and work in $(D+1)$-dimensional Minkowski spacetime (mostly minus convention; signature $1-D$).
The Lagrangian \eqref{eq:lagrangianintro} yields the equations of motion (EoM)
\begin{align}
    \ddot{\Phi} - \nabla^2 \Phi - \mu^2 \Phi + \lambda |\Phi|^2 \Phi &=0\:,    \label{eq:EoM} 
\end{align}
and its complex conjugate; here $\ddot{\Phi} = \partial_t^2\Phi$. 

Analytically, it turns out to be easier to analyse the dynamics of the system by decomposing the field into its amplitude $\rho$ and phase $\theta$ variables defined at \eqref{eq:lagrangianintro}. 
As our intention is to first study the growth of instabilities around a homogeneously evolving background, we perform a perturbative expansion:
\begin{align}
    \rho(t,\bm{x}) &\equiv \rho_0(t) + \epsilon\, \delta\rho(t,\bm{x})\:; \label{eq:FieldDefn2} \\
    \theta(t,\bm{x}) &\equiv \theta_0(t) + \epsilon \, \delta \theta(t,\bm{x})\: ,\label{eq:FieldDefn3a}
\end{align}
where $\epsilon$ is a (formal) small parameter and $\rho_0, \theta_0, \delta \rho, \delta\theta \in \mathbb{R}$.
Then,
\begin{align}
    \Phi &\equiv \Phi_0 + \epsilon \, \delta \Phi + \mathcal{O}(\epsilon^2)\:; \label{eq:FieldDefn3b}\\
    \Phi_0 = \Phi_0(t) &\equiv \rho_0(t)e^{i\theta_0(t)} \:; \label{eq:FieldDefn4}\\
    \delta \Phi = \delta \Phi(t,\bm{x}) &\equiv \Big( \delta \rho(t,\bm{x}) + i \rho_0(t) \delta \theta(t,\bm{x}) \Big) e^{i\theta_0(t)}\:.\label{eq:FieldDefn5}
\end{align} 
Substituting into \eqref{eq:EoM}, equating like powers of $\epsilon$ to zero, and looking separately at the real and imaginary parts of the resulting equations, we arrive at EoM for the background field and the linear perturbations.

In what follows, we examine the evolution of the background field [Sec.~\ref{sec:backgroundevo}] and linear perturbations, first analytically [Sec.~\ref{sec:pert}] and then numerically [Sec.~\ref{sec:linearizednumerics}], before returning to fully nonlinear simulations of \eqref{eq:EoM} [Sec.~\ref{sec:pertfullnonlinear}], and then concluding [Sec.~\ref{sec:Sec2Coda}].

\subsection{Background Field Evolution}
\label{sec:backgroundevo}

The homogeneous background field is governed by two coupled equations:
\begin{align}
    \ddot{\rho}_0 - \rho_0 \left[ \mu^2 + \dot{\theta}_0^2 \right] + \lambda \rho_0^3 &=0\:,    \label{eq:EoM0re} \\    
    \rho_0 \ddot{\theta}_0 + 2 \dot{\theta}_0 \dot{\rho}_0 &=0 \Rightarrow \partial_t\left( \rho_0^2 \dot{\theta}_0 \right) = 0 \Rightarrow \rho_0^2 \dot{\theta}_0 \equiv \mathcal{J}_0 = \text{constant}\:.
    \label{eq:EoM0im}
\end{align}
Substituting this conserved quantity, the $U(1)$ charge density $\mathcal{J}_0$, from \eqref{eq:EoM0im} into \eqref{eq:EoM0re} yields  
\begin{align}
    \ddot{\rho}_0 -  \mu^2 \rho_0 - \mathcal{J}_0^2 / \rho_0^3 + \lambda \rho_0^3 &=0\:,   \label{eq:EoM0re2}
\end{align}
which is equivalent to an effective one-dimensional particle-in-a-potential problem where $\rho_0$ maps to the particle position and the effective potential is 
\begin{align}
    V_{\text{eff}} = \frac{1}{4} \lambda \rho_0^4 - \frac{1}{2} \mu^2 \rho_0^2 + \frac{1}{2} \frac{\mathcal{J}_0^2}{\rho_0^2} + \frac{1}{4} \frac{\mu^4}{\lambda}\:, \label{eq:Veff}
\end{align}
where we have kept the same constant offset as for $V$ in \eqref{eq:V}.
There is also a conserved energy density for the background evolution: $\mathcal{E}_0 = \frac{1}{2} \dot{\rho}_0^2 + V_{\text{eff}}$.
We will also continue to define $v = \mu/\sqrt{\lambda}$; this is the $\rho_0$ field value for which $\ddot{\rho}_0=0$ when $\mathcal{J}_0 = 0$; i.e., the location of the bottom of the wine-bottle potential \emph{in the absence of rotational motion of the field}, which also satisfies $V_{\text{eff}}(\rho_0=v;\mathcal{J}_0=0)=0$. In the presence of a non-vanishing $\mathcal{J}_0$, the minimum of the potential is shifted to $\rho_{\min} > v$.

\begin{figure}
    \centering
    \includegraphics[width=\linewidth]{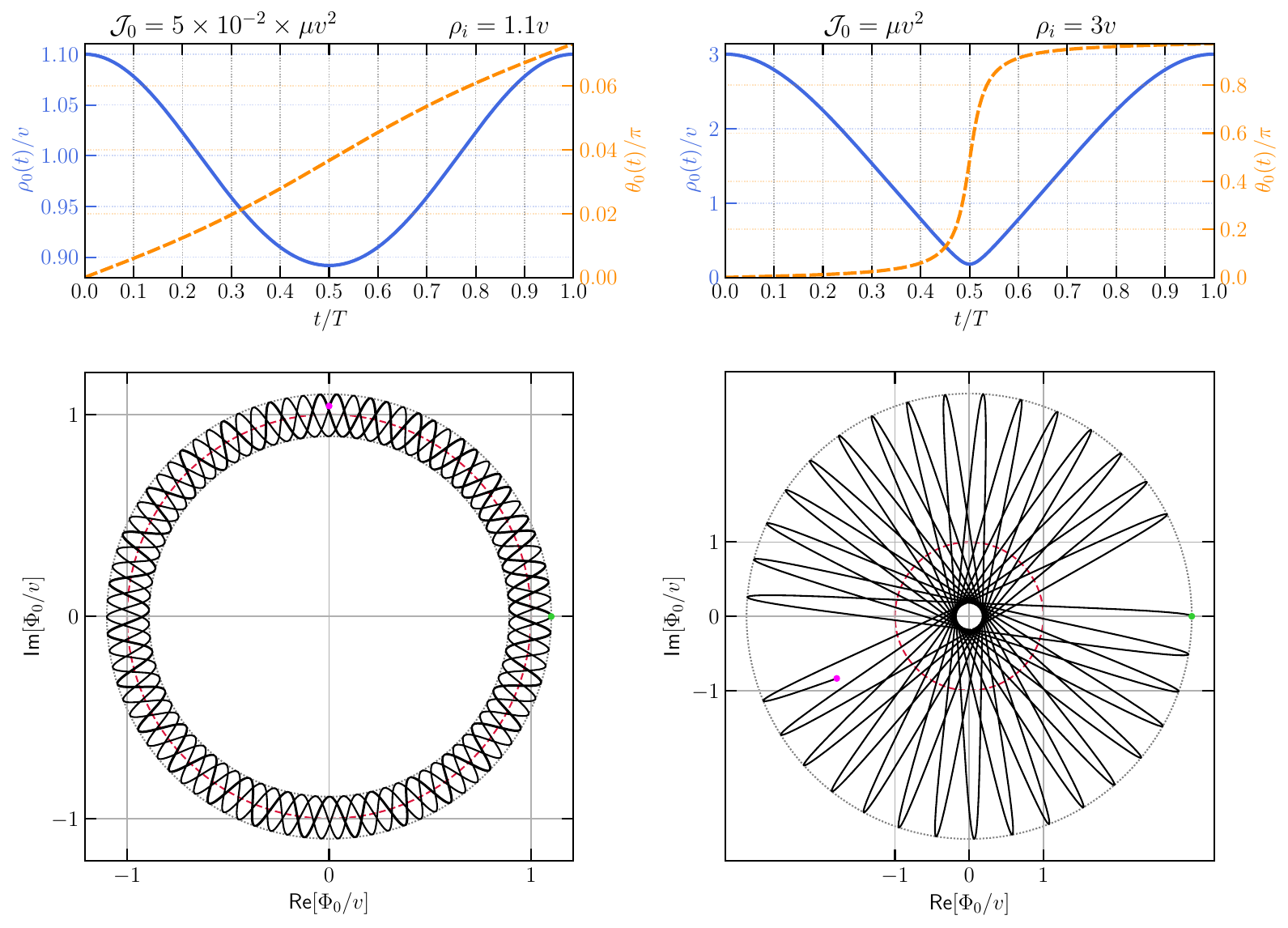}\\
    \caption{%
    \underline{\textsc{Top row:}}~The evolution of the background amplitude $\rho_0(t)$ [solid blue line; left axis scale] and phase $\theta_0(t)$ [dashed orange line; right axis scale] for two different cases: $\rho_i/v = 1.1$ and $\mathcal{J}_0 = 5\times 10^{-2} \times \mu v^2$ (left panel), and $\rho_i/v = 3.0$ and $\mathcal{J}_0 = \mu v^2$ (right panel).
    $T$~is the periodicity of the $\rho_0(t)$ oscillation in each case (which differs between the left and right panels).
    It is important to note that the vertical scales in the left and right panels are different.
    \underline{\textsc{Bottom row:}}~Field-space trajectories of the real and imaginary components of the background field $\Phi_0(t)$, for the same parameters as the panels in the top row. 
    The initial conditions are marked by the green dots, while the final evaluated points after some arbitrary duration of temporal evolution are marked by the purple dots.
    Given infinite time, the field is expected to ergodically explore the entirety of the set of complex field values $\Phi_0$ that lie between the (dotted gray) circles that mark the extremes of the $\rho_0 = |\Phi_0|$ field excursion.
    The red dashed circle indicates $\rho_0 = v$.
    All four plots assume $\lambda = 1$, $\dot{\rho}_0(0) \equiv  \dot{\rho}_i = 0$, and $\theta_0(0) \equiv \theta_i =0$.
   }
    \label{fig:bgndEvol}
\end{figure}

Although the background-field trajectory $\Phi_0(t)$ is completely specified by a choice of the two conserved quantities $\mathcal{J}_0$ and $\mathcal{E}_0$, supplemented with an initial field point on the trajectory fixed by $\rho_0(0) \equiv \rho_i$ and $\theta_0(0) \equiv \theta_i$ (for a total of 4 real initial conditions, as required for a second-order complex equation of motion), the temporal evolution of $\rho_0(t)$, $\theta_0(t)$ must be obtained numerically (but see, e.g., Appendix G of \cite{Gouttenoire:2021jhk} for semi-analytical solutions valid in certain limits); see Fig.~\ref{fig:bgndEvol} for some examples.
As expected, because the effective potential is neither Coulombic nor of harmonic-oscillator form, the evolution of $\Phi_0$ does not form a closed curve in complex field space; in the absence of perturbations, it would eventually densely explore the entire complex field region that lies between the extremes of the $\rho_0$ motion.

\subsection{Perturbations: Analytical Analysis}
\label{sec:pert}

Once a numerical solution for the homogeneous background field is obtained, we can study the perturbations around this solution.

\subsubsection{Governing Equations}

The perturbations are governed by the coupled second-order equations
\begin{align}
    \delta\ddot{\rho} - \nabla^2\delta\rho - 2 \frac{\mathcal{J}_0}{\rho_0} \delta \dot{\theta} - \frac{\mathcal{J}_0^2 + \mu^2 \rho_0^4 - 3\lambda \rho_0^6}{\rho_0^4} \delta \rho &= 0 \label{eq:deltaEoM1}\:,\\
    \delta\ddot{\theta} - \nabla^2\delta\theta - 2 \frac{\mathcal{J}_0\dot{\rho}_0}{\rho_0^4}  \delta \rho + 2 \frac{\dot{\rho}_0}{\rho_0} \delta\dot{\theta} + 2 \frac{ \mathcal{J}_0}{\rho_0^3} \delta \dot{\rho}&= 0 \label{eq:deltaEoM2}\:
\end{align}
where $\delta \ddot{\rho} \equiv \partial_t^2(\delta \rho)$, $\delta \dot{\theta} \equiv \partial_t(\delta \theta)$, etc.
Let us define 
\begin{align}
    \delta \sigma(t,\bm{x}) \equiv \rho_0(t) \delta\theta(t,\bm{x})\:, \label{eq:FieldDefn6}
\end{align}
and at the same time pass to the spatial Fourier domain:%
\footnote{%
    We abuse notation and use the same symbol for both the original perturbation field and its spatial Fourier transform.
    } %
\begin{align}
    \delta\ddot{\rho} + \left[ k^2-\mu^2 - \frac{\mathcal{J}_0^2}{\rho_0^4} +  3\lambda \rho_0^2 \right] \delta \rho + 2 \frac{\mathcal{J}_0}{\rho_0^3} \left[ \dot{\rho}_0 \delta\sigma - \rho_0 \delta\dot\sigma \right]  &= 0 \label{eq:deltaEoM1b}\:,\\
    \delta\ddot{\sigma} + \left[ k^2- \mu^2 - \frac{\mathcal{J}_0^2}{\rho_0^4} +  \lambda \rho_0^2 \right] \delta \sigma - 2 \frac{\mathcal{J}_0}{\rho_0^3} \left[ \dot{\rho}_0 \delta\rho - \rho_0 \delta\dot\rho \right]  &= 0 \label{eq:deltaEoM2b}\:,
\end{align}
where $k^2 \equiv k^ik^i$.
We can remove the single time derivatives on the perturbations by going to an $SO(2)$-rotated field basis:
\begin{align}
    \begin{pmatrix}
        \delta \rho \\
        \delta \sigma
    \end{pmatrix} 
    &\equiv
        \mathcal{R}[\theta_0(t)] 
    \begin{pmatrix}
        \delta \beta \\
        \delta \xi
    \end{pmatrix}\:, &
    \mathcal{R}[\gamma] &\equiv \begin{pmatrix} 
                        \cos\gamma & \sin \gamma \\
                        - \sin\gamma & \cos\gamma       
                    \end{pmatrix}\:, &
    \theta_0(t) &\equiv \mathcal{J}_0 \int_0^t \frac{dt'}{\rho_0(t')^2}\:.
\end{align}
Note that $\mathcal{R}[-\gamma] = (\mathcal{R}[\gamma])^{-1}$ and that the off-diagonal terms in the matrix $\mathcal{R}[\gamma]$ are proportional to $\mathcal{J}_0$ in the limit of small $\mathcal{J}_0$.
Then, taking appropriate linear combinations%
\footnote{%
 The linear combinations are, in obvious shorthand,
    \begin{equation*}
       \mathcal{R}[-\theta_0(t)] \cdot \begin{pmatrix} \eqref{eq:deltaEoM1b} \\ \eqref{eq:deltaEoM2b} \end{pmatrix} \:.   
    \end{equation*}
    } %
of \eqref{eq:deltaEoM1b} and \eqref{eq:deltaEoM2b}, we obtain
\begin{align}
    \partial_t^2 \begin{pmatrix} \delta \beta \\ \delta \xi \end{pmatrix} &+ [\Omega(t) ]^2 \begin{pmatrix} \delta \beta \\ \delta \xi \end{pmatrix} = 0\:, \label{eq:EoMosc}\\
    &[\Omega(t)]^2  \equiv \begin{pmatrix} 
                        k^2 - \mu^2 + \lambda [\rho_0(t)]^2 ( 2 + \cos[2\theta_0(t)] ) & \lambda [\rho_0(t)]^2 \sin[2\theta_0(t)] \\
                        \lambda [\rho_0(t)]^2 \sin[2\theta_0(t)] & k^2 - \mu^2 + \lambda [\rho_0(t)]^2 ( 2 - \cos[2\theta_0(t)] ) 
                        \end{pmatrix}\:. \label{eq:OmegaDefn}
\end{align}     
This is now a system of coupled oscillators with a time-dependent squared-frequency matrix $[\Omega(t)]^2$.
Of course, full understanding of this system can only be obtained by evaluating $[\Omega(t)]^2$ on the numerical background solution and then solving the resulting coupled system numerically. 

Note also that we can write 
\begin{align}
    [\Omega(t)]^2 & \equiv \mathcal{R}[-\theta_0(t)] \cdot \mathcal{D} \cdot \mathcal{R}[\theta_0(t)]\:,& 
    \mathcal{D} &\equiv \begin{pmatrix} k^2 - \mu^2 + 3\lambda [\rho_0(t)]^2 & 0 \\ 0 & k^2 - \mu^2 + \lambda [\rho_0(t)]^2\end{pmatrix}\:, \label{eq:Ddefn}
\end{align}     
where diagonalization by conjugation with the time-dependent $SO(2)$ rotation matrices $\mathcal{R}[\pm\theta_0(t)]$ gives rise to the matrix $\mathcal{D}$ whose diagonal entries are the time-dependent eigenvalues of $[\Omega(t)]^2$.
From this, we note that \eqref{eq:EoMosc} can be re-written as
\begin{align}
        \mathcal{R}[\theta_0(t)] \cdot \partial_t^2 \Big( \mathcal{R}[-\theta_0(t)] \cdot \mathbb{X} \Big) + \mathcal{D} \mathbb{X} &= 0 \qquad \text{ where } \qquad
        \mathbb{X} \equiv \begin{pmatrix} \delta \rho \\ \delta \sigma \end{pmatrix} \:.
        \label{eq:EoMosc2}
\end{align}
Because both the rotation matrices $\mathcal{R}[\pm\theta_0(t)]$ and the entries of $\mathcal{D}$ are time-dependent, we cannot however read off the full solution to \eqref{eq:EoMosc2} in any closed form from this information.

\subsubsection{Simplified System Analytics at Small \texorpdfstring{$\mathcal{J}_0$}{J0}}
\label{sec:SimplifiedAnalytics}

We can make non-trivial analytical progress in understanding some features of the evolution of a simplified version of the coupled system \eqref{eq:EoMosc2} for small values of $\mathcal{J}_0$.

First, we note that \eqref{eq:EoM0re2} makes it clear that when $\mathcal{J}_0 \ll \mu \rho_0^2$, small-amplitude excursions of the background field $\rho_0$ oscillate with an angular frequency $\sim \mathcal{O}(\mu)$.
In fact, as we show more carefully in App.~\ref{app:mathieuAnalysis}, writing $\zeta_0 \approx \rho_0(t)/v - 1 \ll 1$ as a small-amplitude oscillation (we define $\zeta_0$ more precisely in App.~\ref{app:mathieuAnalysis}), and linearizing \eqref{eq:EoM0re2} also in the $\mathcal{J}_0 \ll \mu \rho_0^2$ limit, we have
\begin{align}
    \partial_{\tilde{t}}^2 \zeta_0 + 2 \zeta_0 &\approx 0 \\
    \Rightarrow \zeta_0(\tilde{t}) &\approx \bar{\zeta}_0 \cos( \sqrt{2}\cdot \tilde{t} + \varphi_{\zeta}) \qquad \left[\zeta_0\ll 1,\: \mathcal{J}_0 \ll \mu \rho_0^2\right] \:,
\end{align}
where we defined $\tilde{t} \equiv \mu t$ to be a dimensionless time, $\bar{\zeta}_0$ is the amplitude of the $\zeta_0(t)$ oscillation, and $\varphi_\zeta$ is an arbitrary phase.
In the same limit $\mathcal{J}_0 \ll \mu \rho_0^2$, it is easy to read off from \eqref{eq:deltaEoM1b} and (\ref{eq:deltaEoM2b}) that%
\footnote{%
   Another way to see this is directly from \eqref{eq:EoMosc2}. 
   In the limit $\mathcal{J}_0 \ll \mu \rho_0^2$, the rotation matrices $\mathcal{R}$ exhibit only slow time dependence and can thus be pulled through the time derivatives with an error that vanishes in the $\mathcal{J}_0 \rightarrow 0$ limit.
} %
\begin{align}
    \partial_t^2 \mathbb{X} + \mathcal{D} \mathbb{X} \approx 0 \qquad \left[ \mathcal{J}_0 \ll \mu \rho_0^2\right]\:,
\end{align}
where $\mathbb{X}$ and $\mathcal{D}$ are as defined at \eqref{eq:Ddefn} and (\ref{eq:EoMosc2}).
But $\mathcal{D}$ is diagonal, so in terms of the dimensionless time $\tilde{t}$, this simplified system can be analysed as two independent oscillators, 
\begin{align}
    \partial_{\tilde{t}}^2 \mathbb{X}_i + \tilde{\omega}_i^2 \mathbb{X}_i \approx 0 \qquad [i = \sigma,\rho] \:, \label{eq:XiEqn}
\end{align}
where the dimensionless, time-dependent squared-frequencies are given by (recall: $\lambda \equiv \mu^2/v^2$)
\begin{align}
    \tilde{\omega}^2_i(\tilde{t}) &= \tilde{k}^2 - 1 + c_i \lambda [ \rho_0(t)/\mu ]^2 \label{eq:omegaTildeDefn} \\
    &\approx \left( \tilde{k}^2 - 1 + c_i \right) + 2 c_i \bar{\zeta}_0 \cos( \sqrt{2} \cdot \tilde{t} + \varphi_{\zeta}) \:,
\end{align}
where we have defined $\tilde{k} \equiv k/\mu$, employed $\bar{\zeta}_0 \ll 1$ for small-amplitude oscillations in the second line to drop the $(\bar{\zeta}_0)^2$ term, and defined
\begin{align}
    c_\rho &= 3\:;& c_\sigma &= 1 \:. \label{eq:ciDefn}
\end{align}

We have thus been able to cast the approximate EoM for the perturbations into the form $\partial_{t'}^2 f + [\omega'_i]^2 f= 0 $ where $[\omega_i']^2 \equiv \omega_{0,i}^2 + 2\epsilon_i \cos( \sqrt{2} (t' - t'_0) )$, otherwise known as the Mathieu equation,%
\footnote{%
    This is a slightly non-standard form of the Mathieu equation; it can easily be re-cast to the standard forms by means of rescaling the time variable and the parameters.
} %
which famously exhibits both narrow and broad parametric resonance instability phenomena, on which a vast body of literature exists (see, e.g., \cite{Dolgov:1989us,Traschen:1990sw,Kofman:1994rk,Kofman:1997yn,MagnusWinkler,Kovacic:2018weg}).
We undertake a detailed analysis of the resulting Fourier-mode instabilities in App.~\ref{app:mathieuAnalysis}. 

The results of that analysis, which we will now summarize, are that both the $\delta \sigma$ and $\delta \rho$ perturbations can exhibit instabilities, depending on the value of $\tilde{k}$.
For $\bar{\zeta}_0 \ll 1$, the $\delta \sigma$ perturbation is unstable via a narrow parametric resonance for (at least) the following band:
\begin{align}
    \sqrt{ \tfrac{1}{2} - \bar{\zeta}_0 } \leq \tilde{k} \leq \sqrt{ \tfrac{1}{2} + \bar{\zeta}_0  } \qquad [\delta \sigma \text{ unstable}]\:. \label{eq:sigmaUnstable}
\end{align}
The largest instability $e$-folding growth rate in this band is $\tilde{\Gamma} \sim \bar{\zeta}_0/\sqrt{2}$ and it should occur for $\tilde{k} = 1/\sqrt{2}$.
On the other hand, the $\delta \rho$ perturbation is unstable via a narrow parametric resonance for (at least) the band
\begin{align}
    \Rightarrow 0 &\leq \tilde{k} \leq \sqrt{\frac{15}{2}} \cdot \bar{\zeta}_0\qquad [\delta\rho \text{ unstable}]\:. \label{eq:rhoUnstable}
\end{align}
The largest instability $e$-folding growth rate in this band is $\tilde{\Gamma} \sim (3\bar{\zeta}_0/2)^2/\sqrt{2}$ and it should occur for $\tilde{k} = \sqrt{3}\cdot\bar{\zeta}_0$.
Both perturbations may also exhibit other instability bands for $\bar{\zeta}_0\ll 1$; moreover, at larger $\bar{\zeta}_0$, both perturbations can instead exhibit broad parametric resonance.

As the field excursion $\bar{\zeta}_0$ increases from very small values, the $\tilde{k}$ ranges that define the narrow resonance bands where $\delta \rho$ and $\delta \sigma$ are unstable will grow in size and eventually merge when the upper limit of the $\delta \rho$ resonance band hits the lower limit of the $\delta \sigma$ band.
This happens when $\sqrt{15/2}\cdot\bar{\zeta}_0 = \sqrt{1/2-\bar{\zeta}_0} \Rightarrow \bar{\zeta}_0 = 1/5 = 0.2$, with the merger occurring at $\tilde{k} = \sqrt{3/10} \sim 0.55$.
Note that this conclusion is marginally inconsistent with the $\delta \sigma$ resonance remaining narrow ($\bar{\zeta}_0 \lesssim 1/6$, as discussed in App.~\ref{app:mathieuAnalysis}), but we expect that this analysis should still be approximately correct.
For larger field excursions, we expect that ranges of $\tilde{k}$ will exist in which both $\delta \rho$ and $\delta \sigma$ will be simultaneously parametrically resonant (possibly in a mix of narrow and broad resonances).

Let us summarize the important qualitative lessons we have learned from the analysis of the simplified system in the limit where the field excursion is small ($\bar{\zeta}_0\ll1$) and the background field Noether charge is also small ($\mathcal{J}_0\ll \mu v^2$): (a) one or other of $\delta \rho$ (at small $\tilde{k}$) or $\delta \sigma$ (for $\tilde{k} \approx 1/\sqrt{2}$) perturbations can grow via narrow  parametric resonance when $\bar{\zeta}_0$ is sufficiently small; but (b) both perturbations may be simultaneously subject to parametric resonance growth for certain ranges of $\tilde{k}$ for larger (although still absolutely small) field excursions $\bar{\zeta}_0$.
Additionally, (c) all modes $\tilde{k} \lesssim 1$ can be unstable in the latter regime. 
Moreover, (d) there are broad parametric instabilities present that will appear and persist when $\bar{\zeta}_0$ is larger.
Although our understanding developed here is predicated on the simplified system evolution, in the next section we show numerically that these qualitative features also appear for the full system of perturbations in the limit of small $\mathcal{J}_0$ and $\bar{\zeta}_0$, and that these instabilities also persist for larger $\mathcal{J}_0$ and $\bar{\zeta}_0$.

A final note: because the full field evolution is nonlinear, once the perturbations grow large enough, the presence of even a single unbounded growing Fourier mode is generally sufficient to conclude that the homogeneous / background field evolution is not a stable solution for the simplified system.

\subsection{Numerical Analysis of the Full Linearized System}\label{sec:linearizednumerics}

Let us now examine how well the qualitative expectations and intuition we have developed via the analysis of the simplified system in the preceding sub-section, is actually borne out when we analyze the linearized system \eqref{eq:EoMosc} numerically.

We numerically integrate \eqref{eq:EoMosc} for various values of $k$, starting from $\delta \beta(0) = \delta \xi(0) = 10^{-3} v$ and $\partial_t \delta\beta(0) = \partial_t \delta\xi(0) = 0$, subject to background field evolution for a variety of choices of $\rho_0(0)\equiv \rho_i$ and $\mathcal{J}_0$ assuming always that $\dot{\rho}_0(0) \equiv \dot{\rho}_i = 0$ and $\theta_0(0)\equiv \theta_i =0$.
We also set $\mu=1$ and $v=1$ (i.e., $\lambda = 1$). 
We integrate these equations until one of two conditions is met: either $|\delta \Phi| = \sqrt{(\delta \beta)^2 + (\delta\xi)^2} = 10^{-1} v$ (i.e., sufficient growth for unstable Fourier modes), or $\tilde{t} = 10^3$ (i.e., a time cutoff for slowly growing or stable Fourier modes).
In either case, let us call the temporal end point of the integration $\tilde{t}_{\text{max}}$.
Even for unstable modes, the evolution of $|\delta \Phi|$ is not monotonic; it instead oscillates in amplitude within a growing envelope. 
In order to diagnose perturbatively unstable modes and extract their exponential growth rates, we proceed as follows: we locate all the \emph{local} temporal maxima of $|\delta \Phi|$ that occur within the duration $\delta < \tilde{t}/\tilde{t}_{\text{max}} \leq 1$ where $\delta =0.5, 0.75$ (the value used depends on the exact case and $|\delta \Phi|$ behavior).
This procedure yields data $(\tilde{t}_j,\ln|\delta \Phi|_j)$ for each local maximum $j$; we fit a straight line to these local maxima data to extract the late-time natural-log exponential growth rate $\tilde{\Gamma}$, measured in units of $\mu$.
That is, the e-folding rate: $|\delta \Phi| \propto \text{exp}[ \tilde{t} \tilde{\Gamma} ] = \text{exp}[t \mu \tilde{\Gamma}] = \exp[t \Gamma]$ at late times, where $\Gamma = \mu \tilde{\Gamma}$.
The larger the value of $\tilde{\Gamma}$, the more unstable the mode; however, for instances where we find $\tilde{\Gamma} \sim 0$, our computational approach precludes us from excluding the possibility that an instability would develop some time after $\tilde{t} \gtrsim 1000$, so results with $\tilde{\Gamma} \ll 10^{-3}$ should not necessarily be interpreted as completely stable.

\begin{figure}[p]
    \centering
    \includegraphics[width=0.75\linewidth]{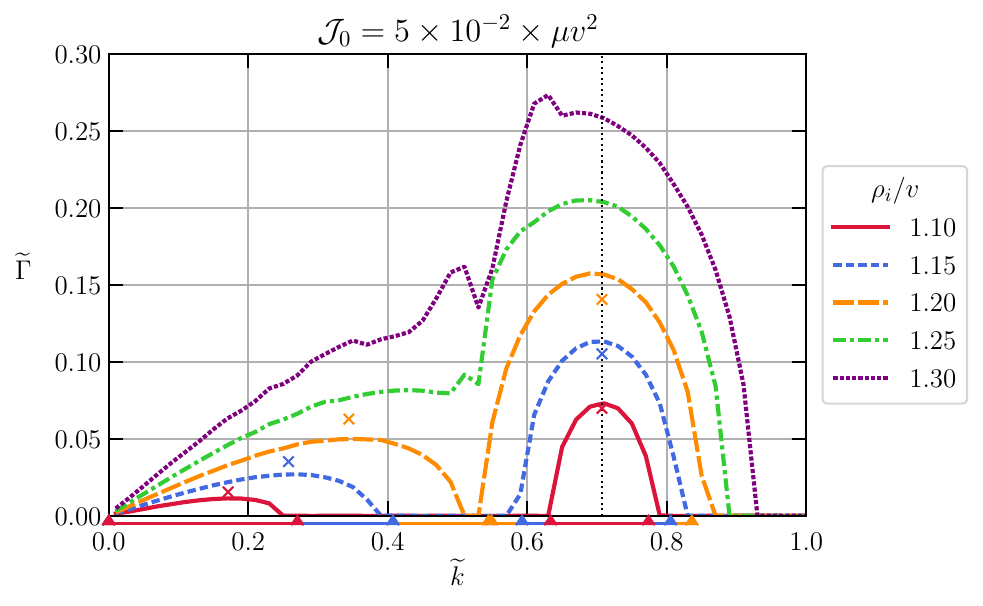}
    \caption{%
    The $e$-folding growth rate $\tilde{\Gamma}$ for the magnitude of unstable perturbations $\delta \Phi$ around the background field $\Phi_0$, plotted as a function of the Fourier-mode momentum $\tilde{k}$ assuming $\lambda = 1$ and $\mathcal{J}_0 = 5\times 10^{-2} \times v^2\mu$.
    These results are plotted for various values of $\rho_i/v$, as denoted in the legend. 
    The triangular carats joined with lines just below the horizontal axis show the simple analytical predictions discussed in the main text for the range(s) of $\tilde{k}$ that are unstable for the cases of $\rho_i/v\in\{ 1.10,1.15,1.20\}$, while the crosses give the corresponding predictions for the locations of the peak growth rates in each of the unstable bands.
    The vertical dotted black line denotes $\tilde{k} = 1/\sqrt{2}$.
    Regions with exponentially unstable perturbations that are identifiably growing prior to $\tilde{t} = 10^3$ are plotted with $\tilde{\Gamma}>0$.
    As discussed in the main text, unstable regions grow in size with increasing $\rho_i/v$.
    Note however that $\tilde{\Gamma}=0$ here means only that a mode has not exhibited identifiable exponential instability before $\tilde{t}=10^3$; such modes may in fact be absolutely stable, but we cannot make that conclusion based on these numerical results.
    The small-scale variability in the curves at larger values of $\tilde{\Gamma}$ (i.e., fast growth) is a numerical artifact associated with our extraction technique for the rate of exponential growth of the $|\delta \Phi|$ envelope.
    }
    \label{fig:growthRateSmall}
\end{figure}

\paragraph{Small Field Excursions, \texorpdfstring{$\bm{\mathcal{J}_0\ll \mu v^2}$}{J0 << mu v^2}}
We first look at a limit in which we expect the simplified analytical analysis of Sec.~\ref{sec:SimplifiedAnalytics} to be accurate, by fixing $\mathcal{J}_0= 5\times 10^{-2} \mu v^2 \ll \mu v^2$ and taking various values of $\rho_i/v \in \{ 1.1,1.15,1.2,1.25,1.3\}$, corresponding respectively to $\bar{\zeta}_0 \in \{ 0.099, 0.15, 0.20 , 0.25 ,\linebreak 0.30 \}$ (see App.~\ref{app:mathieuAnalysis} for the precise definition of $\bar{\zeta}_0$). 
The results for the numerical analysis of the Fourier-mode stability for this case are shown in Fig.~\ref{fig:growthRateSmall} for $10^{-2} \leq \tilde{k} \leq 1$. 

\begin{figure}[p]
    \centering
    \includegraphics[width=0.32\linewidth]{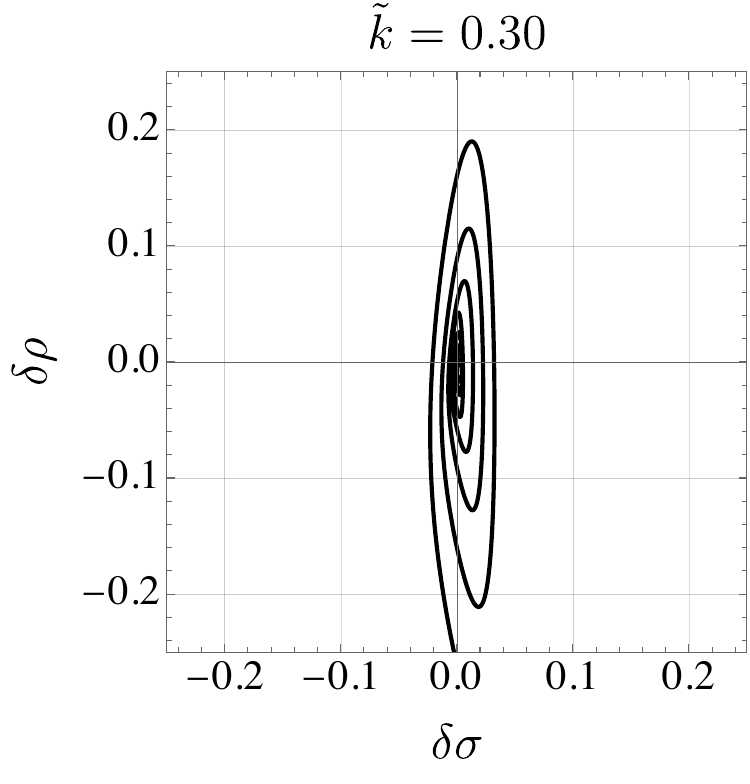}
    \includegraphics[width=0.32\linewidth]{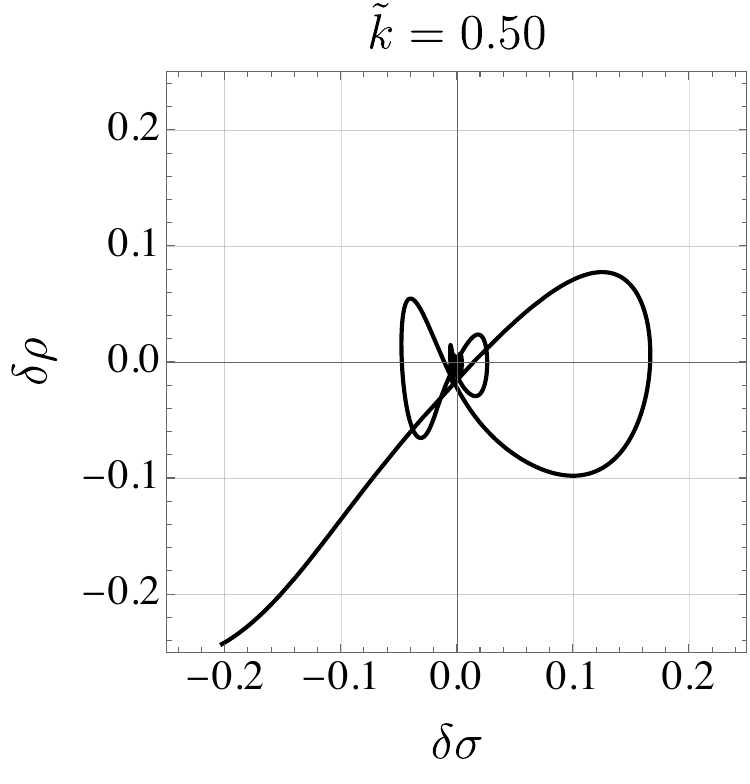}
    \includegraphics[width=0.32\linewidth]{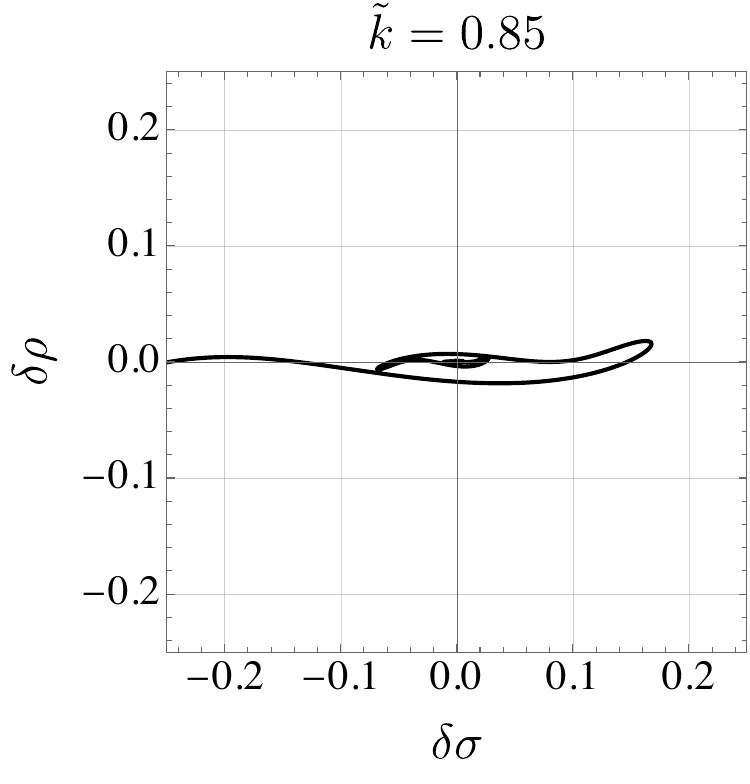}
    \caption{%
    Configuration-space trajectories for the $\delta\sigma(t)$ and $\delta\rho(t)$ components of the $\delta \Phi$ perturbation [see \eqref{eq:lagrangianintro}, \eqref{eq:FieldDefn2}--\eqref{eq:FieldDefn5}, and \eqref{eq:FieldDefn6}], for three different spatial Fourier modes (as annotated, from left to right) $\tilde{k} = 0.30$, $0.50$, and $0.85$.
    These results assume parameters $\lambda = 1$, $\rho_i/v = 1.3$, and $\mathcal{J}_0=5\times 10^{-2} \times \mu v^2$.
    As explained in the main text, these plots demonstrate that the $\delta\sigma(t)$ and $\delta\rho(t)$ perturbation components exhibit qualitatively different relative growth rates depending on which parametric instability band is accessed by a given spatial Fourier mode of the perturbation.
    The vertical and horizontal scales in each panel are the same in order to facilitate visual comparison.}
    \label{fig:ConfigSpacePerturbation}
\end{figure}

For $\bar{\zeta}_0 \lesssim 0.2$, we see the existence of two separate resonance bands for $\tilde{k}$: one for all $\tilde{k}$ less than some threshold, and the other around $\tilde{k} \sim 1/\sqrt{2}$, in agreement with expectations from Sec.~\ref{sec:SimplifiedAnalytics}.
Also note that the latter band has a larger maximum growth rate than the former, which is in qualitative agreement with the fact that the primary narrow resonance band (accessed in the latter band) is more efficient at driving growth than the secondary band (accessed in the former band); see App.~\ref{app:mathieuAnalysis}.

Specifically, consider $\rho_i/v=1.10$.
The simplified analytical analysis tells us to expect that the lower band will span $0 \lesssim \tilde{k} \lesssim 0.27$ with $\delta \rho$ more unstable, while the upper band should span the range $0.63 \lesssim \tilde{k} \lesssim 0.77$ with $\delta \sigma$ more unstable; additionally, the lower band is expected to have a peak growth rate $\tilde{\Gamma} \sim 1.6\times 10^{-2}$ at $\tilde{k} \sim 0.17$ while the upper band is expected to have a peak growth rate $\tilde{\Gamma} \sim 7\times 10^{-2}$ at $\tilde{k} \sim 0.7$.
This all matches well with the numerical results in Fig.~\ref{fig:growthRateSmall}, which also shows these predictions for the cases of $\rho_i/v=1.15,1.20$ graphically; they are also in reasonable agreement (although $\bar{\zeta}_0\sim 0.2$ is beginning to probe the limits of the $\bar{\zeta}_0\ll1$ assumption used to derive the simplified analytics).
For $\bar{\zeta}_0 \gtrsim 1.20$, the bands have merged, with the merger happening near $\tilde{k}\sim 0.5$, again in reasonable agreement with the merger values of $\bar{\zeta}_0$ and $\tilde{k}$ predicted in Sec.~\ref{sec:SimplifiedAnalytics}.

Moreover, for the $\bar{\zeta}_0 \sim 0.3$ case, we show the configuration-space evolution of the two perturbation components for $\tilde{k} = 0.3, 0.50, 0.85$ in Fig.~\ref{fig:ConfigSpacePerturbation}.
At this parameter point, the simplified analytics of Sec.~\ref{sec:SimplifiedAnalytics} indicate that $\delta \sigma$ should be unstable for $0.45 \lesssim \tilde{k} \lesssim 0.90$, while $\delta \rho$ should be unstable for $0\lesssim \tilde{k} \lesssim 0.82$, so that all modes with $\tilde{k} \lesssim 0.9$ should be unstable [cf.~Fig.~\ref{fig:growthRateSmall}].
The numerical results are in reasonable qualitative agreement with these predictions: for $\tilde{k} = 0.30$, Fig.~\ref{fig:ConfigSpacePerturbation} shows that the $\delta\rho$ perturbation exhibits larger growth than the $\delta \sigma$ one.
For $\tilde{k} = 0.50$, both $\delta \rho$ and $\delta \sigma$ show similar instability.
Finally, for $\tilde{k} = 0.85$, $\delta \sigma$ exhibits larger growth as compared to $\delta \rho$. 

\paragraph{Small Field Excursions, \texorpdfstring{$\bm{\mathcal{J}_0\gtrsim \mu v^2}$}{J0 >~ mu v^2}; and Large Field Excursions}
We now turn to the numerical analysis of two cases where we do not expect good agreement with the simplified analytics of Sec.~\ref{sec:SimplifiedAnalytics}, in order to see if instability band(s) persist.

\begin{figure}
    \centering
    \includegraphics[width=0.49\linewidth]{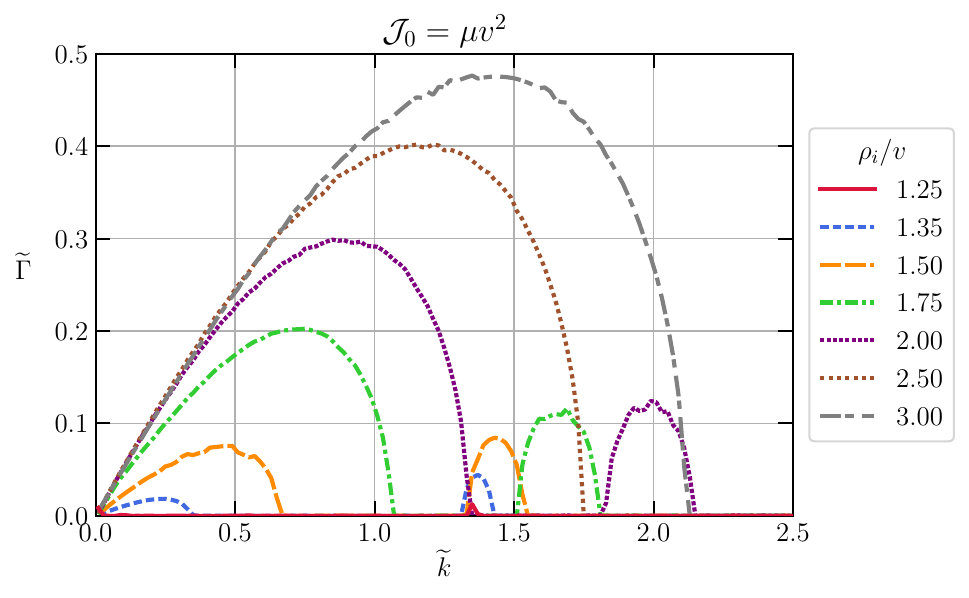}
    \includegraphics[width=0.49\linewidth]{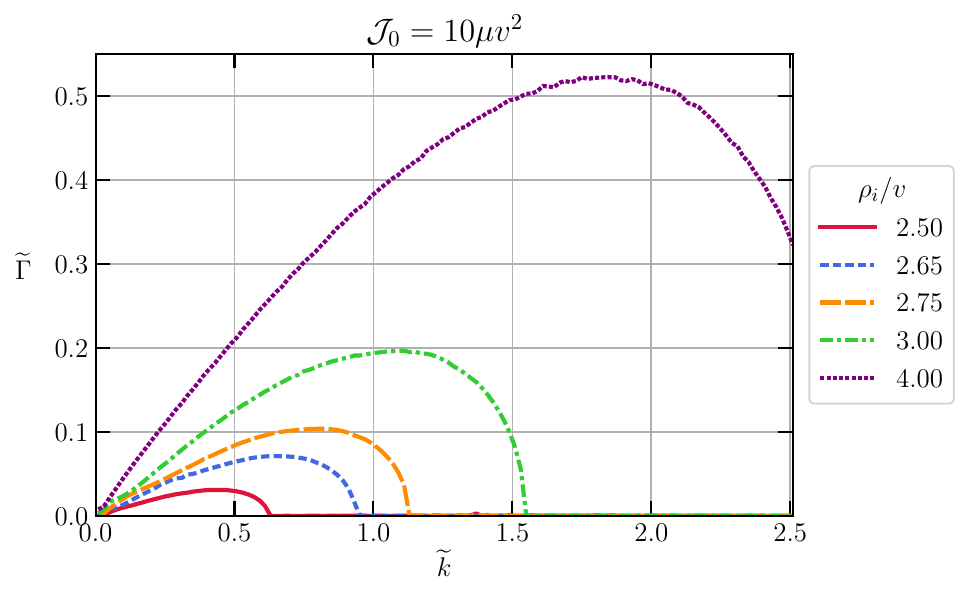}
    \caption{As for Fig.~\ref{fig:growthRateSmall}, but for $\mathcal{J}_0 = v^2\mu$ (left panel) and $\mathcal{J}_0 = 10 v^2\mu$ (right panel), and the values of $\rho_i/v$ denoted in each legend. 
    }
    \label{fig:growthRateLarge}
\end{figure}

Results for $\mathcal{J}_0 = \mu v^2$ with $\lambda = 1$ and various values of $\rho_i/v$ are shown in the left panel of Fig.~\ref{fig:growthRateLarge} for $10^{-2} \lesssim \tilde{k} \lesssim 2.5$.
The locations of the instability bands here are not well predicted by the simplified analytics of Sec.~\ref{sec:SimplifiedAnalytics}; however, importantly, we see that there are instabilities for certain values of $\tilde{k} \lesssim \text{few}$, even for this larger value of $\mathcal{J}_0$.
Moreover, the instability bands for certain low-lying values of $\tilde{k}$ persist to even larger $\mathcal{J}_0 = 10 \mu v^2$, as seen in the right panel of Fig.~\ref{fig:growthRateLarge}, albeit for larger $\rho_i/v$ (note: for $\rho_i/v\sim 2.50$, $\rho_0(t)$ is actually executing small-amplitude oscillations about a large average value $\bar{\rho}_0 \approx \rho_{\min}\gg v$; see also the discussion about the case of large $\mathcal{J}_0$ in Sec.~\ref{sec:LargeJ0discussion} later). 
Although not evidenced by the results shown here, we find that such unstable bands persist as $\mathcal{J}_0$ increases even further.

The methodology developed in this subsection can also be used to obtain two useful limiting behaviors at large $\mathcal{J}_0$, which we wish to highlight before closing this subsection. 
First, as can be already seen from Fig.~\ref{fig:growthRateLarge}, for a large (but fixed) $\mathcal{J}_0$ (and hence fixed $\rho_{\min}$), the value of $\tilde{k}$ for which the growth rate is the fastest increases as $\rho_i/v$ increases. 
Second, for a fixed $(\rho_i - \rho_{\min})/v$, the value of $\tilde{k}$ for which the growth rate is the fastest increases, albeit slowly, as $\mathcal{J}_0$ (or, almost equivalently, $\rho_{\min}/v$) increases.

\subsection{Fully Nonlinear Simulations}
\label{sec:pertfullnonlinear}

Finally, we complete the analysis of the parametric resonance by performing time-domain numerical simulations of the full nonlinear system \eqref{eq:EoM}.
To that end, we discretize the classical field equation \eqref{eq:EoM} in Cartesian coordinates on a two- or three-dimensional spatial grid covering a box of side length $2L$ and impose periodic boundary conditions. 
The initial conditions, $\Phi_i(\bm{x})$ and $\dot{\Phi}_i(\bm{x})$, are chosen to describe phase-coherent field configurations with average radial displacement $\rho_i$ and non-vanishing, spatially homogeneous charge density $\mathcal{J}_0$. 
This average field configuration is supplemented by a scale-invariant Gaussian random field $\mathcal{N}(\bm{x})$ with variance $\sigma_{\rho_i}$ and vanishing mean. 
Ultimately, this implies%
\footnote{%
    These initial conditions are of course slightly tuned in that we have assumed that all field values start with the same phase, but with different radial field values and initial angular speeds, such that $\mathcal{J}_0$ is homogeneous.
    More general perturbed initial conditions would also consider an initial small perturbation to the phase (i.e., the initial $\Phi_i$ field values would be sampled in a ball of complex field values centered on $\rho_i$).
    Because of nonlinearities and the different angular speeds initially imposed, we expect such phase differences to appear naturally from these initial conditions after a short amount of temporal evolution.
    Overall, these slightly tuned initial conditions are unlikely to yield qualitatively different results from the more general case.
    } %
$\Phi_i(\bm{x})=\rho_i+\mathcal{N}(\bm{x})$ and $\dot{\Phi}_i(\bm{x})=i\mathcal{J}_0[\rho_i+\mathcal{N}(\bm{x})]^{-1}$.
While our simulations are performed in the two real variables $\text{Re}[\Phi(t,\bm{x})]$ and $\text{Im}[\Phi(t,\bm{x})]$ (and hence are robust against issues associated with the ill-defined multi-valued angular degree of freedom at the location of vortices, where $|\Phi|\rightarrow 0$), we do also find it convenient to reconstruct the radial $\rho(t,\bm{x})\equiv(\text{Re}[\Phi(t,\bm{x})]^2+\text{Im}[\Phi(t,\bm{x})]^2)^{1/2}$ and angular $\theta(t,\bm{x})\equiv\arctan(\text{Im}[\Phi(t,\bm{x})]/\text{Re}[\Phi(t,\bm{x})])\in [-\pi,\pi)$ degrees of freedom. 

The addition of the Gaussian%
\footnote{%
    The initial conditions we assume are of course scale-invariant, but this is likely not strictly necessary.
    Recall that the parametric resonance instabilities drive \emph{exponential} growth of the unstable modes.
    So long as the unstable modes are not exponentially suppressed in the initial conditions, they will thus still grow to dominate the perturbation spectrum in short order, and we would expect similar results, even if the initial perturbation spectrum is not white noise (so long as all initial perturbation modes are still small).
    Note however that while our specific initial conditions are inspired by similar initial conditions observed in the axion kinetic misalignment / kination literature, we do not specify a microphysical model for them here.
} %
random field $\mathcal{N}(\bm{x})$ ensures that all unstable spatial Fourier modes are excited with (roughly) the same initial amplitude and the same initial complex phase.%
\footnote{%
    On a numerical grid, the finite box size and spatial resolution impose lower and upper bounds, respectively, on resolved Fourier modes. 
    We ensure that the fastest growing (i.e., most relevant) modes are well-resolved in all cases considered. 
    See App.~\ref{app:simdetails} for further details.} %
The evolution of the system from the initial conditions follows the background evolution, discussed in Sec.~\ref{sec:backgroundevo}, for sufficiently small $\sigma_{\rho_i}$ at early times. 
That is, early on in the evolution, $\langle\rho(t,\bm{x})\rangle\approx\rho_0(t)$ and $\langle\theta(t,\bm{x})\rangle\approx\theta_0(t)$ as defined in the previous section.
Here, we denote the spatial average by~$\langle\ldots\rangle$; furthermore, for brevity, we drop the explicit spatial and time dependencies of $\rho(t,\bm{x})\rightarrow \rho$ and $\theta(t,\bm{x})\rightarrow \theta$ unless stated otherwise. 
The energy densities of these respective degrees of freedom, 
\begin{align}
    e_\theta=k_\theta+g_\theta, & & e_\rho = k_\rho+g_\rho+V(\rho)\: ,
\end{align}
split into kinetic, $k_{\theta,\,\rho}$, and gradient energies, $g_{\theta,\,\rho}$:
\begin{align}
    k_\theta=\frac{1}{2}\rho^2\dot{\theta}^2, & & k_\rho=\frac{1}{2}\dot{\rho}^2, & & g_\theta=\frac{1}{2}\rho^2\delta^{ij}\partial_i\theta\partial_j\theta, & & g_\rho = \frac{1}{2}\delta^{ij}\partial_i\rho\partial_j\rho \:.
\end{align}
Note also that the potential energy density, $V(\rho)=-\mu^2\rho^2/2+\lambda \rho^4/4+\mu^4/(4\lambda)$, is defined such that $V(\rho)\geq 0$. 
Finally, the volume-averaged energy densities in $D\in\{2,3\}$ spatial dimensions are then
\begin{align}
    K_\theta= (2L)^{-D}\int_{V_b} d^D\!x\, k_\theta \: ,
    \label{eq:energydensities}
\end{align}
and analogously for the other energy densities, yielding the averages $K_\rho, G_\theta, G_\rho$, and $V_\rho$;
here, $V_b=(2L)^{D}$ is the volume of the simulation box. 
See App.~\ref{app:simdetails} for further details on the numerical implementation, convergence behavior, and parameter choices.
We summarize for reference in Tab.~\ref{tab:systems} the parameter values for which we have run simulations (most of which are only described in later sections of the paper), as well as the figures in which the results for each choice can be found.

\begin{table}[t]
    \centering
    \begin{tabular}{c|c|c|c|c|c|c}
        Figure(s) & Animation & $D$ & FLRW & $\mathcal{J}_0/(v^2\mu)$ & $\rho_i/v$ & $\sigma_{\rho_i}/\rho_i$ \\ \hline
        \ref{fig:2dexpgrowth}, \ref{fig:counterrot}, \ref{fig:vortexformation}, \ref{fig:smallcharge}, \ref{fig:stringmove}, \ref{fig:largecirculation_energy}, \ref{fig:merger_conter_rot} & \cmark & $2$ & \xmark & $0.1$ & $5$ & $10^{-2}$ \\
        \ref{fig:cosmo2d_rho3}, \ref{fig:cosmo_counter} & \xmark & $2$ & \cmark &$0.1$ & $5$ & $10^{-2}$ \\
        \ref{fig:stringsmallcirtulation}, \ref{fig:string_energy_small_circulation} & \xmark & $2$ & \xmark & $0.1$ & $\sqrt{2}+0.1$ & $10^{-2}$ \\
        \ref{fig:cosmo_largec0} & \xmark & $2$ & \cmark & $10^2$ & $6$ & $10^{-5}$ \\
        \ref{fig:2dexpgrowthLargec0}, \ref{fig:diff_rot}, \ref{fig:phidotlargec0}, \ref{fig:largec0energies} & \cmark & $2$ & \xmark & $10^3$ & $11.5$ & $10^{-6}$ \\ \hline
        \ref{fig:cosmo3d_rho3} & \xmark & $3$ & \cmark & $0.1$ & $3$ & $10^{-2}$ \\
        \ref{fig:formation3D}, \ref{fig:3d_string_anni} & \xmark & $3$ & \xmark & $0.1$ & $5$ & $10^{-2}$ \\
        \ref{fig:string_energy_small_circulation} & \cmark & $3$ & \xmark & $0.1$ & $\sqrt{2}+0.1$ & $10^{-2}$ \\ \hline
    \end{tabular}
    \caption{%
    For reference, we collect here information regarding the parameter sets used in the various simulations we consider in this work.
    `Figure(s)' indicates where graphical results can be found, `Animation' indicates whether~(\cmark) or not~(\xmark) a relevant animation is available at \cite{ZenodoVideos}, $D$ is the number of spatial dimensions, `FLRW' denotes whether the system is simulated in Minkowski spacetime~(\xmark) or expanding FLRW spacetime~(\cmark), $\mathcal{J}_0$ is the global $U(1)$ charge density for the homogeneous part of the initial conditions as defined at~\eqref{eq:EoM0im}, $\rho_i$ is the homogeneous initial radial-field displacement, and $\sigma_{\rho_i}$ is the amplitude of the Gaussian noise added to the initial radial-field displacement (see Sec.~\ref{sec:pertfullnonlinear} for discussion of initial conditions).}
    \label{tab:systems}
\end{table}

Since all Fourier modes are initially excited with the same amplitude, the fastest-growing mode dominates the time-domain dynamics. 
The unstable modes exhibit qualitatively different behavior in the small and large $\mathcal{J}_0$ regimes.
We consider these cases in turn in the following.

\subsubsection{Small Initial Charge Density \texorpdfstring{$\mathcal{J}_0$}{J0}}
\label{eq:Sec2smallC0}

\begin{figure}[t]
    \centering    
        \includegraphics[width=0.48\linewidth]{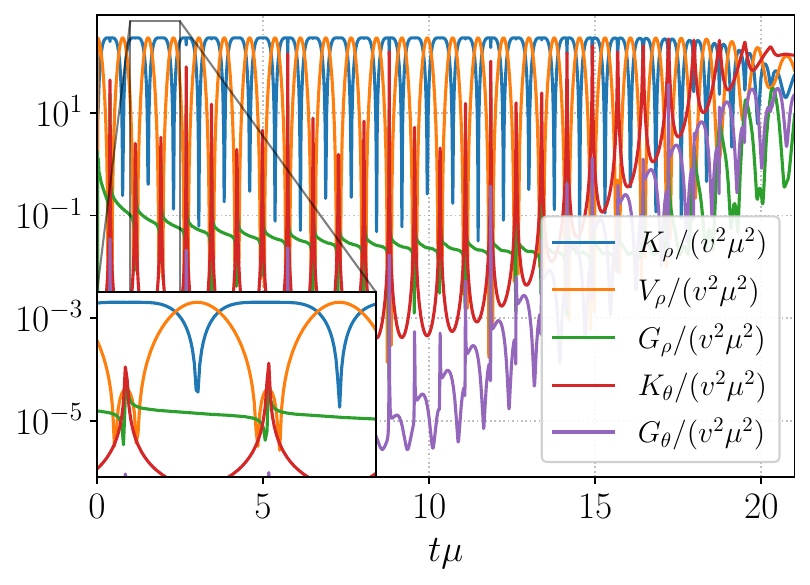}
    \hfill
    \includegraphics[width=0.49\linewidth]{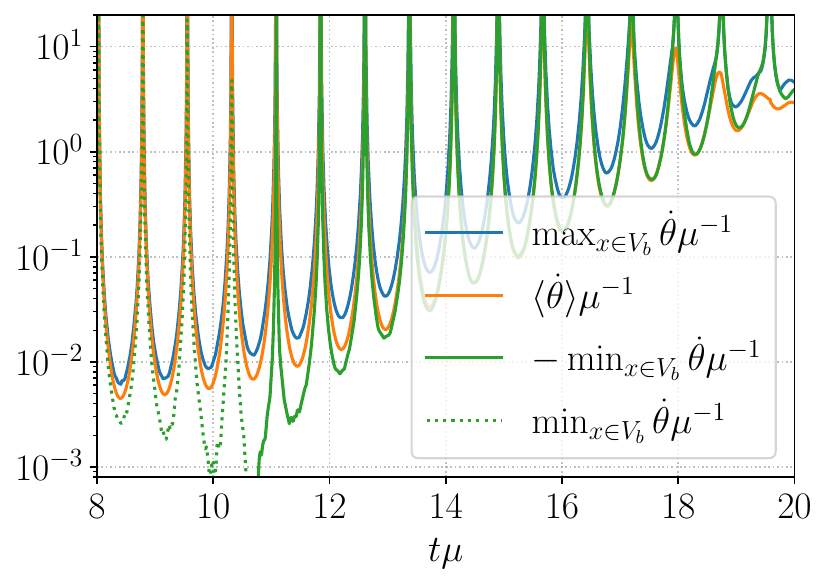}
    \caption{%
    Simulation of the 2-dimensional case ($D=2$) with a small initial angular field motion ($\mathcal{J}_0=0.1\mu v^2$), a large initial field displacement ($\rho_i/v=5$), and small spatial fluctuations ($\sigma_{\rho_i}=10^{-2}\rho_i$).
    \underline{\textsc{Left panel:}}~%
    The evolution in time of the different components of the volume-averaged energy densities of the system, both early and during the exponential growth of unstable perturbations. 
    The various colored lines are appropriately normalized plots of the kinetic (blue), potential (orange), and gradient (green) energies in the radial mode $\rho$; and the kinetic (red) and gradient (purple) energies in the angular mode $\theta$. 
    The inset shows the initial coherent rotation of the field around the field-space origin. 
    \underline{\textsc{Right panel:}}~%
    The evolution in time of the spatial average of the time derivative of the angular mode $\langle\dot{\theta}\rangle$ (orange), as well as its spatial maximum (blue) and the magnitude of its spatial minimum (green; dotted where positive, solid where negative).
    \ZenodoComment}
    \label{fig:2dexpgrowth}
\end{figure}

As a representative example in this regime illustrating the main qualitative features of the unstable modes, we focus on $D=2$, $\mathcal{J}_0=0.1\mu v^2$, $\rho_i/v=5$, and $\sigma_{\rho_i}=10^{-2}\rho_i$. 
The evolution of the various terms contributing to the total system energy is shown in the left panel of Fig.~\ref{fig:2dexpgrowth}. 
Due to the small variance $\sigma_{\rho_i}$, the radial and angular degrees of freedom are highly spatially homogeneous at early times. 
Here the field evolves coherently as described in Sec.~\ref{sec:backgroundevo} (see also the inset of the left panel of Fig.~\ref{fig:2dexpgrowth}): the radial mode starts out with large potential energy since $\rho_i/v=5$ [corresponding to $V(\rho_i)/V(0)=576$], rolls down the potential and passes close to the origin in field space. 
There, the angular mode changes rapidly from $\langle\theta\rangle\approx 0$ to $\langle\theta\rangle\approx \pi$, and the process repeats. 
The Gaussian noise in the initial conditions leads to non-vanishing, but initially subdominant, gradient energies.%
\footnote{\label{ftnt:KOdissipation}%
    The initial polynomical decay of $G_\rho$ in Fig.~\ref{fig:2dexpgrowth} is an artifact due to numerical dissipation of modes with wavelength comparable to the grid spacing; this has negligible impact on the system's evolution and can be ignored.
    See App.~\ref{app:simdetails} for further details on the numerical implementation.} %
As is evident from Fig.~\ref{fig:2dexpgrowth}, this coherent evolution is eventually broken up by an exponentially growing component (for $t\mu\gtrsim 10$ in this case).
The parametric resonance drives Fourier modes with wavenumber $\tilde{k}\sim\mathcal{O}(1)$, exponentially increasing spatial inhomogeneities in both the radial and angular degrees of freedom, and therefore in their gradient energies. 
As an indicator, in the right panel of Fig.~\ref{fig:2dexpgrowth} we show the average and spatial minima/maxima of $\dot{\theta}$ throughout the unstable growth. 
During the coherent phase of the system's evolution (i.e., for $t\mu\lesssim 9$), the average follows $\langle\dot{\theta}\rangle\approx\dot{\theta}_0(t)$ and $\min_{\bm{x}\in V_b}\dot{\theta}\approx \max_{\bm{x}\in V_b} \dot{\theta}$. 
Each passage close to the field-space origin is associated with a rapid increase and subsequent decrease of $\langle\dot{\theta}\rangle$ as the angular mode flips by $\approx\pi$ (recall: $\mathcal{J}_0 = \rho_0^2 \dot{\theta}_0$ is approximately conserved in the presence of small spatial field gradients). 
However, as energy is injected into $\tilde{k}>0$ modes, these indicators exhibit a clear exponential growth. 
As the unstable perturbations have no definite sign and the Gaussian random field excites the latter with both positive and negative amplitudes initially, a state of \textit{differential rotation} develops, where $\max_{\bm{x}\in V_b}[\dot{\theta}-\langle\dot{\theta}\rangle]>0>\min_{\bm{x}\in V_b}[\dot{\theta}-\langle\dot{\theta}\rangle]$. 
Moreover, not only do regions of differential rotation develop, but in this case the system rapidly evolves into a state with \textit{counter-rotating} regions, since $\max_{\bm{x}\in V_b}\dot{\theta},-\min_{\bm{x}\in V_b}\dot{\theta}>0$; this can be seen for $t\mu\gtrsim 11$ in the right panel of Fig.~\ref{fig:2dexpgrowth}.

\begin{figure}[t]
    \centering
    \includegraphics[width=0.32\linewidth]{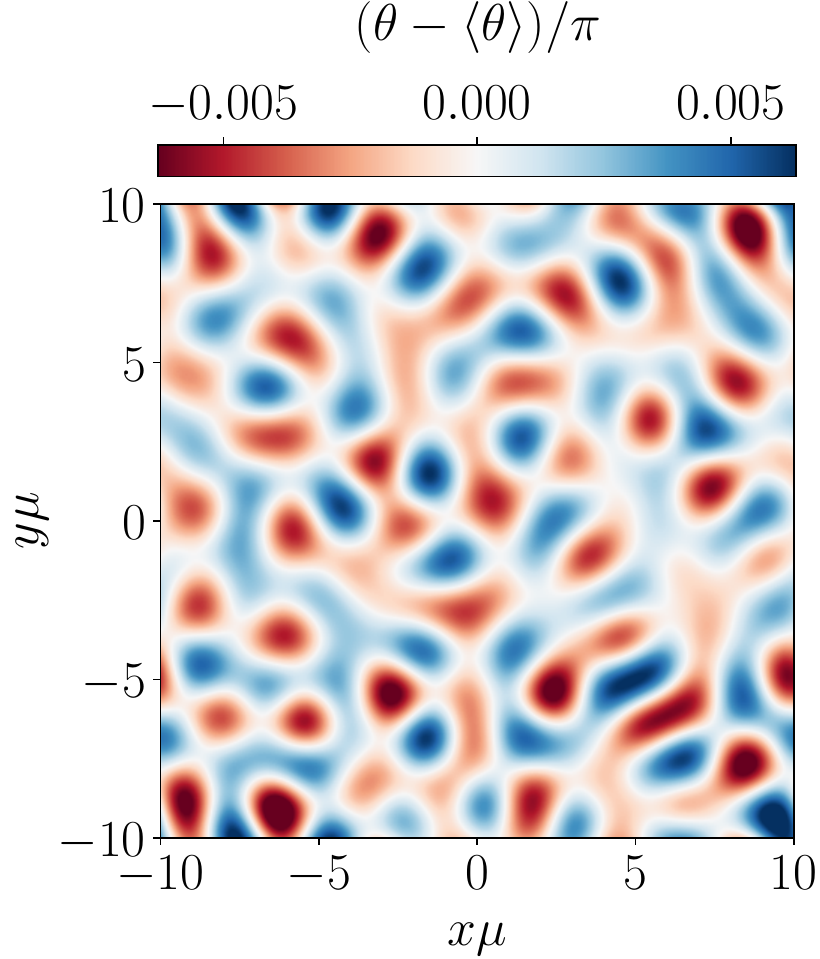}
    \hfill
    \includegraphics[width=0.32\linewidth]{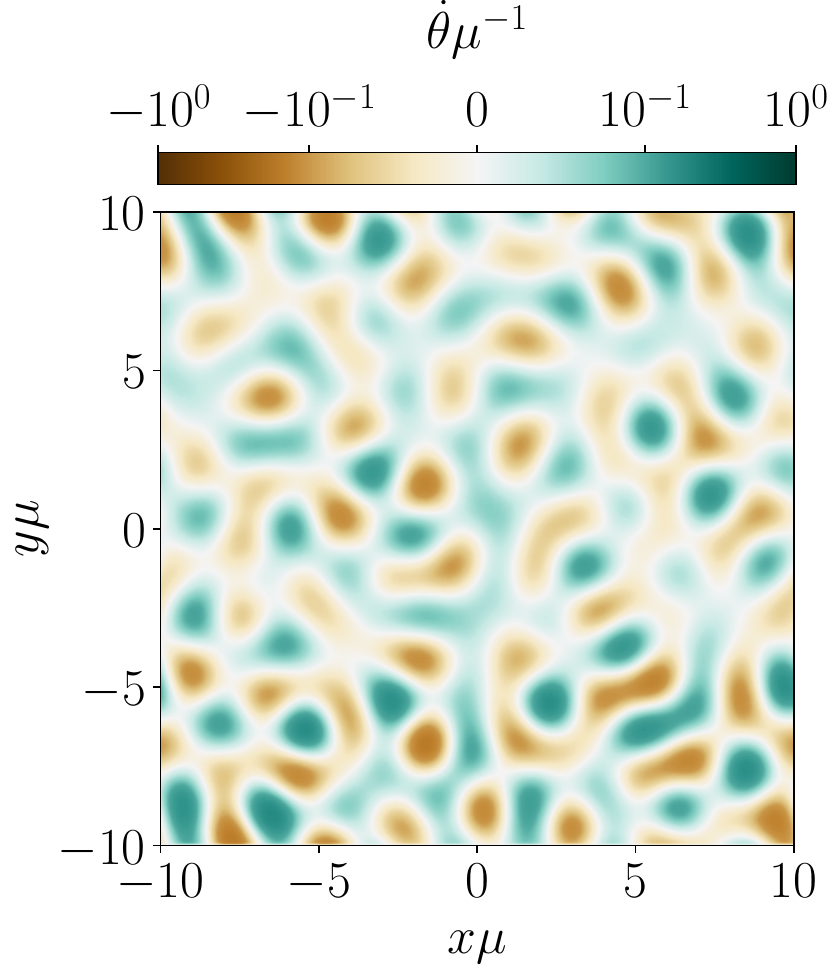}
    \hfill
    \includegraphics[width=0.32\linewidth]{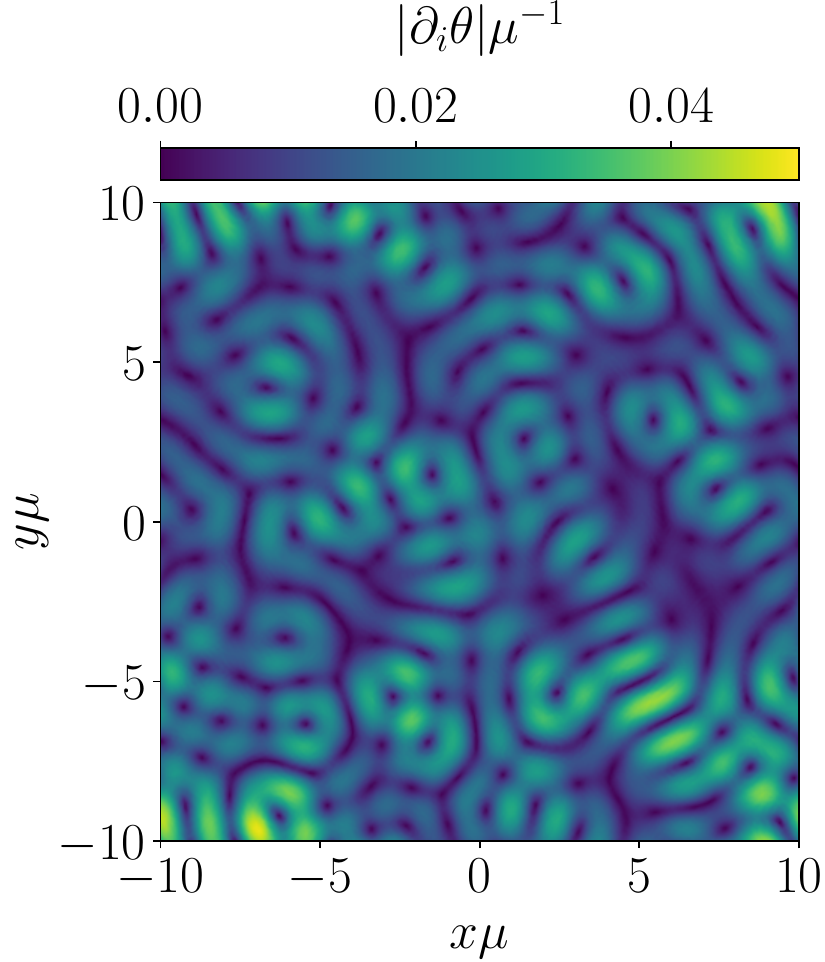}
    \caption{%
    A snapshot at $\mu t=14.6$ of a 2-dimensional simulation with the same parameters as in Fig.~\ref{fig:2dexpgrowth}.
    From left to right, the panels show the spatial dependence of the angular degree of freedom $\theta$ (spatial-average subtracted), its time derivative $\dot{\theta}$, and the magnitude of the associated gradient field $|\partial_i\theta|$, all appropriately normalized as annotated.
    \ZenodoComment
    }
    \label{fig:counterrot}
\end{figure}

To illustrate this and the spatial dependence of the most unstable mode further, we show in Fig.~\ref{fig:counterrot} a snapshot at a point in time (taken during the exponential growth of the modes) of the spatial dependence of the angular degree of freedom, as well as its temporal and spatial gradients. 
The dominant Fourier mode of the perturbation growing around $\langle\theta\rangle$ has wavenumber $\tilde{k}\sim 1$ (consistent with an analysis using methods introduced in Sec.~\ref{sec:pert}).
At this chosen instant of time, counter-rotating regions have already formed (as is evident from the central panel) and are separated by \textit{domain boundaries} of size $\sim 1/\mu$. 
Here and in the following, we define the domain boundaries to be the codimension-1 surfaces separating counter-rotating regions (i.e., the white lines separating green and brown regions in the central panel). 
The difference of the angular degree of freedom on either side of these domain boundaries, which we denote $\Delta\theta$, is growing exponentially in time around the temporal snapshot shown. 
As a result of this difference, the domain boundaries are also characterized by maximized spatial gradient of the angular degree of freedom, as can be seen in the right panel of Fig.~\ref{fig:counterrot}. 
Finally, while we find the amplitude of $\dot{\theta}-\langle\dot{\theta}\rangle\approx \dot{\theta}$ (shown in the central panel of Fig.~\ref{fig:counterrot}) to grow exponentially, the pattern remains stationary over time scales that are much longer than $\sim 1/\mu$.

\subsubsection{Large Initial Charge Density  \texorpdfstring{$\mathcal{J}_0$}{J0}}
\label{sec:largec0resonance}

\begin{figure}[t]
    \centering    
    \includegraphics[width=0.32\linewidth]{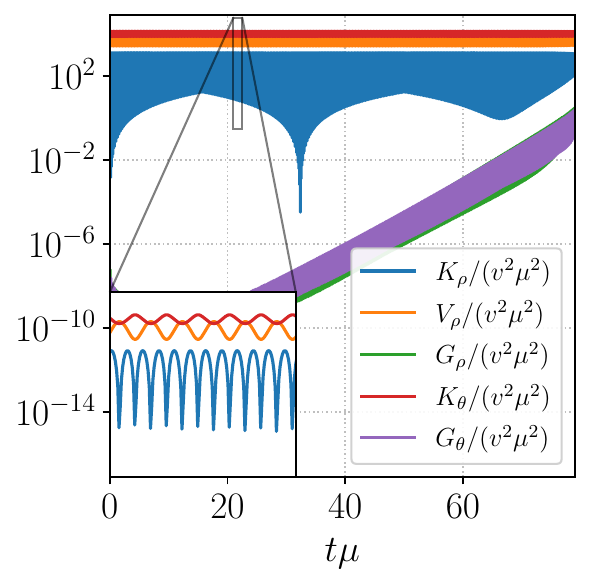}
    \hfill
    \includegraphics[width=0.315\linewidth]{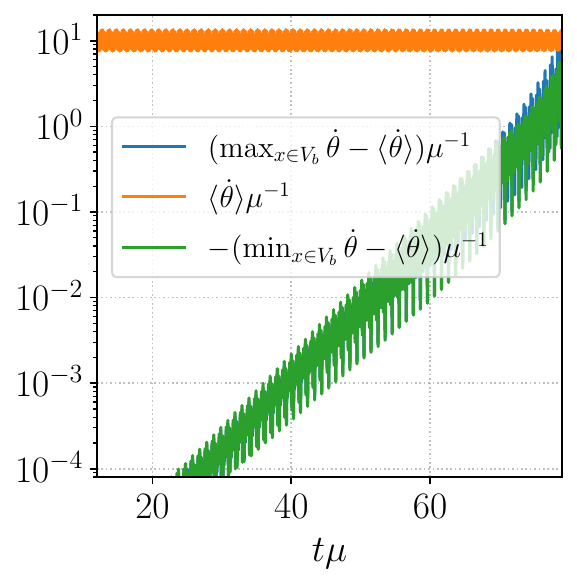}
    \hfill
    \includegraphics[width=0.335\linewidth]{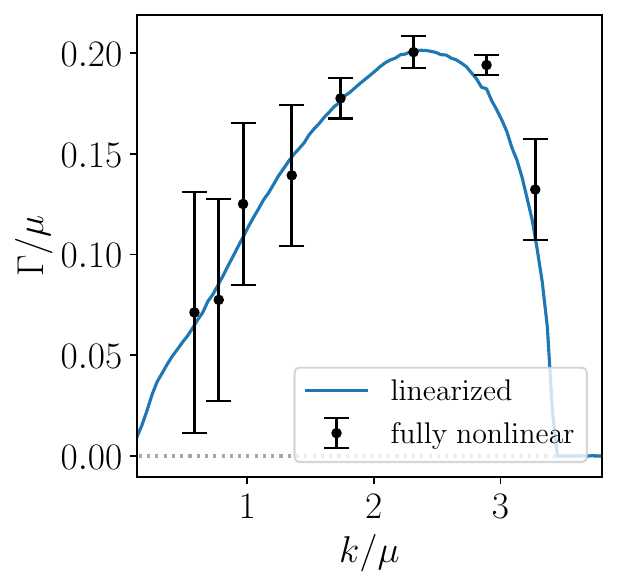}
    \caption{%
    Simulation of the 2-dimensional case ($D=2$) with a large initial angular motion ($\mathcal{J}_0=10^3\mu v^2$), a large initial field displacement ($\rho_i/v=11.5$), and small spatial fluctuations ($\sigma_{\rho_i}=10^{-6}\rho_i$).
    The left and center panels are similar to Fig.~\ref{fig:2dexpgrowth}, albeit for different parameters.
    \underline{\textsc{Left panel:}}~%
    The evolution in time of the different components of the volume-averaged energy densities of the system, both early and during the exponential growth of unstable perturbations. 
    The various colored lines are appropriately normalized plots of the kinetic (blue), potential (orange), and gradient (green) energies in the radial mode $\rho$; and the kinetic (red) and gradient (purple) energies in the angular mode $\theta$. 
    The inset shows the initial coherent rotation of the field around the field-space origin. 
    \underline{\textsc{Center panel:}}~%
    The spatially-averaged time derivative of the angular mode, $\langle\dot{\theta}\rangle$ (orange), as well as the difference between the maxima (blue) and minima (green) of the time derivative from this average. 
    We focus here on the phase before the perturbations grow into the nonlinear regime. 
    \underline{\textsc{Right panel:}}~%
    The $e$-folding growth rate $\Gamma$ (in units of $\mu$) of various spatial Fourier modes, as a function of wavenumber $k$ (also in units of $\mu$). 
    The black data points (labeled ``fully nonlinear'') were extracted from our fully nonlinear numerical simulations, and are compared to the linearized perturbative analysis (blue line, labeled ``linearized'') of Sec.~\ref{sec:linearizednumerics}. 
    We detail in App.~\ref{app:simdetails} how the rates were extracted from the fully nonlinear simulations, and discuss there the errorbars on those rates. 
    \ZenodoComment}
    \label{fig:2dexpgrowthLargec0}
\end{figure}

We now turn to the region of parameter space where $\mathcal{J}_0\gg v^2\mu$ and the amplitude of the initial radial oscillation around the minimum of the effective potential $V_{\rm eff}$ in~\eqref{eq:Veff} is small (i.e., $|\rho_i - \rho_{\text{min}}| \ll \rho_i$). 
An example that captures the main qualitative features in this regime is shown in Fig.~\ref{fig:2dexpgrowthLargec0}; specifically, for $D=2$, $\mathcal{J}_0=10^3v^2\mu$, $\rho_i/v=11.5$, and $\sigma_{\rho_i}=10^{-6}\rho_i$.%
\footnote{%
    The case where $\mathcal{J}_0 \gg \mu v^2$ but $\rho$ executes large-amplitude oscillations initially (i.e., when the quartic term in the effective potential dominates the $\mathcal{J}_0^2/\rho_0^2$ term initially) is expected to be more qualitatively similar to the small-$\mathcal{J}_0$ case, but we do not consider it in detail.
} %
As before, the small variance $\sigma_{\rho_i}$ implies a large degree of homogeneity of the radial and angular modes at early times. 
During this phase, the radial and angular modes oscillate coherently, while the parametric resonance instability becomes effective, causing the inhomogeneities to grow. 
As a result, the gradient energy densities of both the radial and angular degrees of freedom, $G_\rho$ and $G_\theta$, respectively, increase exponentially as energy is transferred from the coherent motion into the fastest growing $\tilde{k}\sim\mathcal{O}(1)$ Fourier mode; this can be seen in the left panel of Fig.~\ref{fig:2dexpgrowthLargec0}. 

As before, up to $t\mu\lesssim 80$, the dynamics are dictated by the coherent motion of the average quantities; i.e., $\langle\dot{\theta}\rangle\approx \dot{\theta}_0(t)$. 
In stark contrast to the case of $\mathcal{J}_0\ll v^2\mu$ considered above, however, the unstable modes are only \textit{differentially} rotating throughout the entire exponential growth of the parametric resonance. 
From the central panel of Fig.~\ref{fig:2dexpgrowthLargec0}, we conclude that $\langle\dot{\theta}\rangle-\min_{\bm{x}\in V_b}\dot{\theta}>0$ up until the modes enter the nonlinear regime at $t\mu\approx 80$. 
Counter-rotating regions (i.e., those with $\min_{\bm{x}\in V_b}\dot{\theta}<0$) begin to form only for $t\mu\gtrsim 80$, as we discuss below. 
Moreover, for large $\mathcal{J}_0 \gg v^2\mu$, the angular speeds of the unstable modes begin surpassing that of the background, $\langle\dot{\theta}\rangle$, only when $\Delta\theta\sim \mathcal{O}(1)$.
In the initial growth phase ($t\mu\lesssim 80$), the spatial dependence of the unstable modes is, as before, determined by the fastest growing Fourier modes, and is qualitatively similar to the case shown in Fig.~\ref{fig:counterrot} (while, of course, not exhibiting any region with $\dot{\theta}<0$).
In the right panel of Fig.~\ref{fig:2dexpgrowthLargec0} we compare the growth rates of various spatial Fourier modes obtained using our fully nonlinear numerical simulations, to the predictions of the linearized analysis in Sec.~\ref{sec:linearizednumerics}, finding excellent agreement between the two methods.

Thus far, we discussed the properties of exponentially growing linear perturbations on a coherently oscillating background, but have postponed discussion of any effects impacting these perturbations as they enter the nonlinear regime. 
Ignoring nonlinearities, the instability would tend to continue to drive the system from a state of differential rotation into one with counter-rotating regions separated by domain boundaries, across which the angular mode difference $\Delta\theta$ will continue to grow in size.
Whether or not the exponential growth is halted either before counter-rotating regions form, or (if they do form) before $\Delta\theta$ reaches $\sim \mathcal{O}(2\pi)$, depends on nonlinear effects that will be discussed in further detail in later sections.

\subsection{Summary}
\label{sec:Sec2Coda}

To conclude this section, let us recap what we have shown.
In Sec.~\ref{sec:SimplifiedAnalytics}, we showed in an analytical analysis in the spatial Fourier (i.e., momentum) domain of a simplified system that holds when $\mathcal{J}_0$ is small (formally, vanishing), that when an initially homogeneous, phase-coherent background field is executing small-amplitude oscillations around the minimum of the effective potential, parametric resonance instabilities (either narrow or broad) give rise to instabilities in spatial field perturbations, which causes them to grow exponentially.
We then undertook a perturbative (i.e., linearized) numerical analysis of the spatial Fourier modes in the time-domain in Sec.~\ref{sec:linearizednumerics} that showed that these instabilities also occur in the actual system of interest for small but non-vanishing $\mathcal{J}_0$, and that they persist at larger $\mathcal{J}_0$ (albeit in different regions of parameter space).
These analyses thus showed at the level of linear theory that initially small perturbations around the background field evolution are unstable to growth.
Finally, turning to numerical simulations of the fully nonlinear system of interest, we demonstrated in Sec.~\ref{sec:pertfullnonlinear} that these unstable modes form either differential- or counter-rotating regions that are separated by thin domain boundaries possessing large spatial field gradients. 
In the next section, we detail the system's behavior as the unstable spatial perturbations grow to the reach the nonlinear regime.

\section{Vortex Formation}\label{sec:formation}

\begin{figure}
    \centering
\includegraphics[width=0.8\linewidth]{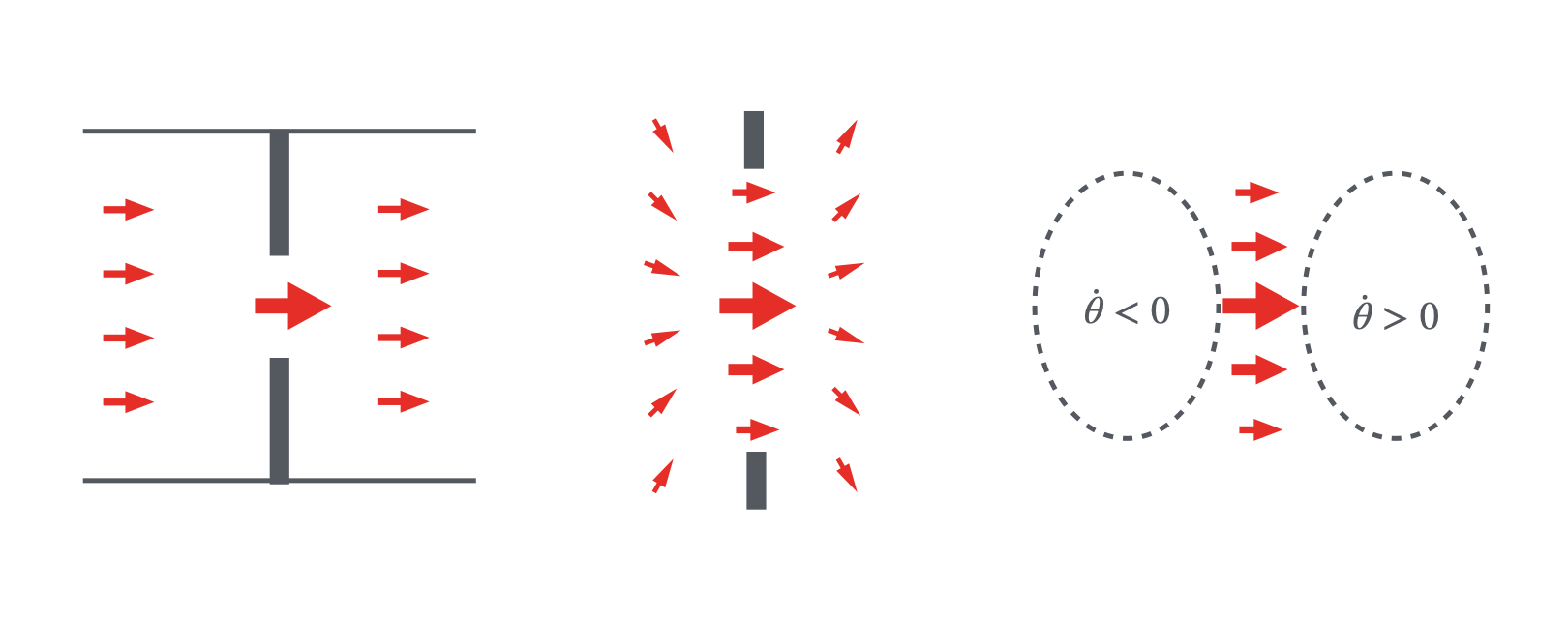}
    \caption{A schematic figure showing the similarities between the velocity field of a superfluid flowing through a thin, narrow orifice (left and center panels) and the gradient field $\partial_i \theta$ on the domain boundary separating counter-rotating regions immediately prior to vortex formation (right panel) in our case. 
    The direction and size of the red arrows is intended to indicate the direction and magnitude of the velocity or gradient field, as appropriate. 
    \underline{\textsc{Left and center panels:}}~%
    For a superfluid, the flow velocity is largest in the middle of a narrow orifice in a thin wall (solid black line) and smallest at the edges of the orifice owing to boundary conditions, as shown in these zoomed-out (left) and zoomed-in (center) views.
    \underline{\textsc{Right panel:}}~%
    A schematic representation of the gradient field on a domain boundary between two selected counter-rotating regions; cf.~the left and right panels of Fig.~\ref{fig:counterrot} and the top row of Fig.~\ref{fig:vortexformation}. 
    }
    \label{fig:flow}
\end{figure}

In this section, we will present our understanding of the phenomenon of global-string production that occurs at the domain boundaries of the counter-rotating regions we discussed in the previous section.
Let us first summarize again the field profile before string formation occurs.
As shown in Fig.~\ref{fig:counterrot}, the initial field evolution leads to the development of counter-rotating regions: regions inside of each of which there is a large value of $|\dot{\theta}|$, but where neighboring regions have opposite signs of $\dot{\theta}$.
This is despite an almost vanishing spatially averaged $\dot{\theta}$ in the cases with small initial $\mathcal{J}_0$. 
These regions are separated by thin domain boundaries, with thicknesses of order $1/\mu$. 
The $\theta$ field varies spatially across these domain boundaries, with a gradient $\partial_i\theta$ that is large and that grows over time due to the differential rotation between the regions. 
In this section, we show that when the difference in $\theta$ across the domain boundary grows to $\Delta \theta\sim\mathcal{O}(2\pi)$, vortices begin to form.

This phenomenon of vortex formation is analogous to that which occurs in a superfluid flow, first studied by Feynman~\cite{feynman1955chapter}, who identified a critical minimum velocity for vortex formation in superfluid flow through a thin, narrow orifice.
In the field-theory description~\cite{Ginzburg:1950sr}, this corresponds to a critical phase gradient $\partial_i \theta$ above which vortices form (see App.~\ref{app:cmtvortex}).%
\footnote{%
    In the superfluid context, a static flow velocity is proportional to the (static) phase gradient. 
    Except when analogizing our case to the superfluid case, we avoid this terminology as we have additional explicit time-dependence of the phase not present in the stationary superfluid case, and it would be confusing to refer to a spatial gradient of the $\theta$ field as a velocity field as the latter usually denotes time dependence.
    Instead, in our case, we will refer to $\partial_i \theta$ as the gradient field.
} %
Feynman's results motivate us to consider whether a similar critical $\partial_i \theta$ supplies a necessary condition for string formation in our case. 
However, because the $\partial_i \theta$ field profile across domain boundaries separating counter-rotating regions that is shown in, e.g., Fig.~\ref{fig:counterrot} only qualitatively resembles the velocity field profile for a static superfluid flowing through a thin, narrow orifice (see Fig.~\ref{fig:flow} for a schematic plot which showing this resemblance), the detailed numerical studies we present in the paper are required to refine this analogy.
The differences include both the geometry of the systems (see Fig.~\ref{fig:flow}), as well as the fact that, unlike in a superfluid, the radial mode in our case is significantly displaced from the potential minimum. 
Nevertheless, whenever possible and helpful throughout the description of our numerical results, we will provide qualitative connections of our findings to the superfluid case.%
\footnote{%
    Condensed matter systems where similar defect formation has been identified include the classical rotor model~\cite{sachdev1999quantum,RevModPhys.59.1001} and networks of Josephson junctions~\cite{tinkham2004introduction}.
    } %
In what follows, we will describe the dynamics of vortex formation based on results of two-dimensional simulations and identify precise conditions for vortex--anti-vortex pair creation, before demonstrating similar conditions for string-loop formation in three dimensions.

\subsection{Global-String Profile}
\label{sec:globalstring}

In two spatial dimensions, an isolated (anti-)vortex at $r=0$ in an otherwise unexcited system has a field profile $\Phi(\bm{x}) = \rho(r) e^{\pm i \theta}$ with boundary conditions $\rho(0) = 0$ and $\rho(\infty) = v$, where the $\pm$ sign choice specifies positive (vortex) or negative (anti-vortex) vorticity. 
Note that $\theta$ is multi-valued at $r=0$, and $\partial_i \theta$ diverges as a result. 
The radial profile of a vortex has been solved explicitly; see~\cite{Weinberg:1992hc} and references within. 
Schematically, the string has a `core' region in which $\rho(r)$ increases from $0$ to $\mathcal{O}(v)$ over a distance of $\mathcal{O}(1/\mu)$.

\begin{figure}
    \centering
    \includegraphics[width=0.5\linewidth,trim={0 2.5cm 0 3cm},clip]{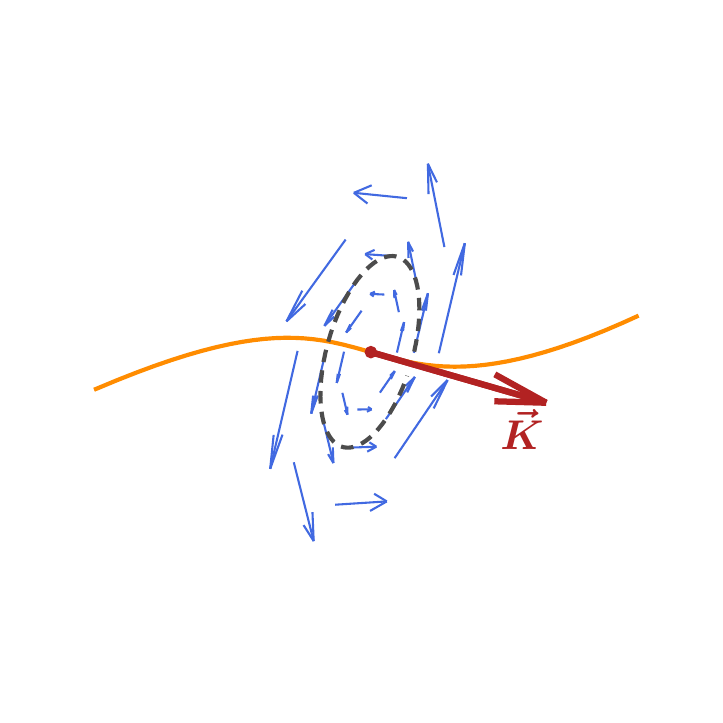}
    \caption{A schematic picture of a string (solid orange line) in three spatial dimensions.
    The circulation vector $\vec{K}$ is indicated with the red arrow, while the blue arrows indicate the direction of the gradient field $\partial_i \theta $ in the plane normal to $\vec{K}$. 
    The line integral of $\partial_i \theta $ along any positively oriented contour around the string (e.g., the grey dashed line) is $2\pi$. 
    }
    \label{fig:stringK}
\end{figure}

The two-dimensional vortex solution generalizes in three dimensions to a string, or vortex line, which is in general a closed loop in the absence of boundaries. 
The binary choice of vorticity ($\pm$) generalizes to a circulation vector $\vec{K}$, which points along the string core (i.e., the locus of points where $\rho=0$), with an orientation that is set by the sense of the circulation of $\theta$ around the core; see Fig.~\ref{fig:stringK}. 
See App.~\ref{app:cmtvortex} for more details regarding the significance of this circulation vector $\vec{K}$ for vortex dynamics in superfluid and superconductor contexts.

The vortices in our analysis differ from isolated vortices in an otherwise-unexcited system in two major ways. 
First, for most of our simulations, our vortices reside in a background-field configuration for which the radial mode is excited (i.e., displaced significantly from the minimum of the potential); specifically, the average radial-mode displacement is $\rho > v$ in the counter-rotating regions and $\rho < v$ in the domain-boundary regions. 
Second, the background-field configuration in which the vortices reside can have $\dot{\theta}$ and/or $\partial_i{\theta}$ significantly different from zero. 
Both of these differences modify the vortex structure in ways that are analytically intractable to compute. 
Luckily, however, their topological nature still makes vortices or strings easy to identify, even absent knowledge of their spatial profile: they are simply points (in 2D) or lines (in 3D) where $\rho=0$ and around which $\theta$ changes by $\pm 2\pi$, even on an arbitrarily small loop.
As a result, it is possible to count the number of vortices (and anti-vortices) in 2D, or measure the length of strings in 3D.

Note, however, that because all field excitations are simply excitations of the single complex scalar field $\Phi$ in a nonlinear theory, it is not in general possible to unambigiously partition the field energy into energy that is stored in vortices vs.~in the background-field profile.

\subsection{Vortex Formation Dynamics}
\label{sec:vortexformationdynamics}

\begin{figure}[t]
    \centering
    \includegraphics[width=1\linewidth]{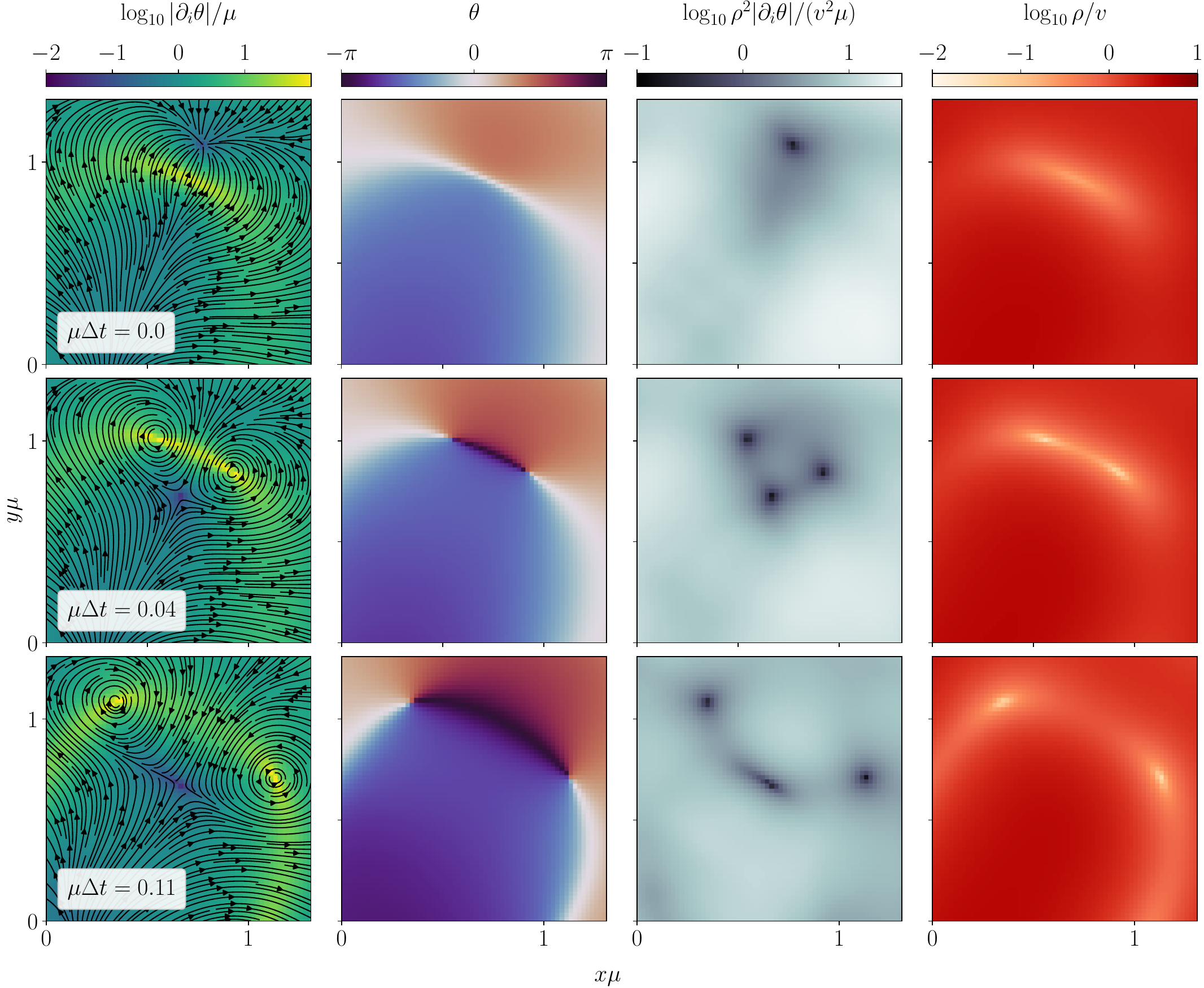}
    \caption{A series of snapshots of the spatial configuration of (from left to right) the spatial gradient of the angular mode, the angular mode, the spatial Noether current, and the radial field amplitude, from early to late times (time increases from top to bottom), tracking the system through the formation of a pair of vortices. 
    The left column shows the flow lines of $\partial_i\theta$, in addition to its magnitude. 
    The spatial region shown here corresponds to a small zoom-in of the evolution of the system shown in Fig.~\ref{fig:counterrot} at $t\mu\approx 28$.
    The attentive reader will note that the vortices move superluminally after production (second and third rows); we explain \protect\hyperlink{text:superluminal}{in the main text} that this is not a violation of causality, but rather a manifestation of the fact that the vortices here should not be thought of as weakly interacting particle-like objects that exist independently of the background field, but rather as a feature of the overall nonlinear $\Phi$ field evolution. 
    \ZenodoComment} 
    \label{fig:vortexformation}
\end{figure}

In this subsection, we describe the vortex-formation dynamics that occur once the exponentially growing spatially inhomogeneous modes reach the nonlinear regime. 
While the conditions on $\rho_i$ for vortex formation differ across the $\mathcal{J}_0$ parameter space (as discussed in more detail in Sec.~\ref{sec:formation_conditions}), we find the overall formation dynamics (i.e., instability leading to counter-rotating regions and vortex formation on domain boundaries) to be universal if they occur.
Here, we will first describe these dynamics in the example introduced in Sec.~\ref{sec:pertfullnonlinear}; that is, we set the initial charge density $\mathcal{J}_0=0.1\mu v^2$ to be homogeneous in the simulation box, take the initial radial-mode displacement to be $\rho_i/v=5$, and set its variance to $\sigma_{\rho_i}=10^{-2}\rho_i$. 
We will then move on to discuss other cases.

As we saw in Sec.~\ref{sec:pertfullnonlinear}, parametric resonance drives the growth of modes with wavenumber $k\sim \mu$, eventually leading to the formation of counter-rotating regions. 
These regions of large $|\dot{\theta}|$ are separated by domain boundaries
characterized by a vanishing angular speed, $\dot{\theta}\approx 0$, and large spatial gradients of the angular mode $\partial_i\theta$ (see also Fig.~\ref{fig:counterrot}). 
This corresponds to the top row~in~Fig.~\ref{fig:vortexformation}. 
Prior to vortex formation, the location where $|\partial_i \theta|$ is large coincides with locations where $\rho/v$ is small, while the magnitude of the spatial component of the Noether current, $|j_i| = \rho^2 |\partial_i \theta|$, is regular (in particular, non-zero) across the domain boundary. 
Note in general that the gradient field $\partial_i\theta$ is not divergence-free; i.e., $\partial^i\partial_i\theta\neq 0$.
However, in the absence of a vortex, this gradient field is curl-free (by definition). 
Similar to the superfluid case~\cite{RevModPhys.59.1001}, vortex production eventually occurs in the presence of the large spatial phase gradients at these domain boundaries~(second row in Fig.~\ref{fig:vortexformation}).
At these locations, vortex pair production may be facilitated by, for example, a passing field perturbation with a non-vanishing phase-gradient divergence.
This perturbation can be seen explicitly in both the first and third columns of the first row in Fig.~\ref{fig:vortexformation} (see also the animation at \cite{ZenodoVideos}). 
The interaction between this perturbation and the large-gradient domain boundary produces a pair of vortices (second row in Fig.~\ref{fig:vortexformation}). 
As a pair of vortices is formed, the radial mode and magnitude of the current density drop to zero at the production site, while the angular mode becomes multi-valued and $\partial_i\theta$ diverges at the vortex locations.
The gradient field $\partial_i\theta$ swirls around the vortex core locations, implying non-vanishing vorticity%
\footnote{%
    Here, $\epsilon^{\mu\nu\lambda\rho}$ is the completely antisymmetric $(3+1)$-dimensional Minkowski spacetime Levi-Civita symbol with $\epsilon^{0123}\equiv+1$. 
    In $(2+1)$ dimensions, we would simply drop the index $i$ in the expression in the text (2D spatial vorticity is a pseudoscalar); see also the definitions at footnote \ref{ftnt:LeviCivita2Ddefn}.
} %
$\epsilon^{0ijk}\partial_j\partial_k\theta$ at the center of the vortex core. 
Subsequently, this vortex--anti-vortex pair separates and begins propagating along the domain boundary separating the relevant counter-rotating regions (bottom row in Fig.~\ref{fig:vortexformation}). 
We will come back to the reason for this movement in Sec.~\ref{sec:dynamics}. 
In the following subsections, we will first discuss the conditions for vortex formation in different regimes. 

\hypertarget{text:superluminal}{Before} we do, we pause to note that the vortices in Fig.~\ref{fig:vortexformation} actually move superluminally~(along the domain boundaries) at speeds $v\sim\mathcal{O}(3)$, at least for these parameter values; their speed would decrease for smaller $\mathcal{J}_0$ and for $\rho\approx v$, which is the limit in which familiar weakly coupled vortex results from condensed matter systems should hold.
We stress that this superluminality is however \emph{not} a violation of causality in our numerical simulations.
Instead, this is a clear signal that one should not think of the vortices here as weakly coupled particle-like objects with a definite identity that exist independently of the background field in which they reside; rather, they are configurations that appear within the overall nonlinear causal evolution of the single field~$\Phi$.
We note that this kind of apparent superluminality is actually familiar and happens even in completely prosaic \emph{linear} wave-interference phenomena, if one fixes attention on specific field features instead of looking at the field evolution as a whole.
For example, consider a pair of Gaussian line disturbances $f_{\pm}$ of amplitude $A$ and width $\sigma$, defined (in two dimensions) as
\begin{align}
    f_{\pm}(t,\bm{x}) &\equiv A \exp\left[ - \frac{( t - \bm{x}\cdot \bm{\hat{n}}_\pm )^2}{2 \sigma^2} \right]\:, & \bm{\hat{n}}_\pm &\equiv (\sin\theta,\pm \cos\theta) \:.
\end{align}
Each of $f_{\pm}$ satisfies the linear 2D wave equation and propagates in the direction $\bm{\hat{n}}_\pm$ at speed $v=1$~(i.e., causally).
Their sum $f(t,\bm{x}) \equiv f_{+}(t,\bm{x}) + f_{-}(t,\bm{x})$ is of course also a linear 2D wave-equation solution and has a maximum $f_{\text{max}} = 2A$ at $\bm{x}_{\text{max}}=(t/\sin\theta,0)$.
Were one to track only that maximum of the field and attempt to think of it as some particle-like object of definite identity, it would appear to move down the $x$ axis at a speed $|v| = 1/|\sin\theta| \geq 1$, despite the fact that the overall field evolution is causal by construction. 
Note that this example of apparent superluminality in a familiar system is provided purely as a reminder to the reader than apparent superluminality is not equivalent to a causality violation; we note specifically that we do not intend, by giving this example, to suggest that linear interference is the mechanism solely or even primarily responsible for the superluminal motion of vortices in this case we are considering in this work.

\subsection{Formation Conditions}
\label{sec:formation_conditions}

As we detailed in the previous section, vortex--anti-vortex pairs form on the domain boundaries where the spatial gradient of the angular mode is large. 
In our setting, this is realized qualitatively differently across the $\mathcal{J}_0$ parameter space. 
Recall, we found in Sec.~\ref{sec:growth} that parametric resonance drives the growth of different $k$-modes (and, correspondingly, spatial scales), depending on both the initial charge density $\mathcal{J}_0=\rho_i^2\dot{\theta}_i$ and the energy density $V_{\rm{eff}}(\rho_i)$ of the coherent component.
This has two major implications for the vortex production, as we detail below: (i) the system contains enough free energy density for vortex production only for a sufficiently large initial radial-field displacement away from $\rho_{\min}$.
Here, the free energy is defined as $V_{\text{eff}}(\rho=\rho_i) - V_{\text{eff}}(\rho=\rho_{\min})$ with $\mathcal{J}_0$ fixed to the initial value, and $\rho_{\min}$ is the minimum of the effective potential with $\mathcal{J}_0$ fixed to the initial value, as defined below \eqref{eq:Veff}. 
And, (ii) the length (in 2D) or area (in 3D) of the vortex-forming domain-boundary surfaces decreases with increasing charge density $\mathcal{J}_0$ for a fixed $\rho_i\gtrsim \rho_c > v$.
Here, $\rho_c$ is the critical initial radial field amplitude for vortex formation to occur, as we discuss below.
We demonstrate these dependences in this section by first considering the case of small initial $\mathcal{J}_0$ in Sec.~\ref{sec:SmallJ0discussion}, after which we move on to large initial $\mathcal{J}_0$ (the precise definition of which will be made clear in Sec.~\ref{sec:LargeJ0discussion}).

\subsubsection{Small Initial Charge Densities \texorpdfstring{$\mathcal{J}_0$}{J0}}
\label{sec:SmallJ0discussion}

Consider the case where the initial charge density $\mathcal{J}_0$ is small, taking again as a concrete example the case we considered in Sec.~\ref{eq:Sec2smallC0}:%
\footnote{Recall, $\sigma_{\rho_i}$ sets the initial amplitude of the unstable modes; the production time of the first vortices depends logarithmically on $\sigma_{\rho_i}$.
    } %
$D=2$, $\mathcal{J}_0=0.1\mu v^2$, $\rho_i/v=5$, and $\sigma_{\rho_i}=10^{-2}\rho_i$.
The initial charge density is small here in the sense that the $\mathcal{J}_0^2/(2\rho^2)$ term in the effective potential \eqref{eq:Veff} is small compared to the other terms when evaluated at $\rho = \rho_i$, as well as in the sense that the magnitude of the typical angular-field temporal gradient $|\dot{\theta}|$ in the relevant counter-rotating regions immediately prior to vortex formation at their domain boundary is much larger than the magnitude of the initial global field rotation speed, $|\dot{\theta}_i|$.
In this case, this initial charge density plays a limited role in the evolution of the system, and conditions for vortex formation can be found independent of the exact value of $\mathcal{J}_0$. 
We will discuss two conditions for vortex formation: a microscopic condition requiring a sufficiently large gradient field $\partial_i \theta$ at the location of vortex formation, and a global condition on the initial energy density in the system.
We will also comment on the connections between these criteria. 

\begin{figure}[t]
    \centering
    \includegraphics[width=0.32\linewidth]{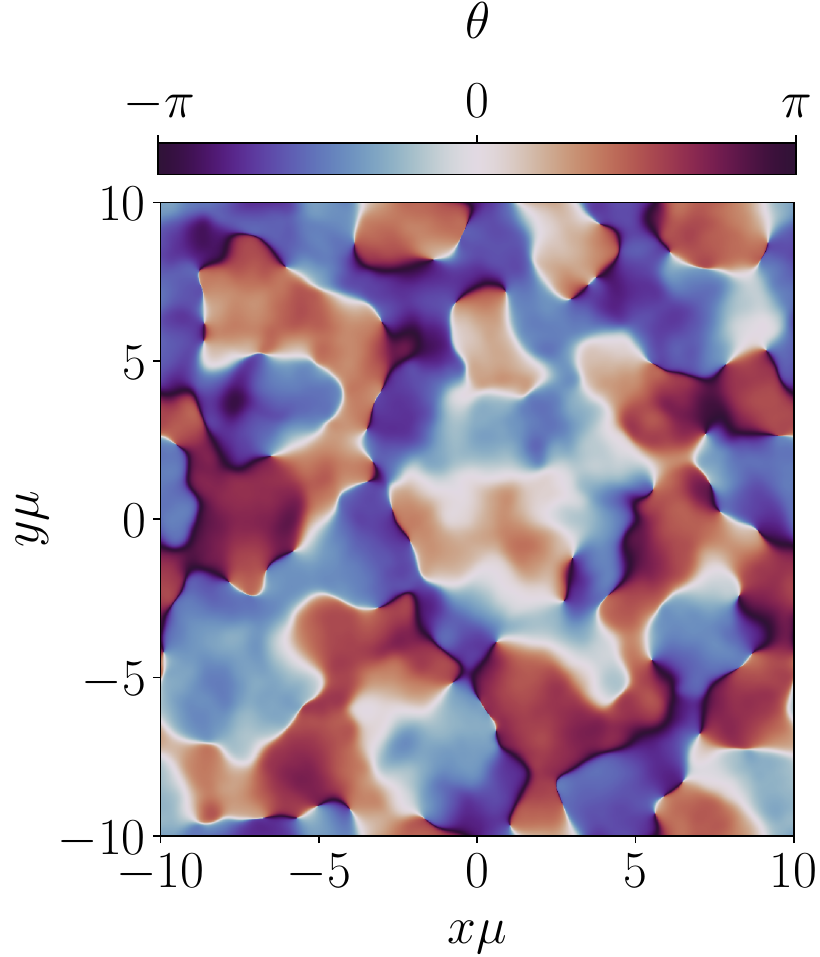}
    \hfill
    \includegraphics[width=0.32\linewidth]{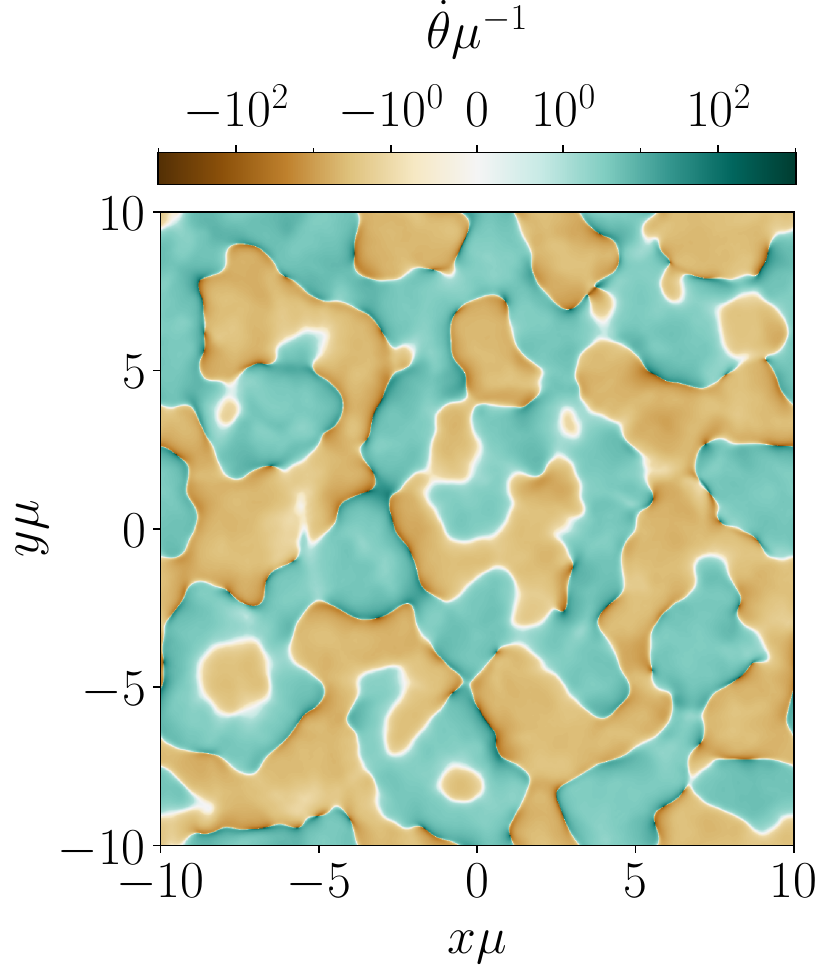}
    \hfill
    \includegraphics[width=0.32\linewidth]{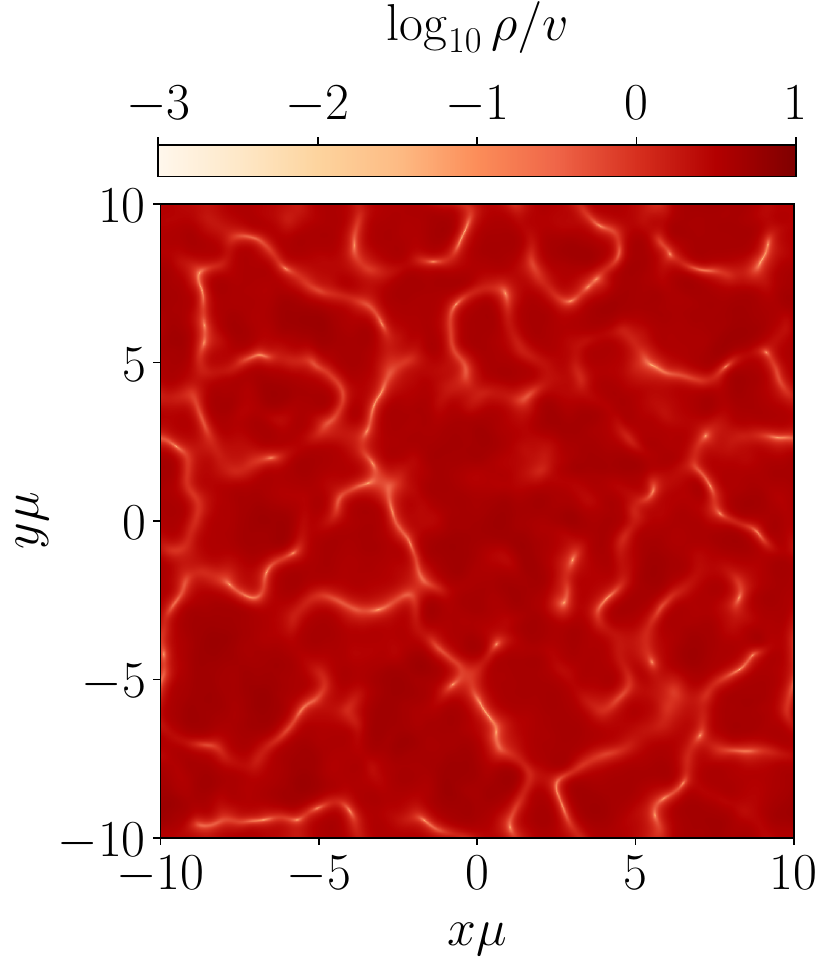}
    \caption{The field configuration shortly after vortex pair production became energetically favorable along domain boundaries in the $\mathcal{J}_0=0.1\mu v^2$, $\rho_i/v=5$, and $\sigma_{\rho_i}=10^{-2}\rho_i$ case considered so far. 
    The snapshot corresponds to time $t\mu\approx 31$. 
    \ZenodoComment 
    }
    \label{fig:smallcharge}
\end{figure}

In this example, counter-rotating regions form rapidly once the parametric resonance is active~(see right panel of Fig.~\ref{fig:2dexpgrowth}). 
In particular, the total spatial area of co- and counter-rotating regions~(where the sense of rotation is defined with respect to the initial coherent state) is roughly the same during much of the exponential growth phase of the resonance~(see the center panel of Fig.~\ref{fig:counterrot}). 
Counter rotation persists through saturation~(i.e., termination of the exponential growth caused by the parametric resonance) and vortex pair production. 
In Fig.~\ref{fig:smallcharge}, we show the state of the system shortly after the critical gradient $|\partial_i \theta|$ was surpassed\emph{along the domain boundaries}, and (anti-)vortices have formed in the system. 
This critical gradient that is reached locally on the domain boundaries prior to vortex formation is parametrically of order $|\partial_i \theta| \sim 2\pi / m^{\textsc{crr}}_\rho$, where $m^{\textsc{crr}}_{\rho}$ is the mass of the radial mode in the immediately adjacent CRR [parametrically, $m^{\textsc{crr}}_{\rho} \sim \mathcal{O}(\mu)$].
The field configuration in Fig.~\ref{fig:smallcharge} is qualitatively representative of the post-saturation states of systems with $\mathcal{J}_0\lesssim v^2\mu$ and $\rho_i/v\gg 1$, when vortex formation actually occurs abundantly on domain boundaries whenever a critical gradient $|\partial_i \theta|$ is reached.

Qualitatively, the existence of a critical $|\partial_i \theta|$ can be understood with the help of particle--vortex duality in $2+1$ dimensions (see App.~\ref{app:cmtvortex} and \cite{Peskin:1977kp,Dasgupta:1981zz,Senthil:2018cru,Tong:2016kpv}), where the vortices are dual to electric particles and the velocity field $\partial_i\theta$ is mapped onto an electric field via%
\footnote{\label{ftnt:LeviCivita2Ddefn}%
    The $(2+1)$-dimensional Minkowski spacetime Levi-Civita symbol is defined to be the completely anti-symmetric tensor with $\epsilon^{012}\equiv +1$.
    In a $(2+1)$-dimensional spacetime with mostly minus signature, we have $\epsilon^{012}=+\epsilon_{012}$.
    } %
\begin{align}
    \frac{1}{2 \pi} \epsilon_{0 i j } F^{0i} =  \partial_j \theta \: .
\end{align}
In the dual picture, the existence of a critical $|\partial_i\theta|$ beyond which vortex pair production occurs~(classically) maps onto the existence of a critical electric field strength beyond which Schwinger pair production is no longer exponentially suppressed. 
The duality offers a qualitative explanation of the critical field strength for string/vortex formation that we expect, as well as a plausible reason for our observation that string/vortex formation occurs only on domain boundaries, where the gradient field is large (we elaborate on this in the next subsection). 

\begin{figure}
    \centering
    \includegraphics[width=0.475\linewidth]{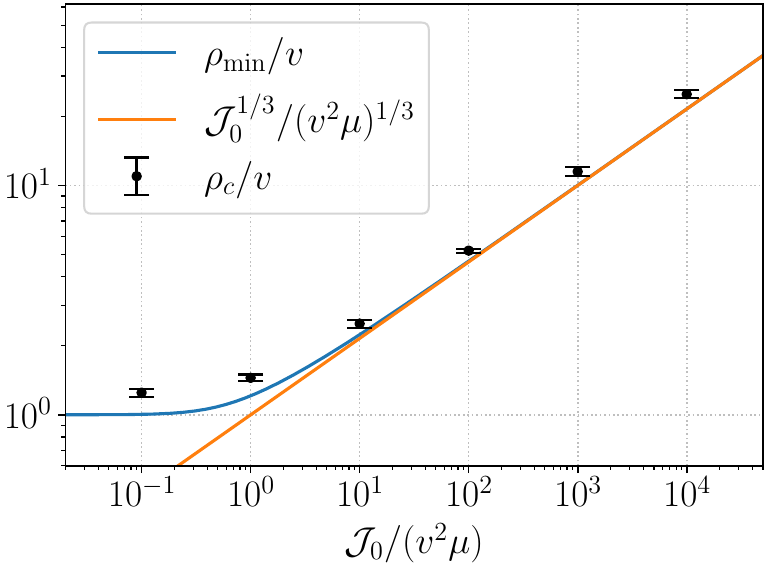}
    \hfill
    \includegraphics[width=0.49\linewidth]{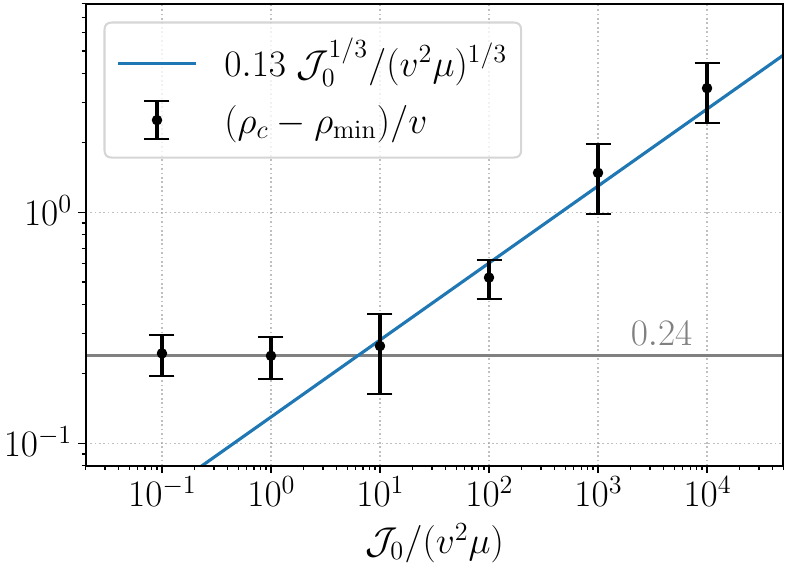}
    \caption{%
    \underline{\textsc{Left panel:}}~%
    The critical initial radial field amplitude for vortex formation, $\rho_c$, as a function of $\mathcal{J}_0$, determined numerically here in $D=2$ with associated uncertainties [black points with errorbars]; see App.~\ref{app:simdetails} for further details. 
    This is confronted with the radial field amplitude $\rho_{\min}$ minimizing the effective radial-mode potential $V_{\rm eff}$ defined at \eqref{eq:Veff} [blue line]. 
    For large $\mathcal{J}_0$, the $\rho_{\min}$ scales as $\sim \mathcal{J}_0^{1/3}$, as shown [orange line]. 
    Recall also that there is a second critical field value at $0<\rho_c^{(-)} < \rho_{\text{min}}$ which we do not show here; see discussion in footnote~\ref{ftnt:rhoC}.
    \underline{\textsc{Right panel:}}~%
    The difference between $\rho_{\min}$ and the critical field amplitude $\rho_c$.
    For large $\mathcal{J}_0$, this difference is roughly $\rho_c-\rho_{\min}\approx 0.13 \ \mathcal{J}_0^{1/3}(v/\mu)^{1/3}$ [blue line]. 
    We also indicate the value that the critical field amplitude takes in the $\mathcal{J}_0\rightarrow 0$ limit, $\rho_c/v\approx 1.24$ [horizontal gray line].
    }
    \label{fig:veffmin}
\end{figure}

In the small-$\mathcal{J}_0$ limit, vortex formation may depend on the initial displacement $\rho_i$, which is practically the only free parameter. 
In fact, as we described in Sec.~\ref{sec:growth}, the initial displacement $\rho_i$ determines which $k$-modes would undergo parametric resonance and their growth rate.
For large enough $\rho_i$, as is the case in the example we are presenting here, the growth of parametric resonance saturates when vortices form. 
However, decreasing $\rho_i$ down to $\rho_{\min}\approx v$, eventually causes the system to be unable to reach the critical field gradient $|\partial_i\theta|$ necessary for vortex production; instead, the instability saturates by only producing radiation at frequencies set by the wavenumber of the most unstable mode.%
\footnote{%
    Note, for $\mathcal{J}_0\ll v^2\mu$, the coherent rotation in systems with $\rho_c>\rho_i\gtrsim\rho_{\min}$ is still unstable and these systems may still fragment to produce counter-rotating regions, with associated spatial field gradients at domain boundaries in the process.} %
If we let $\rho_c$ be the critical initial radial-mode amplitude below which%
\footnote{\label{ftnt:rhoC}%
    Since the potential is by definition quadratic in small field displacements around $\rho = \rho_{\text{min}}$ (and it is moreover actually divergent $V\rightarrow +\infty$ as $\rho \rightarrow 0$ or $\rho \rightarrow \infty$ for $\lambda >0$ and $\mathcal{J}_0\neq 0$), it is always the case for $\lambda >0$ and $\mathcal{J}_0\neq 0$ that there are two $\rho >0$ solutions to $V_{\text{eff}}(\rho=\rho_c) - V_{\text{eff}}(\rho=\rho_{\min}) = \Delta V_{\text{c}}$, where $\Delta V_{\text{c}}$ is the critical free energy required for vortex formation.
    One solution will have $\rho^{(+)}_c>\rho_{\min}$ and the other will have $0<\rho^{(-)}_c<\rho_{\min}$.
    Throughout this work, we conventionally fix $\rho_c \equiv \rho^{(+)}_c$ in the main text, in order to streamline the presentation; the reader should however remember that there is a second solution, that $0<\rho_i < \rho_c^{(-)}$ would also suffice, and that the condition we are describing is really one on the free energy.
    Note that for $\mathcal{J}_0 = 0$ and $\lambda>0$, there are in principle either one or two $\rho>0$ solutions to $V_{\text{eff}}(\rho=\rho_c) - V_{\text{eff}}(\rho=\rho_{\min}) = \Delta V_{\text{c}}$, depending on whether $\Delta V_{\text{c}}>V(0)$ [1 solution] or $0<\Delta V_{\text{c}}< V(0)$ [2 solutions]; however, since we find that $ c_D < 1$ [see \eqref{eq:conditionV0}], we should always have 2 solutions in practice.
} %
no vortex pairs are produced during the parametric resonance (or after it saturates), we find numerically that $\rho_c/v = 1.24$ in the small $\mathcal{J}_0$ limit and in two spatial dimensions; see Fig.~\ref{fig:veffmin}. 

The observation that this critical $\rho_c$ is smaller than $\sqrt{2} v$ is surprising because, one might naively expect that sufficient initial potential energy density would be required for the field to be able to reach $\rho=0$; i.e., $V(\rho_i) \geq V(\rho=0) \Rightarrow \rho_i \geq \sqrt{2} v$.
This suggests that spatial inhomogeneities are important to vortex formation.
To see this most clearly, we rewrite this condition on $\rho_i$ in terms of the initial free energy density in the system $V(\rho_i) = V(\rho_i)-V(v)\approx V_{\text{eff}}(\rho=\rho_i,\mathcal{J}_0\ll \mu v^2)-V_{\text{eff}}(\rho=\rho_{\min},\mathcal{J}_0\ll \mu v^2)$. 
In two spatial dimensions ($D=2$), and given the initial conditions introduced in Sec.~\ref{sec:pertfullnonlinear}, we find vortices will form during the saturation of the parametric resonance, if
\begin{align}
    V(\rho_i) \geq c_D V(0), & & c_{D=2}\approx 0.29 \: ,
    \label{eq:conditionV0}    
\end{align}
where $c_D$ is a constant that may depend on the spatial dimensionality of the system, as well as the exact form of the (effective) potential; generally, however, we expect that $c_D\leq 1$ regardless of the exact form of the (effective) potential. 
One way to understand the parameter $c_D$ is as a geometrical factor accounting for the small fractional size of the regions of critical $|\partial_i\theta|$ (i.e., the domain boundaries), as compared to the counter-rotating regions where $|\partial_i\theta|$ is small; in other words, the system does not have to satisfy the vortex-formation conditions homogeneously/globally, in order to permit individual vortex pairs to form locally.

The two critical conditions [i.e., sufficiently large $|\partial_i \theta|$, and \eqref{eq:conditionV0}] are clearly related. 
As can be seen in Fig.~\ref{fig:2dexpgrowth}, the energy that sustains counter rotation ($K_{\theta}$) arises via the parametric resonance instability from the initial potential energy $V(\rho_i)$.
And a large $K_{\theta}$ (equivalently, a large $\dot{\theta}$ in the counter-rotating regions) is of course essential to drive the growth of the gradient $|\partial_i\theta|$ on the domain boundaries to the critical value for vortex formation. 
Moreover, as we will see in further examples below, with sufficient initial energy, counter-rotating regions persist even after vortices first form.
This ensures that $|\partial_i \theta|$ can repeatedly reach the critical value as the difference of $\theta$ across the domain boundary periodically reaches $\Delta \theta \sim \mathcal{O}(2\pi)$, thereby causing repeated epochs of vortex formation.

\subsubsection{Large Initial Charge Densities \texorpdfstring{$\mathcal{J}_0$}{J0}}
\label{sec:LargeJ0discussion}

Vortex formation also occurs when the initial charge density is large: $\mathcal{J}_0 \gtrsim v^2\mu$.
As $\mathcal{J}_0$ increases beyond $v^2\mu$, the minimum of the effective potential $V_{\rm eff}$ (i.e., $\rho_{\min}$) deviates from the minimum of the bare potential $V$; that is, $\rho_{\text{min}} \gtrsim v$. 
In the left panel of Fig.~\ref{fig:veffmin}, we show the value of $\rho$ that minimizes $V_{\rm eff}$ [i.e., $\rho_{\text{min}}$ such that $V'_{\text{eff}}(\rho_{\text{min}}) = 0$] as a function of $\mathcal{J}_0$. 
Towards large $\mathcal{J}_0$, the minimum goes as $\rho_{\min}\propto\mathcal{J}_0^{1/3}(v/\mu)^{1/3}$.

There are essentially two qualitatively distinct cases that can arise for $\mathcal{J}_0 \gtrsim v^2\mu$, which we can distinguish on the basis of which terms in the effective potential are important initially:~(i)~the quartic term in the potential dominates the $\mathcal{J}_0^2/(2\rho_i^2)$ term, and~(ii)~the $\mathcal{J}_0^2/(2\rho^2_i)$ term in the effective potential is comparable to the quartic term at the initial displacement $\rho_i$.
In both cases, the quadratic term is subdominant.
It turns out that case (i) has behavior qualitatively similar to the case of small $\mathcal{J}_0$, just with a displaced potential minimum, so we omit detailed consideration of it here.
On the other hand, case (ii) displays qualitatively distinct behavior to the small-$\mathcal{J}_0$ case, so we examine it in more detail.
More precisely, for this case we assume ${(\rho_{i}/v)^3 \gtrsim  (\rho_{\text{min}}/v)^3 \sim \mathcal{J}_0/(\mu v^2)\gg 1}$ and also $\rho_i \gtrsim \rho_{\min} \gg \rho_i - \rho_{\min}, v$; this implies that $\rho$ initially exhibits small-amplitude oscillations around $\rho_{\text{min}}$.
For the rest of the paper, we refer to this set of assumptions as the ``large-$\mathcal{J}_0$ case''.

As described in Secs.~\ref{sec:linearizednumerics} and~\ref{sec:largec0resonance}, the radial mode initially oscillates around $\rho_{\min}$ and the angular mode rotates in field space with large $\dot\theta$, triggering the parametric resonance, which leads to growing modes.
As these modes are perturbations away from the average coherent motion of the angular mode, they induce {\it differential rotation} (i.e., regions of larger and smaller $\dot{\theta}$, as compared to the average). 
However, unlike the case of small $\mathcal{J}_0$, this differential rotation does not, at least initially, grow to give rise to regions of {\it counter rotation} (i.e., regions where the sense of rotation in field space is reversed as compared to the initial condition).
Rather, the differentially rotating regions are separated initially by boundaries that possess non-vanishing gradients $|\partial_i\theta|$ \textit{and} non-vanishing $\dot{\theta}$.
Indeed, the angular speed on these boundaries is roughly equal to the coherent motion in the system: $\dot{\theta}|_{\rm bdry}\sim\langle\dot{\theta}\rangle$.
For sufficiently large $\mathcal{J}_0$ and small enough $\rho_i - \rho_{\min}$, this $\dot{\theta}|_{\rm bdry}$ can be much larger than the differences between the rotation speeds on the two sides of this boundary, so the whole system still rotates in the same sense in field space as the initial conditions.

\begin{figure}
    \centering
    \includegraphics[width=1\linewidth]{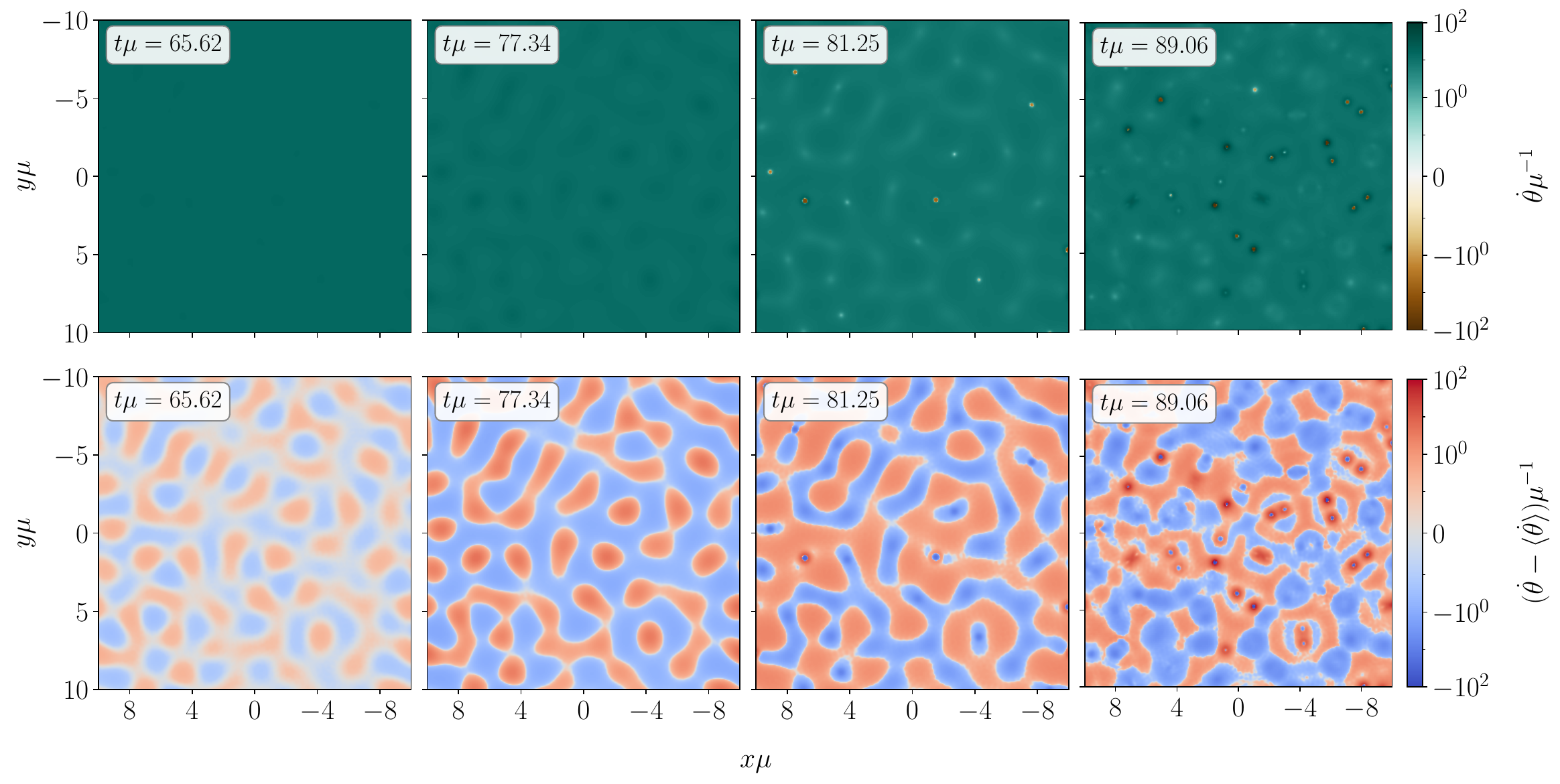}
    \caption{A few snapshots of the state of the system with parameters as chosen in Fig.~\ref{fig:2dexpgrowthLargec0} during the parametric resonance phase ($t\mu\lesssim 80$) and the vortex formation phase ($t\mu\gtrsim 80$). 
    \underline{\textsc{Top row:}} The angular speed $\dot{\theta}$, revealing the emergence of counter-rotating regions (isolated orange regions) at late times ($t\mu \gtrsim 80$, last two columns).
    \underline{\textsc{Bottom row:}} The spatial-average-subtracted angular speed, revealing the earlier existence ($t\mu \lesssim 80$) of a global pattern of differentially rotating regions (larger alternating red and blue patches).
    \ZenodoComment}
    \label{fig:diff_rot}
\end{figure}

As shown in Fig.~\ref{fig:diff_rot}, no vortex formation has yet occurred prior to $t \mu \sim 80$ on the boundaries of the differentially rotating regions,%
\footnote{%
    Note, the pattern of $\dot{\theta}-\langle\dot{\theta}\rangle$, shown in the bottom row of Fig.~\ref{fig:diff_rot}, remains fixed, while its amplitude grows exponentially and oscillates with a frequency smaller than the growth rate; i.e., the amplitudes behaves as $\sim\exp[-i(\omega_r+i\Gamma)t]$, with $\omega_r>\Gamma>0$.} %
despite the large differential rotation and the large $|\partial_i \theta|$ that exists there. 
Rather, the system continues to evolve until counter-rotating regions form around $t \mu \sim 80$ (in between the second and third columns of Fig.~\ref{fig:diff_rot}). 
These counter-rotating regions contain opposite sign, but similar magnitude, charge densities ${j}_0$, and are therefore unequal in size in the large-$\mathcal{J}_0$ regime: the regions that co-rotate with $\dot\theta_i$ are significantly larger due to the large non-zero $\mathcal{J}_0$. 
Indeed, the ratio of the sizes of the co-rotating and counter-rotating regions stays roughly constant towards late times since $\int d^D\!x\, j_0$ is conserved.
After they form, these counter-rotating regions are now separated by domain boundaries on which $\dot{\theta}=0$. 
Regardless of their area, once counter-rotating regions possessing a large $\dot{\theta}$ of the opposite sign to the average/initial $\langle\dot{\theta}\rangle$ have formed, vortices can start to pair produce on the $\dot\theta=0$ domain boundaries that surround them (last column of Fig.~\ref{fig:diff_rot}).

As we show in Fig.~\ref{fig:diff_rot}, the counter-rotating regions are much smaller in size compared to the differential-rotating regions from which they emerge. 
Nonlinear interactions are likely responsible for generating these smaller-scale (higher-$k$) modes from the lower-$k$ modes that initially grew due to parametric resonance. 
These higher-$k$ modes (and hence, the appearance of counter-rotating regions) do not appear to have the same level of spatial and temporal coherence than is exhibited in the small-$\mathcal{J}_0$ case (cf.~Fig.~\ref{fig:counterrot}), a fact that will be important for our discussions in the next sections.

In terms of conditions for vortex formation, these findings suggest that a large $\partial_i \theta$ is actually \emph{not} sufficient for triggering vortex formation in regions where $\dot{\theta}$ is nonzero, and perhaps a Lorentz-invariant generalization of the condition that $|\partial_i\theta|$ be sufficiently large, is required.
Specifically, motivated by the condition for Schwinger pair production in an electromagnetic field, we make an informed hypothesis for a generalization of this critical gradient $\partial_i \theta $ condition (cf.~the critical velocity in a superfluid~\cite{feynman1955chapter}): a critical value of $\kappa \equiv -|\partial_\mu \theta|^2 \equiv (\partial_i \theta)^2 - \dot{\theta}^2$, above which vortex pair production may occur.
Such a generalization could be understood also from the perspective of the particle--vortex duality (see App.~\ref{app:cmtvortex} and~\cite{iengo1995vortex}), in which $\dot \theta$ maps onto a magnetic field%
\footnote{
    The $(2+1)$-dimensional Levi-Civita tensor $\epsilon$ is as defined in footnote \ref{ftnt:LeviCivita2Ddefn}.
} %
\begin{align}
    \frac{1}{2 \pi} \epsilon_{0 i j } F^{ij} =  \dot \theta \: .
\end{align}
From this perspective, our hypothesized generalization of the condition for vortex formation is analogous to the condition $F_{\mu\nu}F^{\mu\nu} = E^2 - B^2 > E_c^2$ for Schwinger pair production in general electromagnetic fields (instead of simply $E^2>E_c^2$ in a pure electric field) where $E_c = m_e^2/e$ is the critical field strength for un-suppressed Schwinger electron--positron pair production. 
We do not however rigorously demonstrate this.

We can also find a minimum condition for vortex production on the initial field value $\rho_i$ for the large-$\mathcal{J}_0$ case (but, recall footnote \ref{ftnt:rhoC}). 
For each decade in $\mathcal{J}_0/(v^2\mu)$ between $10^{-1}$ and $10^4$, we determine the critical radial-field amplitude $\rho_c$, beyond which vortices form in our 2-dimensional simulations; see the right panel of Fig.~\ref{fig:veffmin}.
This critical initial field amplitude plateaus below $\mathcal{J}_0\sim v^2\mu$, but begins following a $\rho_c \propto \mathcal{J}_0^{1/3}$ trend towards large $\mathcal{J}_0$. 
Specifically, we find that a good fit to our data is provided by
\begin{align}
    \rho_c\approx 1.13\left(\frac{v}{\mu}\right)^{1/3}\mathcal{J}_0^{1/3} \: ,
    \label{eq:rhoclargeC0}
\end{align}
in the range of $\mathcal{J}_0\gg v^2\mu$ considered here. 
As before, those initial conditions with $\rho_c>\rho_i>\rho_{\min}$ are parametrically unstable, but do not form vortices; once the unstable modes reach the nonlinear regime, they simply dissipate to radiation. 

Naturally, this change in the scaling of $\rho_c$ with $\mathcal{J}_0$ modifies the simple geometric condition that was introduced in \eqref{eq:conditionV0} for small $\mathcal{J}_0$. 
A condition similar to \eqref{eq:conditionV0} cannot be easily generalized to the large $\mathcal{J}_0$ case since $V_{\rm eff}(0)$ is actually infinite; this reinforces the important role played in vortex formation by nonlinear dynamics causing spatial rearrangement of (free) energy densities (regions of $\rho=0$ would naively be forbidden were the Noether charge and free energy to remain homogeneous).
At the critical radial field amplitude, the available energy is given by $V_{\rm eff}(\rho_c)-V_{\rm eff}(\rho_{\min})$, which scales as $\sim \lambda^{1/3}\mathcal{J}_0^{4/3}$ for $\mathcal{J}_0\gg v^2\mu$ when assuming \eqref{eq:rhoclargeC0}. 

The $\mathcal{J}_0^{1/3}$ scaling of $\rho_c-\rho_{\min}$ is not unexpected, for the following reasons.
In the large-$\mathcal{J}_0$ limit, the EoM for the unperturbed radial mode~\eqref{eq:EoM0re2} can be made dimensionless and approximately $\mathcal{J}_0$-independent with the redefinitions $\rho_0 = \hat{\rho}_0\lambda^{-1/6}\mathcal{J}_0^{1/3}$ and $t= \hat{t} \lambda^{-1/3}\mathcal{J}_0^{-1/3}$, and an appropriate overall rescaling:
\begin{align}
    \partial_{\hat{t}}^2 \hat{\rho}_0 - \hat{\mu}^2 \hat{\rho}_0 - \frac{1}{\hat{\rho}_0^3} + \hat{\rho}_0^3 &= 0\:; & \hat{\mu} &\equiv \frac{\mu}{\lambda^{1/3}\mathcal{J}_0^{1/3}}\:.  \label{eq:nonDimEoM}
\end{align}
The place where $\mathcal{J}_0$ appears in this dimensionless equation is the dimensionless coefficient $\hat{\mu}^2 \propto \mathcal{J}_0^{-2/3}$, which is subdominant in the large $\mathcal{J}_0$-limit. 
This suggests that, in this limit, the evolution of the system in terms of the fields $\hat{\rho}$ and $\hat{t}$ should be $\mathcal{J}_0$-independent during the growth of the perturbations; this in turn implies that the behavior of the system during this period should be almost invariant under scalings of the initial conditions that obey $\rho_i \propto \lambda^{-1/6}\mathcal{J}_0^{1/3}$.
However, at the onset of the formation of the counter-rotating regions, the domain boundaries emerge as regions where $\dot \theta \approx 0$, implying both that $j_0$ [cf.~\eqref{eq:noethercurrent}] becomes small  on the boundaries, and also that there is a large field perturbation at the boundary location, which invalidates the naive application of the perturbative approach that led to \eqref{eq:EoM0re2} and thence to \eqref{eq:nonDimEoM} in the understanding of the nonlinear field dynamics. 
Moreover, to the extent that \eqref{eq:EoM0re2} or \eqref{eq:nonDimEoM} do still give some guidance, a term morally of the form $-\hat{\mu}^2 \hat{\rho}_0$ will become relevant on the boundaries, invalidating the scaling we found above at large $\mathcal{J}_0$.
Unfortunately, the locations where these failures occur are precisely the domain boundaries, which is where vortices form in pairs.
As a result of all of these issues, we cannot prove analytically that $\rho_c -\rho_{\min}$ should scale as $\lambda^{-1/6}\mathcal{J}_0^{1/3}$.
Nevertheless, the above argument is suggestive that we should, in the large $\mathcal{J}_0$ limit, expect similar system behavior for the perturbations as they grow toward the counter-rotating regions that precede vortex formation under the rescaling $\rho_i \propto \lambda^{-1/6}\mathcal{J}_0^{1/3}$ as $\mathcal{J}_0$ is varied.
It is therefore not unexpected that similar vortex formation dynamics would be at play under such rescaling.
And, indeed, the numerical results in fact indicate that the scaling $\rho_c -\rho_{\min}\propto \mathcal{J}_0^{1/3}$ does obtain in the large $\mathcal{J}_0$ limit.

\subsection{Vortex Formation in Three Spatial Dimensions}
\label{sec:VortexFormation3D}

In two spatial dimensions, pairs of vortices are efficiently produced on domain boundaries, where the magnitude of the spacetime gradient of the angular mode is sufficiently large. 
In three spatial dimensions, on the other hand, strings in our setting must form in loops.%
\footnote{
    By contrast, in condensed matter systems, strings may end on material boundaries~\cite{tinkham2004introduction,Galaiko1966FormationOV}.
} %
Therefore, it is important that we verify the string-production mechanism in three spatial dimensions behaves similarly to its 2D counterpart. 
To that end, we numerically solve the full nonlinear scalar field evolution, determined by \eqref{eq:EoM}, in $D=3$ spatial dimensions.
As before, we impose periodic boundary conditions and set initial conditions as described in Sec.~\ref{sec:pertfullnonlinear}. 
We focus on the example case $\mathcal{J}_0=0.1\mu v^2$, $\rho_i/v=5$, and $\sigma_{\rho_i}=10^{-2}\rho_i$.

\begin{figure}[t]
    \centering
    \includegraphics[width=0.99\linewidth]{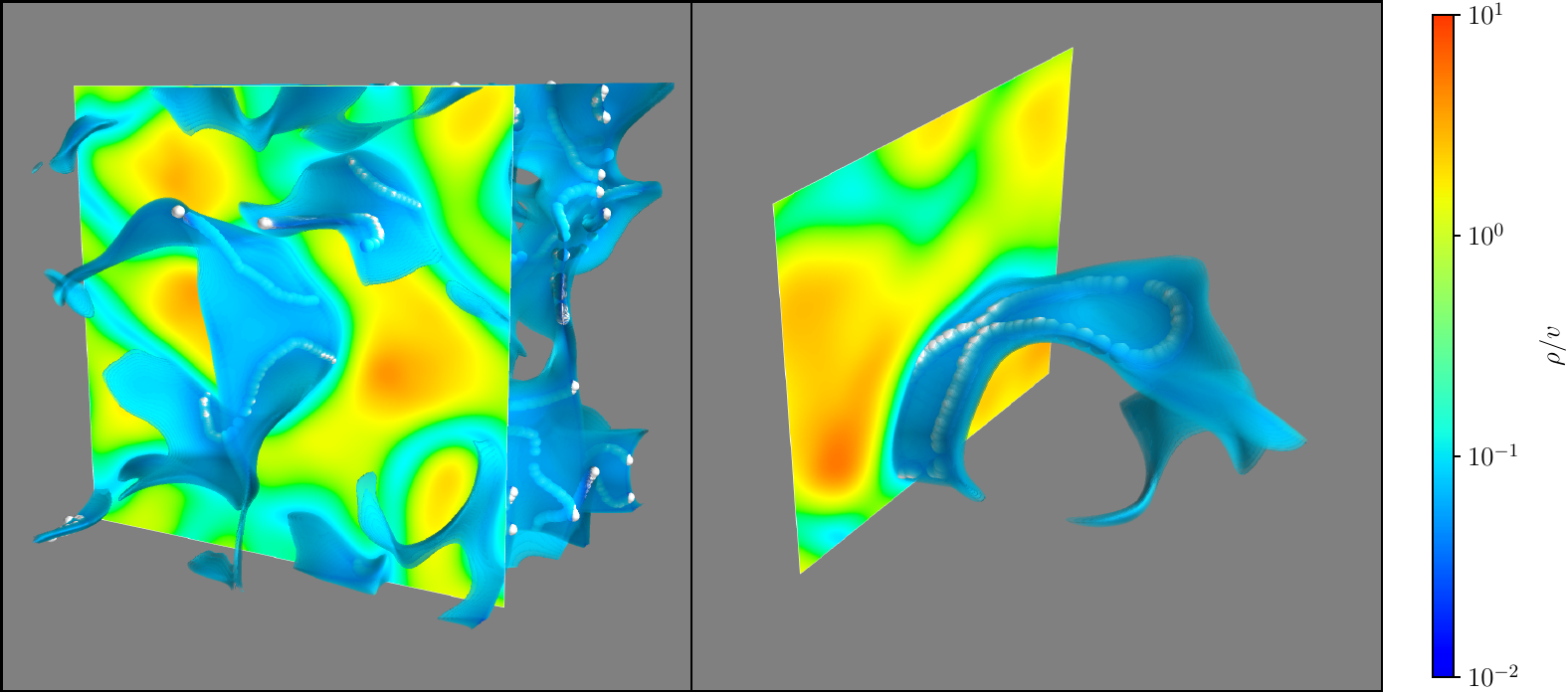}
    \caption{A three-dimensional isosurface plot of the magnitude of the radial mode; the outermost surface corresponds to $\rho/v=0.1$. 
    Additionally, we plot this magnitude on a single plane that slices the volume (see legend) for visual aid. 
    Lastly, the strings are continuous lines that thread through the white dots indicated in the plots (the white dots indicate where a cell of the discretized simulation grid is pierced by a string; see App.~\ref{app:simdetails} for details). 
    The left panel shows a subset of the simulation domain, with side length $4/\mu$, while the right panel shows a subset of side length $2/\mu$, with both showing the same temporal snapshot taken at a time after both the saturation of the parametric resonance and the onset of string formation.
    These results are shown for parameters $\mathcal{J}_0=0.1\mu v^2$, $\rho_i/v=5$, and $\sigma_{\rho_i}=10^{-2}\rho_i$.
    }
    \label{fig:formation3D}
\end{figure}

The evolution of the system through the parametric resonance proceeds as in the two-dimensi- onal case. 
Once the unstable modes reach the nonlinear regime, string loops form; in fact, just like the 2D case, the strings form on codimension-1 domain boundaries separating counter-rotating regions of the angular mode. 
In Fig.~\ref{fig:formation3D}, a snapshot of the simulation shows both an isolated string loop, as well as the larger string network, just after string formation has first occurred. 
The domain boundaries, characterized by small $\rho/v$, form a sponge-like filament network with characteristic spatial scale set by the fastest-growing parametrically unstable Fourier mode (as in two dimensions). 
Therefore, we find the \textit{entire} string network to be embedded in the codimension-1 domain boundaries. 
For $\mathcal{J}_0\ll v^2\mu$ (and sufficiently large $\rho_i>\rho_{\min}$), note that the string length can be comparable to the simulation box size $2L$ (which is at least an order of magnitude larger than $1/\mu$ in all simulations).
Due to the similarity between the two- and three-dimensional cases at small $\mathcal{J}_0$, we expect string formation in the $\mathcal{J}_0\gg v^2\mu$ regime to proceed in direct analogy to its two-dimensional counterpart.

\subsection{Summary}
\label{sec:Sec3Summary}

To summarize, we have demonstrated that vortex--anti-vortex pairs (2D) or string loops (3D) form on domain boundaries between counter-rotating regions. 
Specifically, we observe that vortex pair production occurs in 2D simulations for both small and large (but not fine-tuned) initial $ U(1)$ charge density $\mathcal{J}_0$, and that string loops form in 3D simulations for small $\mathcal{J}_0$.
Given the similarities of the underlying formation dynamics that we find in the 2D and 3D cases at small $\mathcal{J}_0$, we also expect that analogous formation dynamics would occur in 3D at large $\mathcal{J}_0$ as occur in 2D for that same case.

We have identified simple conditions for vortex formation. 
Microscopically, formation occurs when the Lorentz scalar $\kappa = -|\partial_\mu \theta|^2 = (\partial_i \theta)^2 - \dot{\theta}^2$ is large enough. 
Practically, this corresponds to when the gradient field $|\partial_i \theta|$ on the domain boundary (where $\dot{\theta}$ is zero or close to zero) is large enough, and, as a result, the difference in $\theta$ across the domain boundary reaches $\Delta \theta \sim \mathcal{O}(2\pi)$. 
This microscopic condition can be understood with analogies to vortex formation in a superfluid, as well as with the help of the particle--vortex duality. 
A large difference in $\theta$ across the domain boundary is generated by large opposite-sign rotation velocity $\dot \theta$ in adjacent counter-rotating regions, whereas the energy that sustains this counter rotation is supplied by the initial potential energy (equivalently, the initial displacement of the radial mode).
This allows us to also identify simple macroscopic criteria for vortex formation on the initial conditions. 
For small $\mathcal{J}_0$, this is a simple condition on the initial potential energy density: surprisingly, we find that vortices can form when $V(\rho_i) \simeq 0.29 V(0)$, suggesting that vortex production occurs, even if there is not enough energy to reach $\rho = 0$ globally. 
For large $\mathcal{J}_0$, similarly, we find that vortices form even with a relatively small-amplitude initial oscillation around the minimum $\rho_{\rm min}$ of the effective potential $V_{\rm eff}$; that is, if $\rho_i\gtrsim 1.13\times\mathcal{J}_0^{1/3}(v/\mu)^{1/3}$.
These criteria resemble the conditions for gauge-string formation in the gauged $U(1)$ case studied in~\cite{East:2022rsi}: vortex formation can occur locally even if there is not enough energy to restore the $U(1)$ symmetry globally.

In fact, the string-formation mechanism described in this section can also be utilized to form gauged strings, and can be understood as one of the underlying mechanisms that leads to formation of strings in a large-initial-amplitude coherent background of the longitudinal mode of the massive gauge boson. 
This mechanism was first identified by Pearl~\cite{1964ApPhL...5...65P} as a means to produce vortices in thin-film superconductors. 
This intuition might suggest that the global vortices/strings would behave relatively similar to the gauged vortices considered in~\cite{East:2022rsi} after they first formed. 
However, as we discuss next, the evolution of vortex--anti-vortex pairs (2D), and string loops (3D), is actually very different from what was found in~\cite{East:2022rsi}.

\section{Vortex Evolution}\label{sec:dynamics}

In this section, we focus on the evolution of the vortices/strings following their formation, considering both the two and three dimensional cases. 
Some qualitative intuition for the dynamics of the vortices/strings after formation can be gained from knowledge of vortex dynamics in superfluid helium, or Pearl vortex dynamics in thin-film superconductors with a large bias supercurrent~\cite{PhysRevB.84.174510}. 
In a superfluid, the interaction between the vortices and the background flow is through the Magnus force~\cite{Ao:1993zz,1985PhRvL..55.2887H,Thouless:1996mt,1993PhyA..200...42T,Sonin_1997} (see also~\cite{Ao:1993zz,Levin:2023ewn}):%
\footnote{\label{ftnt:CrossProductCheat}%
    Note that \eqref{eq:Magnus} is unambiguous in $D=3$, but involves a minor (but transparent) swindle in $D=2$.
    In the latter case, the 2D plane can be thought of as being embedded at $z=0$ in a 3D Euclidean space.
    The circulation vector $\vec{K}$ is then taken to be $\vec{K} \propto +\hat{e}_3$ for vortices and $\vec{K} \propto -\hat{e}_3$ for anti-vortices, both being normal to the 2D plane.
    Both of the velocities are lifted from 2-vectors to 3-vectors via the addition of a conventional zero component entry in the $\hat{e}_3$ direction.
    The cross-product is then computed as normal in $\mathbb{R}^3$, yielding a 3-vector force that automatically has a zero component entry in the $\hat{e}_3$ direction; it can thus be consistently dropped back to a 2-vector force in the 2D plane.
    In other words, in 2D we have $F_M^i \propto \epsilon^{0ij} K ( v_v^j - v_{\text{bkg}}^j)$, where $K$ is the pseudoscalar 2D vorticity ($K >0$ for vortices; $K<0$ for anti-vortices).
} %
\begin{align}
    \vec{F}_M \propto \vec{K} \times (\vec{v}_{v}-\vec{v}_{\rm bkg})\:,  \label{eq:Magnus} 
\end{align}
where $\vec{K}$ is the circulation vector of the vortex as defined in Sec.~\ref{sec:globalstring}, $\vec{v}_{v}$ is the speed of the vortex, and $\vec{v}_{\rm bkg} \simeq \nabla \theta/\mu$ is the velocity of the superfluid flow (see App.~\ref{app:cmtvortex}). 
An analogy, first provided in~\cite{Ao:1993zz}, between vortex motion in a superfluid flow and electron motion in a background magnetic field (i.e., the Hall effect) is instrumental in understanding the vortex motion in two dimensions (see also~\cite{RevModPhys.59.1001} for a review). 
In a superfluid (or thin-film superconductor), the Magnus force transfers energy from the background superfluid flow (respectively, the supercurrent) to the motion of vortices, which eventually leads to thermalization~\cite{RevModPhys.59.1001}. 

In the following, we will discuss how this intuition from the superfluid/superconductor cases can assist us in qualitatively understanding many of the simulation results of vortex dynamics in two dimensions in the system at hand, since the action that describes a superfluid (the Ginzburg--Landau model) is similar to the one studied here (see App.~\ref{app:cmtvortex}). 
However, our system also differs from a superfluid in several key aspects. 
First, after counter-rotating regions have formed, most regions have a large $|\dot{\theta}|$ and timelike gradient $\partial_\mu{\theta}\partial^\mu{\theta}$; by contrast, on the domain boundary, $\dot{\theta}$ is close to zero while the gradient $\partial^i {\theta}$ is large, rendering $\partial_\mu{\theta}\partial^\mu{\theta}$ spacelike. 
Second, in the counter-rotating regions, the radial mode is displaced from the minimum to $\rho > v$, while on the domain boundaries between these regions, we have $\rho \lesssim v$. 
Both these effects may contribute to confinement of vortices (2D) or strings (3D) to the domain boundaries. 
These key differences between the condensed matter systems we are analogizing to, and the system we consider, have interesting dynamical consequences for our vortices, which we elaborate on in this section.
That said, as we have already emphasized, vortices in an excited background field configuration (as found here) should not be thought of as weakly interacting particle-like objects subject to dynamical evolution via interaction with a background field (they can often be thought of in this way in condensed-matter contexts, where the background is quiescent); rather, there is only the $\Phi$ field, which is behaving highly dynamically, and the vortices are a feature that appears in the evolution of that single complex scalar field.

\subsection{Local Vortex Confinement and Annihilation}
\label{sec:ConfinmentAndAnnihilation}

In this subsection, we discuss the vortex dynamics after pair creation has occurred. 
In each of the next few paragraphs, we first describe the important features observed in the numerical simulation, and then provide qualitative understandings of each of these features, sometimes with the help of analogies to the superfluid system, or particle--vortex duality. 
In the following, we first focus on the $\mathcal{J}_0\ll v^2\mu$ regime; however, we observe qualitatively similar behavior for $\mathcal{J}_0\gtrsim v^2\mu$ as well.

\begin{figure}[t]
    \centering
    \includegraphics[width=1\linewidth]{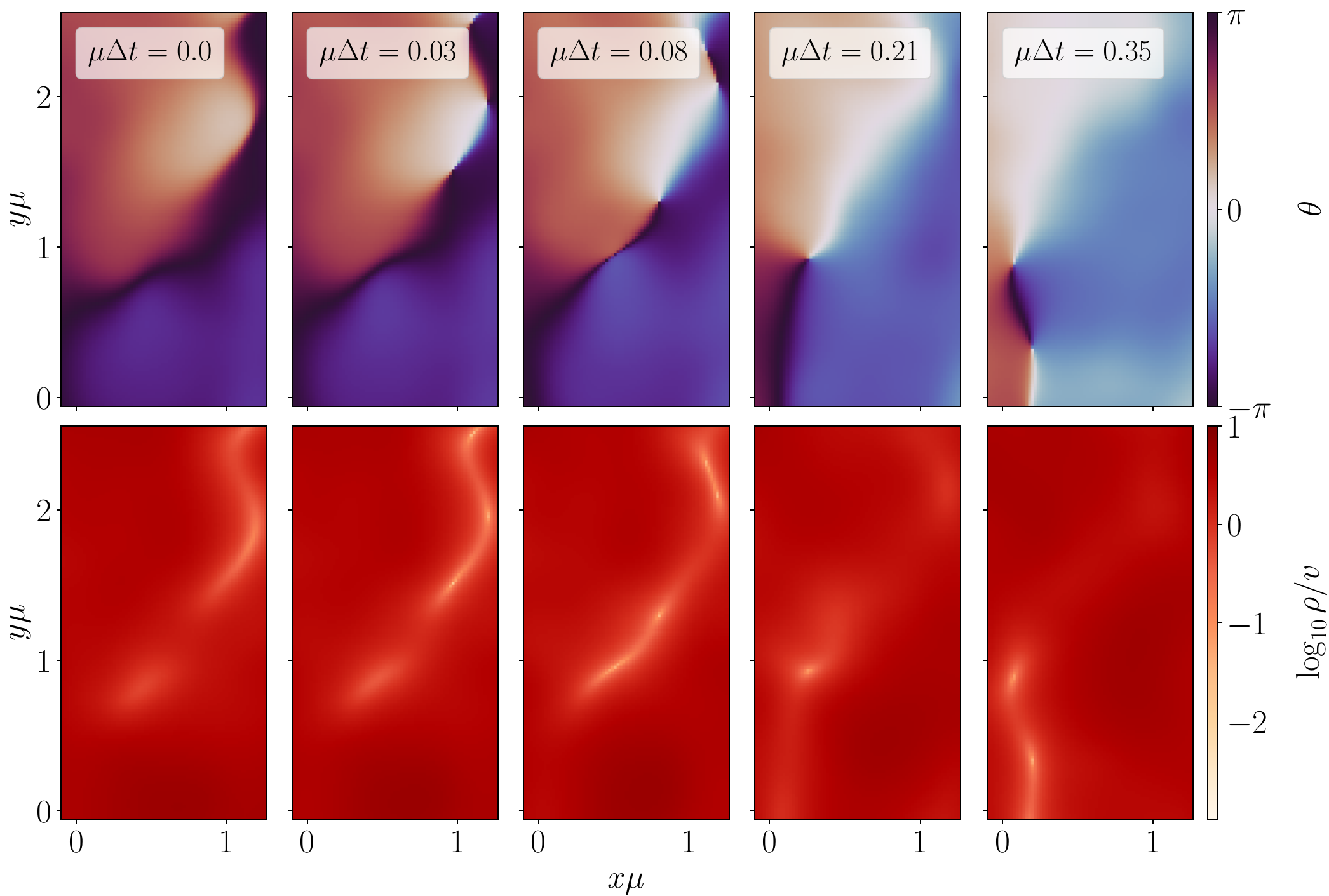}
    \caption{A sequence of temporal snapshots of the state of the system with $\mathcal{J}_0=0.1\mu v^2$, $\rho_i/v=5$, $\sigma_{\rho_i}=10^{-2}\rho_i$, and $D=2$, shown as a function of position in (a zoomed-in subset of) the simulation domain. 
    The sequence shows the production and subsequent propagation of several vortex--anti-vortex pairs along the domain boundary that can be seen in the left-most column extending diagonally from the top-right down to the bottom-left of the spatial region depicted. 
    The time $\Delta t$ indicates the elapsed time since the left-most column. 
    We show both the angular (top row) and radial (bottom row) degrees of freedom. 
    The vortices here again move superluminally, but this is not a violation of causality; see the \protect\hyperlink{text:superluminal}{discussion} in Sec.~\ref{sec:vortexformationdynamics}. 
    \ZenodoComment}
    \label{fig:stringmove}
\end{figure}

In Fig.~\ref{fig:stringmove}, we show (for $D=2$) a segment of a domain boundary on which vortex--anti-vortex pairs are created.
Such events occur at several sites along the boundary.
Once created, vortex--anti-vortex pairs separate and move apart along the domain boundary between the counter-rotating regions.
We can roughly understand this behavior in the condensed-matter analogy, where the motion of the vortices after their birth is determined by their interactions with the background field profile from which they emerge, along with any interactions they have with other vortices, etc. 
While a vortex--anti-vortex pair generally attract~\cite{tinkham2004introduction}, superfluid vortices can also be dynamically driven apart by the background field via the Magnus force~\cite{Ao:1993zz}.
In our setting, the background's field phase gradient, which is analogous to the superfluid flow velocity, is perpendicular to the domain boundaries (see, e.g., Fig.~\ref{fig:vortexformation}), so that this Magnus force would be correctly aligned to drive oppositely handed vortices apart, despite the attractive vortex--anti-vortex interaction.

Furthermore, we observe that the vortices are confined to the domain boundaries.
Naively, this confinement is surprising, since the domain boundaries are generically curved and the vortices are fast-moving.
A few interactions may be responsible for this confinement. 
First, the domain boundary has smaller $\rho$ compared to the nearby regions, and it costs more energy for a vortex to be in a region where the background's radial mode is larger. 
Second, recall that, in the dual picture, the time component of the particle current, $j^0 \sim \dot{\theta}$, maps onto a magnetic field $F^{ij}$. 
Vortices that attempt to move off the domain boundary and enter the counter-rotating regions will therefore be pushed back onto the boundary (similar to charged particles that enter a strong magnetic field). 
Ultimately, it is likely that a combination of both the profile of the radial mode, and the interaction between the vortex and the background field in the counter-rotating regions, contribute to this confinement.

\begin{figure}[t]
    \centering
    \includegraphics[width=0.582\linewidth]{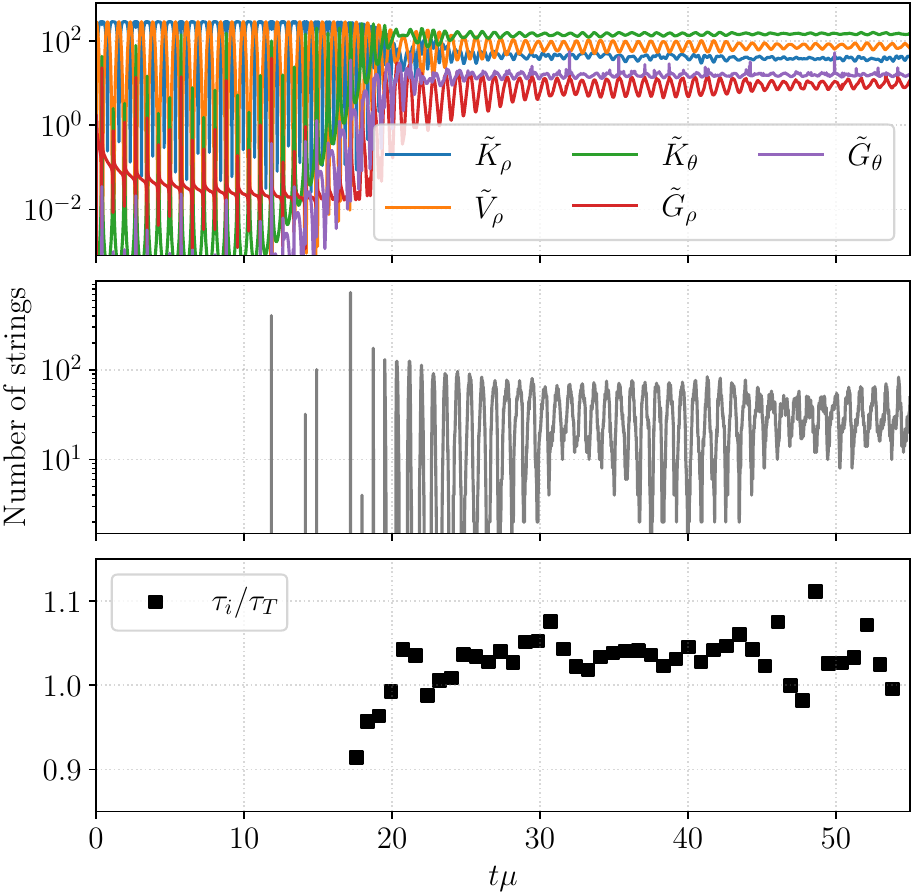}
    \includegraphics[width=0.408\linewidth]{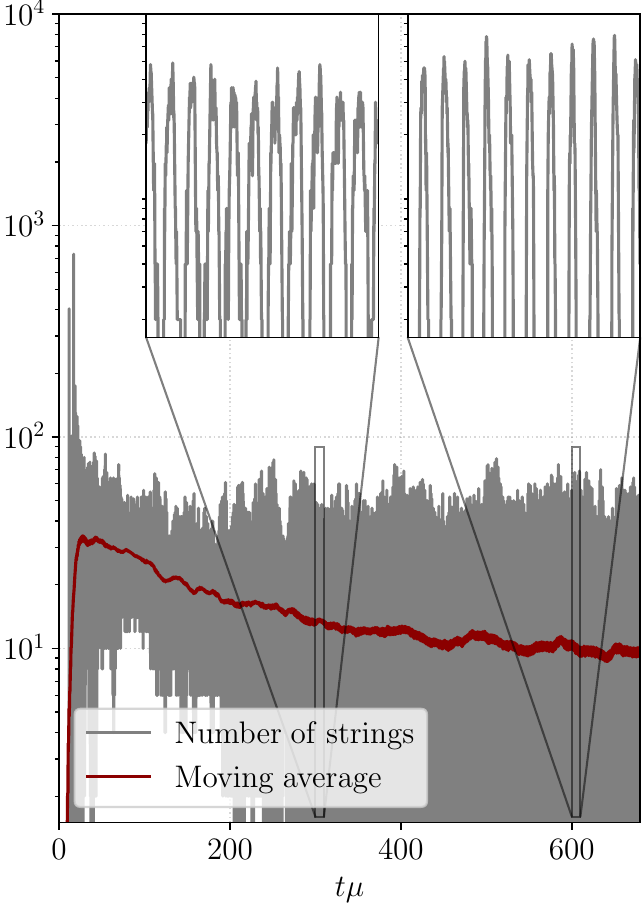}
    \caption{A continuation to later times of the dynamics shown in Fig.~\ref{fig:2dexpgrowth} for the parameters $D=2$, $\rho_i/v=5$, $\sigma_\rho=10^{-2}\rho_i$, and $\mathcal{J}_0=0.1v^2\mu$. 
    \underline{\textsc{Left column, top row:}}~%
    Relevant normalized energy densities $\tilde{K}_\rho=K_\rho v^{-2}\mu^{-2}$, etc.~through the instability saturation and vortex formation. 
    \underline{\textsc{Left column, center row:}}~%
    The total number of strings in the two-dimensional simulation domain of volume $(20/\mu)^2$. 
    Note that the overall scale for the total number of strings depends of course on the volume of the simulation domain; the important information presented here is how the number of strings in a fixed box changes in a relative sense over time.
    \underline{\textsc{Left column, bottom row:}}~%
    The time difference $\tau_i$ between sequential instances of maximal number of strings in units of $\tau_T\equiv \pi/\dot{\vartheta}_{\text{sim}}^{\text{avg}}$, where $\dot{\vartheta}_{\text{sim}}^{\text{avg}}$ is the moving time-average (taken over an interval $T=8/\mu$) of the spatial average of the magnitude of the angular speed, $\langle|\dot{\theta}|\rangle$. 
    Note that the offset of the ratio $\tau_i/\tau_T$ from unity is consistent with the uncertainties of our numerical methods.
    \underline{\textsc{Right column:}}~%
    As for center row of the left column, but extending to much later times.
    \ZenodoComment
    }
    \label{fig:largecirculation_energy}
\end{figure}

Vortices of opposite vorticity that were created in separate, but contemporaneous, pair-creation events occurring at widely separated spatial locations move \textit{towards} each other along the domain boundaries, and annihilate when they encounter each other. 
This can be seen in the last few columns of Fig.~\ref{fig:stringmove}. 
This behavior, which we identify in the two-dimensional simulations, has a close analogue in the 3D case: we find that string segments with locally oppositely oriented circulation vectors find each other and annihilate efficiently in 3D.
In two dimensions, a crucial ingredient for efficient annihilation is that all the vortices born along any domain boundary have alternating vorticity;%
\footnote{%
    The situation in three dimensions is more subtle; in this case, the crucial ingredient is essentially that strings are born as closed loops that live on two-dimensional surfaces whose shape and topology do not evolve significantly on the timescale of the string lifetime, so that a loop can contract back to zero size and annihilate away, instead of undergoing a cascade.
    } %
i.e., along the boundary the vortices alternate as vortex--anti-vortex, or anti-vortex--vortex. 
Such an arrangement is expected from the dual picture, where the dual electric field points \emph{along} the domain boundaries, and the vortex--anti-vortex pairs map onto a particle--anti-particle pair: Schwinger charged-particle pair creation in a large electric field takes place with all the pairs aligned in the same way in the electric field, which maps back to vortices lining up with alternative vorticity along the domain wall.
As a result, efficient vortex (respectively, string) confinement can lead to a \textit{complete} annihilation of the vortex (string) network in two (and even in three) spatial dimensions; this is particularly surprising in three dimensions where strings generally cross each other at an angle, as discussed below. 
To illustrate these annihilation dynamics, we show the evolution of the energy densities, as well as the total number of strings, from the onset of the parametric resonance through vortex formation, in Fig.~\ref{fig:largecirculation_energy} for the same case previously presented only at early times in Fig.~\ref{fig:2dexpgrowth}. 

Towards the end of the parametric resonance, large quantities of vortices form in pair-creation events, separate, and then subsequently annihilate efficiently with vortices formed in neighboring pair-creation events.
This process repeats periodically at a characteristic frequency set by the rate of change of the phase difference between the co- and counter-rotating regions located on either side of a domain boundary; i.e., $\partial_t( \Delta\theta)=\Delta \dot\theta =\dot{\theta}_{\rm co}-\dot{\theta}_{\rm coun}$.
This can be understood as follows: we found in Sec.~\ref{sec:formation_conditions} that vortices form for sufficiently large gradients, $|\partial_i\theta|$, across the domain boundary.
Since the angular mode $\theta$ evolves in opposite senses on either side of a domain boundary, the difference between the value of $\theta$ on either side (and hence, the spatial gradients $|\partial_i\theta|$) is maximized with an oscillation period $\tau\approx 2\pi/\Delta\dot{\theta}$. 
For $\mathcal{J}_0\ll v^2\mu$ (and, in particular, for the case shown in Fig.~\ref{fig:largecirculation_energy}), we have roughly $\Delta\dot{\theta}\approx 2\dot{\theta}_{\rm co}$ because $\dot{\theta}_{\rm coun}\sim - \dot{\theta}_{\rm co.}$.
Therefore, the timescale is $\tau\approx\pi/\langle|\dot{\theta}|\rangle$. 
During the other phase of the cycle, when $|\partial_i\theta|$ is insufficient locally to trigger vortex formation, no new vortex pairs can be produced, and the existing (anti-)vortices travel along the domain boundaries and annihilate efficiently with other, opposite-vorticity defects also confined on the boundaries. 
As can be seen in the center panel of the left column of Fig.~\ref{fig:largecirculation_energy}, this process leads to periods of time, after the first vortex-production epoch, at which all vortices in the domain have annihilated; in particular, this occurs in the first few periods after saturation of the parametric resonance.
In the bottom left panel of Fig.~\ref{fig:largecirculation_energy}, we show that the spacing between successive epochs of peak string formation and annihilation is indeed $\sim \tau$ for $\mu \tau \gtrsim 20$; at earlier times, the system is still in a transient state where the energy densities are redistributing. 
There is then an intermediate epoch in which complete annihilation is less likely (see Sec.~\ref{sec:largedisplacement} for further discussion); however, the remaining coherent motion (on the spatial scale of the counter-rotating regions) in the system still leads to periodic oscillations in the number of vortices with a timescale $\sim\tau$.
And, at later times, durations of complete annihilation re-emerge.
On very long timescales, $\sim\mathcal{O}(10^2)\times \tau$, we observe a slight \textit{decrease} in the vortex density by roughly a factor of two for the case considered in Fig.~\ref{fig:largecirculation_energy} (see the right column); we discuss this in further detail in the following section. 
While we have only explicitly demonstrated phase coherence of the periodic appearance and (at least partial) annihilation of the vortices on spatial scales set by the simulation box that we chose (i.e., $L\sim 20/\mu$ in the case shown in Fig.~\ref{fig:largecirculation_energy}), we expect this coherence also on larger scales as this is already in the regime $L\gg \mu^{-1}$.

\begin{figure}[t]
    \centering
    \includegraphics[width=0.95\linewidth]{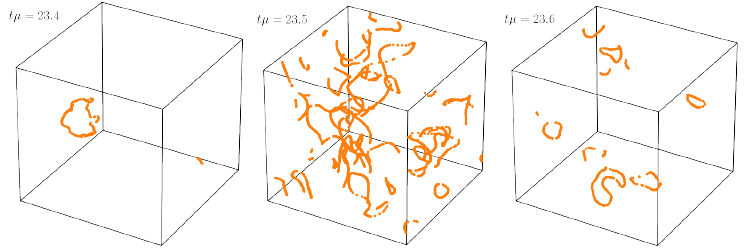}
    \caption{The string network evolution through a single cycle of string loop production (left), further loop production, expansion and interaction (center), and subsequent annihilation (right) in three spatial dimensions.
    The parameters of this case are the same as in Fig.~\ref{fig:largecirculation_energy}, but with $D=3$. 
    The box size shown is $(4/\mu)^3$.
    Each orange dot represents a cell of the discretized simulation volume that is pierced by a string; see App.~\ref{app:simdetails}.
    }
    \label{fig:3d_string_anni}
\end{figure}

In Fig.~\ref{fig:3d_string_anni}, we show the string creation, propagation, and annihilation process in three spatial dimensions, over a full cycle. 
Here, strings are created in loops confined to the domain walls (compare also Fig.~\ref{fig:formation3D}). 
Analogously to the two-dimensional case discussed above, these strings are confined to move on the two-dimensional domain-boundary filaments. 
Naively, one may expect strings in three-dimensions always intersect at an angle, and hence never completely annihilate with each other. 
However, their dynamics are strongly modified due to their confinement to a domain boundary that has a thicknesses comparable to the string core size: they thus effectively live on a surface that is one spatial dimension lower than the spacetime (i.e., the strings' behavior on the domain boundaries is more akin to that of codimension-1 defects, than that of codimension-2 defects).
As a result, we observe that the entire string network formed from loops begins interacting and efficiently annihilating throughout a single period $\sim \tau$, in direct analogy to the two-dimensional setting; see Fig.~\ref{fig:3d_string_anni}. 
The late-time evolution of the system is discussed in detail in the next section.

Before we close this subsection, it is beneficial to contrast the findings regarding the post-formation behavior of the strings in the global $U(1)$ case presented here with those for the gauged $U(1)$ case studied in~\cite{East:2022ppo,East:2022rsi}.%
\footnote{For the avoidance of doubt, we note that the string formation mechanisms in the global and gauged cases under discussion here are very different; we intend here to compare only the \emph{post-formation} behavior of the strings.} %
In the gauged $U(1)$ case, the number of vortices grows almost monotonically after initial formation due to interactions between the vortices and the background field; by contrast, in the global $U(1)$ case here, the number of vortices actually oscillates periodically and slowly decreases over long timescales.
This difference in post-formation behavior mainly comes from the fact that the global vortices are confined to the domain boundaries, and therefore annihilate much more efficiently as compared to the gauged case, whereas the gauged vortices can absorb almost all the energy from the background gauge field, and therefore persist. 
Particularly in the three-dimensional setting, the overall evolution of the string network after formation is different. 
In the gauged $U(1)$ case, the strings, via their interactions with the background gauge fields and each other, quickly cascade to a large number of short strings, whereas the strings in the global $U(1)$ case instead form and annihilate periodically.

\subsection{Late-Time Evolution of the String Network: Small Initial \texorpdfstring{$\mathcal{J}_0$}{J0}}
\label{sec:lateTimeStringEvol}

We have seen that, during the growth of the perturbations driven by parametric instabilities, once the counter-rotating regions have emerged, strings are formed (confined to the domain boundary) 
and annihilated periodically with frequency $\tau^{-1}=\langle|\dot{\theta}|\rangle/\pi$.
In this subsection, we outline how, on longer timescales (i.e., $\gtrsim \mathcal{O}(10^2)\times \tau$), the system exhibits two qualitatively distinct late-time behaviors, depending on the size of the initial radial-mode amplitude $\rho_i$.
While the numerical results in this subsection are obtained with $\mathcal{J}_0=0.1v^2\mu$, the qualitative behaviors discussed are generic in the small-$\mathcal{J}_0$ case (as defined in Sec.~\ref{sec:formation_conditions}).

\subsubsection{Large Initial Displacement}\label{sec:largedisplacement}

\begin{figure}[t]
    \centering
    \includegraphics[width=1\linewidth]{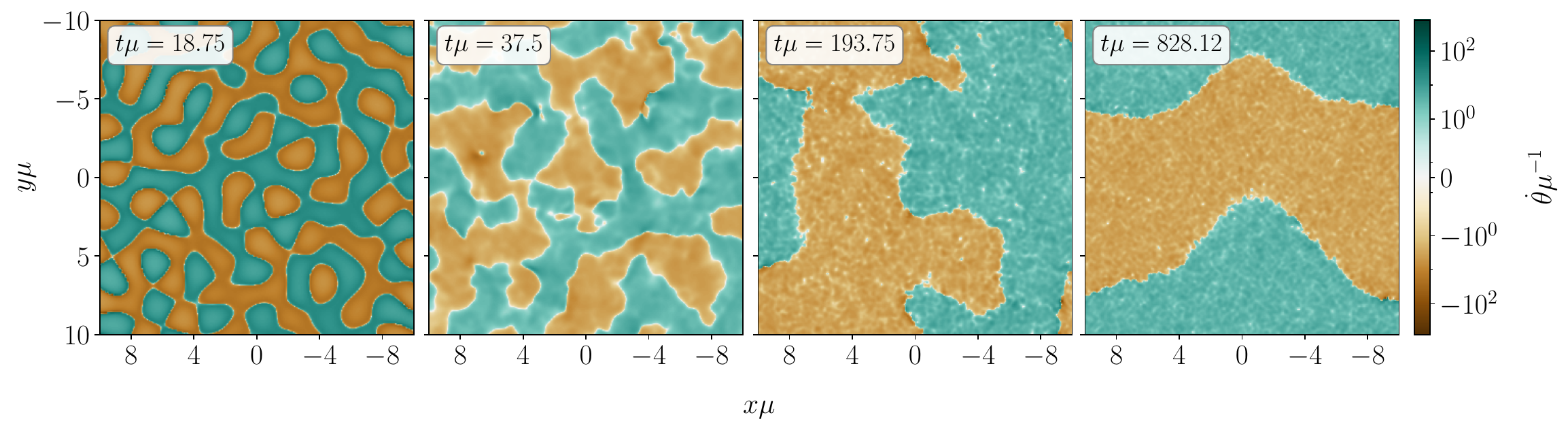}
    \caption{The evolution of the counter-rotating regions from their creation during the parametric resonance phase, through initial vortex formation and annihilation cycles, to the late-time endstate of the system, in two spatial dimensions. 
    These results are for the parameter choices $D=2$, $\rho_i/v=5$, $\sigma_\rho=10^{-2}\rho_i$, and $\mathcal{J}_0=0.1v^2\mu$ that we have considered throughout this work.
    For reference, the (time-averaged) total number of vortices in the domain falls by roughly a factor of two between the second and last panels. 
    \ZenodoComment}
    \label{fig:merger_conter_rot}
\end{figure}

We begin by focusing on those scenarios with large initial radial amplitude $\rho_i$, but small angular speed $\dot{\theta}_i$.  
As we have seen above, vortices are produced and almost completely annihilated twice per period of the $\theta$ evolution around the unit circle in each of the counter-rotating regions; i.e., with period $\tau \sim \pi/\langle|\dot{\theta}|\rangle$. 
Specifically, this is the behavior of the system in the first few $\tau$ after the emergence of the counter-rotating regions.
In Fig.~\ref{fig:merger_conter_rot}, this corresponds to the time between the first and second panels.
The complete annihilation of vortices that occurs during each cycle is made possible by the domain boundaries being (roughly) one-dimensional closed surfaces in $D=2$ (or two-dimensional closed surfaces in $D=3$) that do not evolve significantly in shape (or topology) on timescales $\sim \tau$; hence, any vortex which is dominantly confined to the domain boundary can find a corresponding anti-vortex to completely annihilate on timescales $\ll\tau$. 
On timescales much longer than one such period, the counter-rotating regions begin to merge, as can be seen in Fig.~\ref{fig:merger_conter_rot}. 
During this merger process, the complete, simultaneous vortex annihilation throughout the simulation box within each $\sim \tau$ period can become temporarily inefficient (see e.g., around $t\mu\gtrsim 30$ in Fig.~\ref{fig:largecirculation_energy}).
At very late times, once this merger process completes, the epochs of complete annihilation return.

The merger of counter-rotating regions is active on timescales $\gg \tau$, as can be seen in Fig.~\ref{fig:merger_conter_rot}, lasting until a state consisting of one single connected region with $\dot{\theta}>0$ and another with $\dot{\theta}<0$ forms.
This late-time merging may be expected because the domain boundaries carry energy densities proportional to their length (i.e., surface tension).%
\footnote{%
    Throughout this paragraph, wherever we refer to the ``length'' of a domain boundary or the ``area'' of a region, we are invoking the terminology applicable for two spatial dimensions merely for concreteness and clarity of presentation.
    These statements all apply equally to the case of three spatial dimensions, for which ``area'' and ``volume'', respectively, would be the appropriate language.
}
As the counter-rotating regions merge, the \emph{fully connected} length of the merged domain boundaries will increase, while the \emph{total} length of all domain boundaries in the box decreases.
At the same time, the ratio of the areas of the co-rotating regions to the counter-rotating regions remains roughly constant and approximately unity. 
Moreover, throughout the domain-merger process the (time-averaged) total number of vortices in the domain decreases by only about a factor of two from early [$\mu t\sim \mathcal{O}(30)$] to late [$\mu t \sim \mathcal{O}(800)$] times; see Fig.~\ref{fig:largecirculation_energy}. 
This slight reduction of vortex density is likely driven by the reduction of total length of the domain boundaries where vortices are born. 
Reducing this boundary size and decreasing the vortex density frees up energy, which is pumped into a scalar radiation field with characteristic wavenumber comparable to the mass of the radial mode: $k_{\rm{rad}}\sim \mu$. 
In Fig.~\ref{fig:merger_conter_rot}, this manifests itself as perturbations in $\dot{\theta}$ with spatial scales of the order $\sim \mu^{-1}$, particularly visible in the rightmost two panels. 
We discuss this radiation in the next subsection in more detail. 

We also advance a few plausible (but speculative) reasons for the disappearance at intermediate times of epochs where there are \emph{no vortices anywhere} within the simulation volume, and the re-appearance of these epochs at very late times; see Fig.~\ref{fig:largecirculation_energy}.
As a preliminary statement, we note that the string formation and annihilation are initially both synchronized in the whole simulation volume.
Then, at intermediate times, owing to spatial gradients in the $\theta$ and $\dot{\theta}$ fields across the simulation volume, phasing mismatches can develop that cause the formation times of individual vortices in different spatial locations to begin to move out of temporal synchronization with each other. 
In particular, at intermediate times, the domain boundaries of various distinct counter-rotating regions may also become of unequal length, which may allow for unequal string lifetimes on different-length boundaries (i.e., a longer/larger boundary can mean it takes longer for a pair of opposite-vorticity strings to find each other to annihilate).
However, once larger-scale order is returned to the fields due to mergers of the counter- and co-rotating regions (see Fig.~\ref{fig:merger_conter_rot}), such phasing mismatches may relax, which could explain the reemergence of durations where very few strings are present in the simulation volume at late times (see the left column of Fig.~\ref{fig:largecirculation_energy}).
We have not, however, definitively established whether this rough qualitative mechanism is primarily responsible for the results we see in simulation. 
Note however that vortex production is always approximately locally periodic on the scale of individual co- and counter-rotating regions.

For large initial displacement, the late-time behavior consisting of a quasi-equilibrium state characterized by the coexistence of some coherent motion, vortices, and a radiation field is observed in both two- and three-dimensions. 
It is plausible that this quasi-equilibrium behavior eventually ends as most of the energy is lost into radiation and there is then not enough free energy to produce more vortices, at which time the system would contain only radiation, and maybe some leftover vortices from earlier formation events.
Eventually, we expect only radiation to remain.

In the next sub-section, we consider the late-time evolution of the small initial displacement regime.
We will return to the discussion of the late-time evolution of both cases in the context of an expanding background cosmology in Sec.~\ref{sec:FLRW}.

\subsubsection{Small Initial Displacement}
\label{sec:evolution_small_displacement}

Turning now to the small initial radial amplitude case, we focus on $\rho_i/v=\sqrt{2}+0.1 \approx 1.51$, which implies $V(\rho_i)/V(0)\approx 1.67$.
Recall from Sec.~\ref{sec:formation_conditions} that the critical amplitude for vortex formation is $\rho_c/v\approx 1.24$ in this case because we are assuming $\mathcal{J}_0 = 0.1 \mu v^2 \ll \mu v^2$.
As described in Sec.~\ref{sec:formation}, vortex--anti-vortex pairs form after the emergence of counter-rotating regions once the $\theta$ field difference across the domain boundaries grows sufficiently large; once formed, the vortex--anti-vortex pairs move apart from each other.
Due to the confinement described in Sec.~\ref{sec:largedisplacement}, if the initial displacement satisfies $\rho_i\gg v$, any (anti-)vortex moves primarily along the domain boundary.
However, for smaller initial displacements, the confining effect of the domain boundaries should weaken and the vortices are expected to move more freely, even potentially ``detaching'' from domain boundaries. 

\begin{figure}
    \centering
    \includegraphics[width=0.32\linewidth]{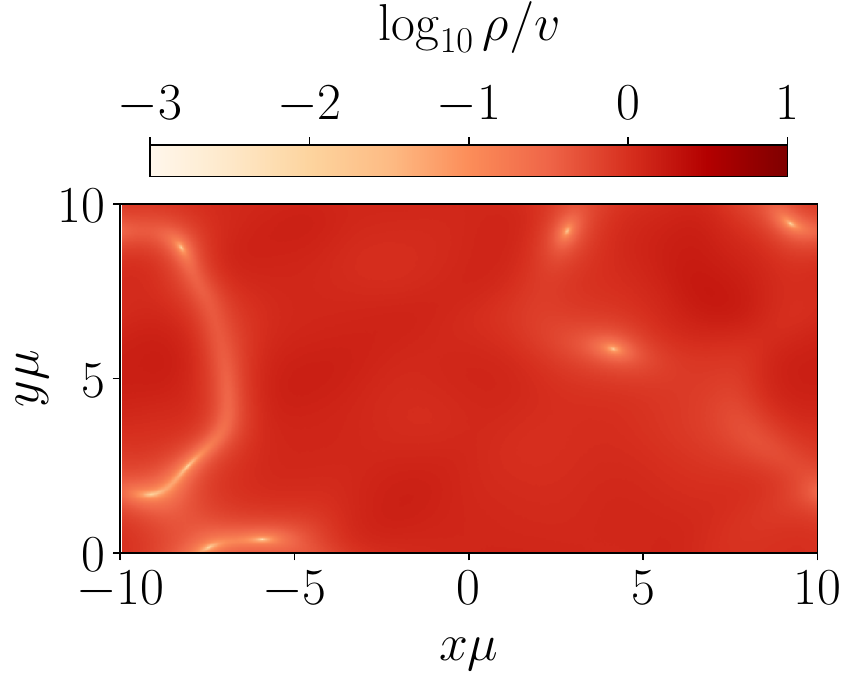}
    \hfill
    \includegraphics[width=0.32\linewidth]{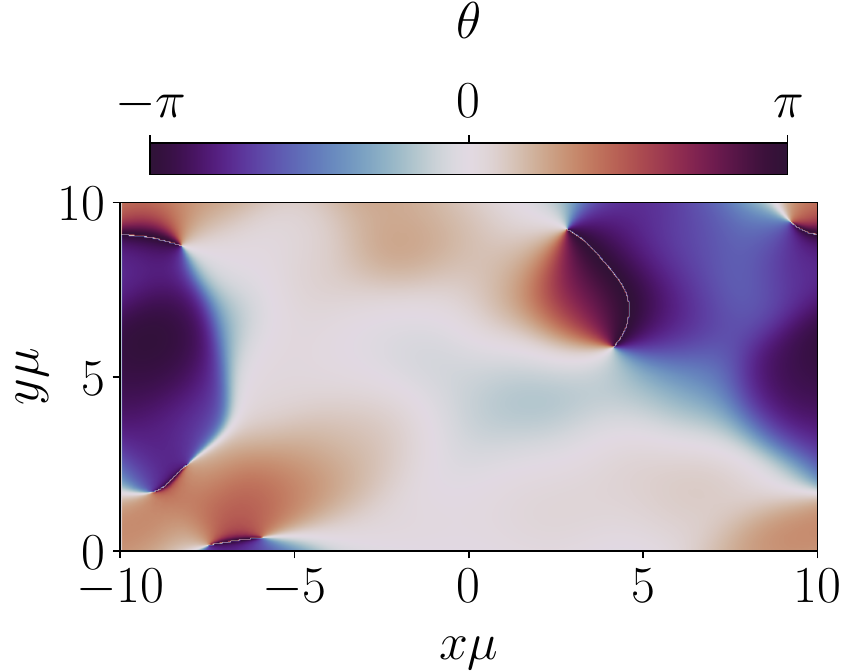}
    \hfill
    \includegraphics[width=0.32\linewidth]{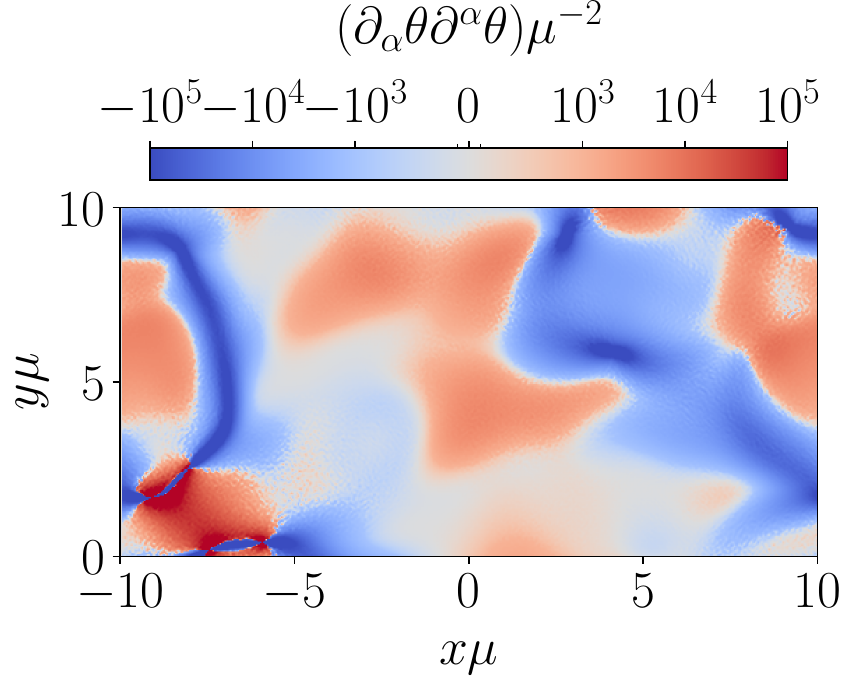}
    \caption{The state of the system with initial displacement $\rho_i/v=\sqrt{2}+0.1$ (corresponding to $V(\rho_i)/V(0)\approx 1.67$) and angular speed $\mathcal{J}_0=0.1v^2\mu$ after vortex pairs have formed along domain boundaries and subsequently separated. 
    Here we show only a selected region from the full two-dimensional simulation box. 
    \underline{\textsc{Left panel:}}~%
    The radial degree of freedom.
    \underline{\textsc{Center panel:}}~%
    The angular degree of freedom. 
    Note that the very thin white lines connecting vortices through otherwise black regions are plotting artifacts; these lines correspond to $\theta=\pi$. 
    \underline{\textsc{Right panel:}}~%
    The spacetime norm of the gradient of the angular mode. 
    Importantly, since $\theta$ is multi-valued at the locations of the vortices, both the temporal and spatial gradient of the angular degree of freedom are ill-defined at the vortex sites. 
    The numerical methods we employ artificially regulate this divergence; the behavior of $\partial_\alpha\theta\partial^\alpha\theta$ should thus be interpreted with caution within a distance of $\lesssim 0.1/\mu$ from the vortex center. 
    Recall, we use the mostly-minus metric signature.
    }
    \label{fig:stringsmallcirtulation}
\end{figure}

We show a state of the system with $\rho_i/v=\sqrt{2}+0.1$ in Fig.~\ref{fig:stringsmallcirtulation} that exhibits both clearly attached and potentially detached vortices. 
The attached vortex pairs (i.e., the pairs located in the $x<0$ part of the domain shown in Fig.~\ref{fig:stringsmallcirtulation}) lie along a domain wall characterized by $\rho\ll v$ and $\kappa = -\partial_{\alpha}\theta\partial^{\alpha}\theta\gg \mu^2$. 
However, these domain boundaries no longer necessarily form closed loop-like structures in two-dimensions, suppressing efficient vortex--anti-vortex annihilation.
Furthermore, this could allow vortices (formed earlier as pairs) to move off the domain boundaries and potentially behave (at least temporarily) similar to an isolated and non-relativistic vortex.
In the $0<x \mu \lesssim 5$ part of the domain shown in Fig.~\ref{fig:stringsmallcirtulation}, two such possible examples can be seen. 
Although it is ambiguous what one means by ``attached to'' vs.\ ``detached from'' a domain boundary,%
\footnote{%
    Perhaps a more unambiguous approach would be to identify isolated points of non-zero vorticity that are separated from other points of non-zero vorticity by region that exhibits $\kappa = - \partial_\alpha \theta \partial^\alpha \theta < 0$. 
    Note however that this is a condition which the prospective candidates in Fig.~\ref{fig:stringsmallcirtulation} would fail.
} %
each member of this vortex--anti-vortex pair exhibits less evidence of being located on some long (length $\gg \mu^{-1}$) filamentary domain-boundary structure on which $\rho \ll v$; that is, at least as compared to the (anti-)vortices that live in the $x<0$ part of Fig.~\ref{fig:stringsmallcirtulation}.
Although the reduced initial field value assumed for the results shown in Fig.~\ref{fig:stringsmallcirtulation} may potentially allow vortices to detach from these more extended field structures, this does not necessarily mean that a larger number of vortices can survive in the simulation domain.
Indeed, the total number of vortex pairs formed also scales down with the initial displacement (down to none, when $\rho_i=\rho_c$).
In particular, as we detail next, vortex annihilation is still sufficiently efficient, even for $\rho_i/v=\sqrt{2}+0.1$, to completely annihilate all vortices.
As such, any detachment effect that may exist is largely irrelevant for the overall dynamics of the system.

\begin{figure}[p]
    \centering
    \textbf{\quad 2D}\\[1ex]
    \includegraphics[width=1\linewidth]{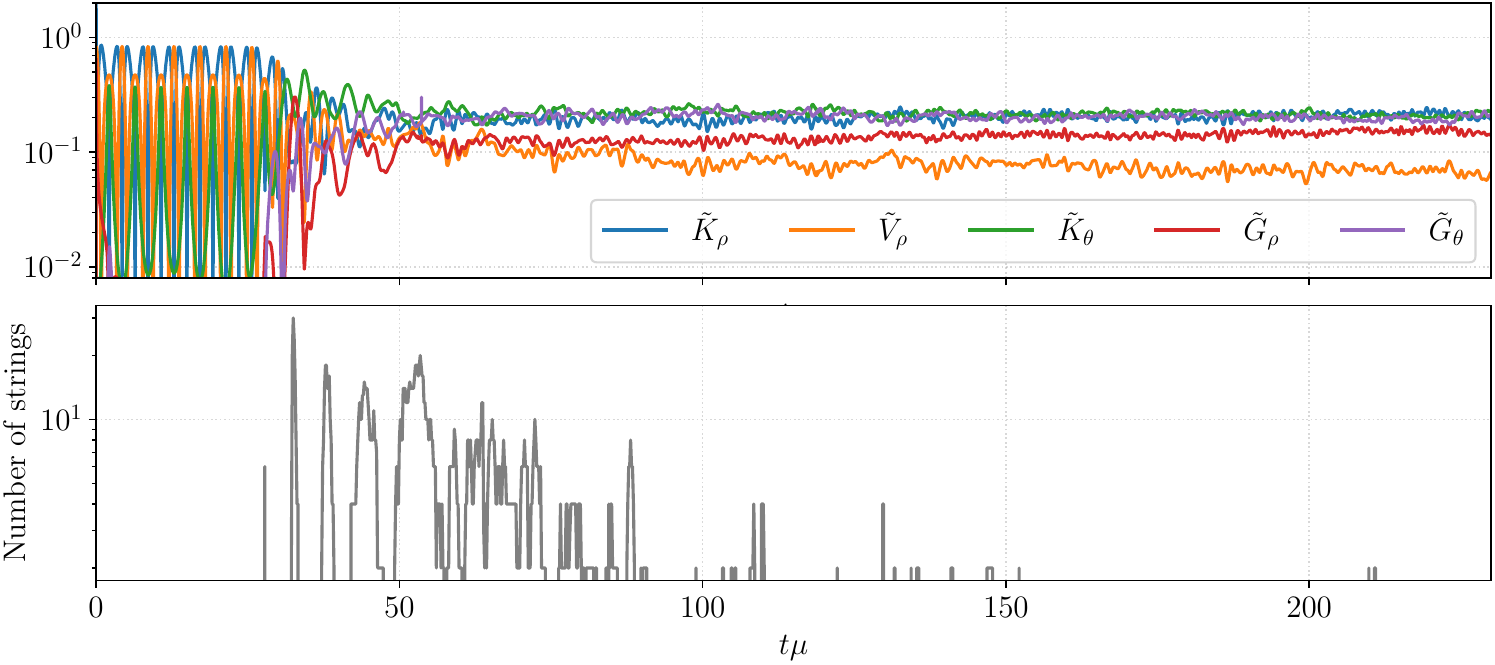}\\[2ex]
    \textbf{\quad 3D}\\[1ex]
    \includegraphics[width=1\linewidth]{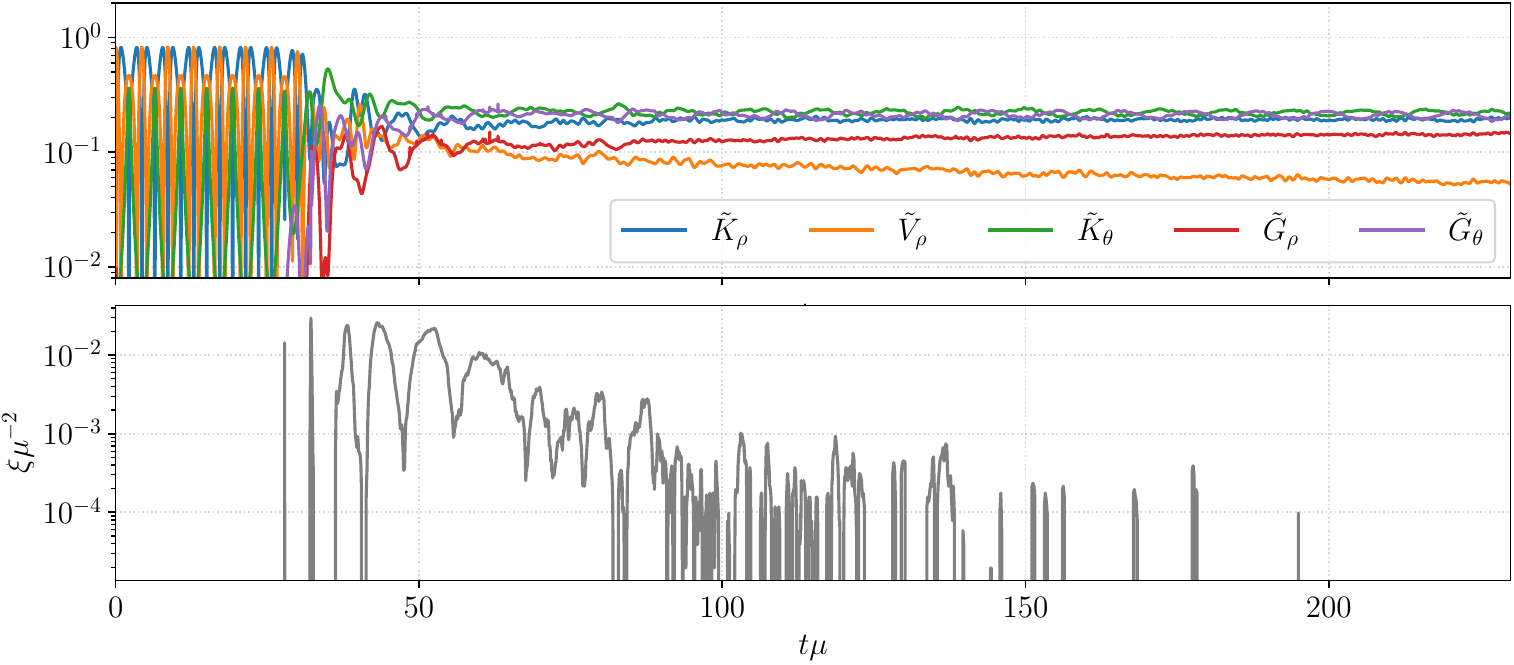}
    \caption{Evolution of the system with the same parameters as in Fig.~\ref{fig:stringsmallcirtulation} in either $D=2$ (the top two panels) or $D=3$ (the bottom two panels) spatial dimensions. 
    \underline{\textsc{2D and 3D, top panels:}}~%
    The normalized energy densities $\tilde{K}_\rho=K_\rho v^{-2}\mu^{-2}$, etc.\ of the angular and radial degrees of freedom.
    \underline{\textsc{2D, bottom panel:}}~%
    The number of vortices in the simulation box as a function of time.
    \underline{\textsc{3D, bottom panel:}}~%
    The string length density $\xi=\ell/V_b$, where $\ell$ is (an estimate for) the total string length in the simulation box of volume $V_b=(20/\mu)^3$; see App.~\ref{app:simdetails} for further details.
    An animation of the dynamics in the style of Fig.~\ref{fig:3d_string_anni}, but for the parameters shown in the bottom panel here, is available at \cite{ZenodoVideos}.}
    \label{fig:string_energy_small_circulation}
\end{figure}

In Fig.~\ref{fig:string_energy_small_circulation}, we show the evolution of this same case from the onset of the parametric resonance through the end state in both two and three spatial dimensions. 
Let us focus first on the 2D results.
Analogous to the large-initial-amplitude setting, vortex--anti-vortex pairs form on domain boundaries and annihilate on timescales $\tau$ associated with $\langle|\dot{\theta}|\rangle$ in the counter-rotating regions once the parametric resonance saturates.
However, in stark contrast to the previously considered scenario, after a few $\tau$ this periodic behavior gives way to slow, but complete, annihilation of all vortices present in the simulation domain. 
In this case, the energy contained in vortices and boundaries of counter-rotating regions is efficiently injected into a scalar radiation state, such that \textit{no} strings or domain boundaries remain after $\sim\mathcal{O}(100)\tau$. 
This occurs despite the partial detachment of some vortices from domain boundaries as described above. 
The final state reached here is one consisting purely of radiation. 
This manifests itself also in the distribution of energies shown in the top panel of Fig.~\ref{fig:string_energy_small_circulation}: the  $\tilde{K}_\rho \approx \tilde{K}_\theta \approx \tilde{G}_\theta > \tilde{G}_\rho$ and $\tilde{V}_\rho\ll 1$ behavior towards late times suggests that this radiation state is characterized by both massless $\theta$ modes and massive semi-relativistic $\rho$ modes.%
\footnote{ 
    The massless-ness of the $\theta$ modes can be seen from $\tilde{K}_\theta\approx \tilde{G}_\theta$ as would be expected from a dispersion relation $\omega_\theta^2 = k_\theta^2$; the massive-ness of the $\rho$ modes can be seen from $\tilde{K}_{\rho} > \tilde{G}_\rho$, as would be expected from a dispersion relation $\omega_{\rho}^2 = \mu^2 + k_{\rho}^2$ for a semi-relativistic mode with mass $\mu$ and $k_\rho \lesssim \mu$.
} %
The results in 3D are analogous (except that strings form in loops on two-dimensional domain boundaries instead of vortex--anti-vortex pairs forming on one-dimensional domain boundaries).

This situation is very different from the gauged $U(1)$ case in~\cite{East:2022ppo,East:2022rsi}, which suggests that confinement to domain boundaries might not be the only reason for a lack of string-length growth over long timescales. 
We attribute this difference between the global string and gauged string to the weakness of the interaction between scalar-field radiation and global strings~\cite{beekman2011electrodynamics}, compared to the electromagnetic force between gauged radiation and gauged strings.
In the gauged $U(1)$ case, the strings absorb the energy stored in the background gauge fields even when the field is very weak, and the string energy density dominates over the background gauge field at late time.  
However, there is no evidence that the global strings in our simulations absorb much energy from the background radiation at late time, and they could behave more like strings in isolation. 
Global strings, once in isolation, are expected to oscillate under their own tension, and release energy in the form of stochastic scalar-field radiation.
What we observe in the global $U(1)$ case that we have simulated is that the system after all global strings disappear contains an almost equal amount of energy stored in semi-relativistic radiation of $\rho$ ($k_{\rm rad} \sim \mu$), and radiation of the massless $\theta$ field, with very little energy stored in the coherent rotation (which can be seen from the approximate equipartition of energy of kinetic and gradient energies in $\theta$ and $\rho$).
This scalar radiation (and potentially gravitational waves; see Sec.~\ref{sec:stringburst}) is the only late-time remnant of the periodic bursts of string formation and annihilation.

\subsection{String Network Endstate: Large Initial \texorpdfstring{$\mathcal{J}_0$}{J0}}
\label{sec:endstatelargec0}

We turn now to the case of large initial angular speeds, $\mathcal{J}_0\gg v^2\mu$, but with relatively small initial radial displacement: $\rho_i\gtrsim\rho_c$ (but also recall footnote~\ref{ftnt:rhoC}).
We focus on the system with $\mathcal{J}_0=10^3v^2\mu$, initial radial displacement $\rho_i/v=11.5$ and $\sigma_{\rho_i}=10^{-6}\rho_i$, in two spatial dimensions, $D=2$. 
For these parameters, we have $\rho_c \approx 11.3$ from \eqref{eq:rhoclargeC0}.

\begin{figure}[t]
    \centering
    \includegraphics[width=1\linewidth]{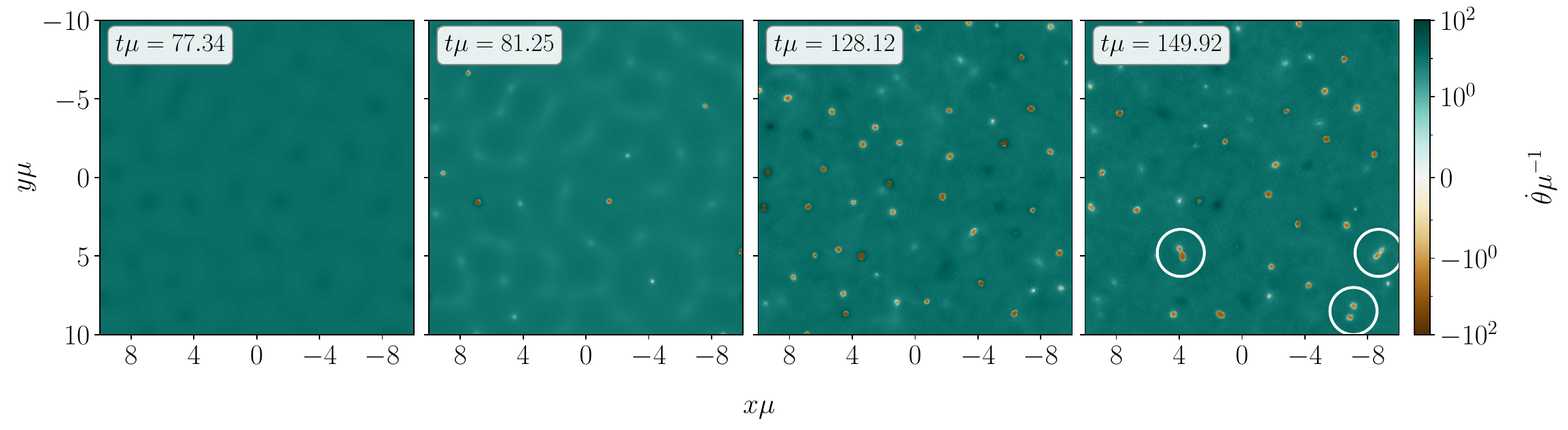}
    \caption{The evolution of the angular speed $\dot{\theta}$ once the parametrically unstable modes, inducing differential rotation (left panel; note subtle variations in colour), reach the threshold for counter-rotating regions to form (center two panels; the isolated orange regions).
    In the last panel, the merger of such regions is indicated by white circles.
    These results are for the parameter choice $\mathcal{J}_0=10^3v^2\mu$ with $\rho_i/v=11.5$, $\sigma_{\rho_i}=10^{-6}\rho_i$, and $D=2$. 
    \ZenodoComment}
    \label{fig:phidotlargec0}
\end{figure}

Recall from Sec.~\ref{sec:largec0resonance} that, during the exponential growth of the most-unstable Fourier mode driven by the parametric resonance, the system exhibits differential rotation, but no counter rotation, until the unstable modes reach the nonlinear regime.
We illustrate this, and the subsequent evolution of the angular speed $\dot{\theta}$, in Fig.~\ref{fig:phidotlargec0}. 
Counter-rotating regions only form once the unstable modes reach amplitudes comparable to $\langle\dot{\theta}\rangle$, which is now a very large quantity owing to the large initial value of $\mathcal{J}_0$; this occurs around $t\mu\sim 80$ in Fig.~\ref{fig:phidotlargec0}. 
The individual counter-rotating regions are smaller than in the small-$\mathcal{J}_0$ case owing to a combination of factors.
First, the unstable modes have higher $k$ when $\rho_i$ is larger, which forces each differentially rotating region to have a smaller spatial extent.
Because a counter-rotating region can only form inside a differentially rotating region when the differential rotation becomes large enough, this forces the counter-rotating regions to also become smaller. 
Second, the ratio of the areas of the regions that are counter-rotating to those that are co-rotating (where co- and counter- are as compared to the initial angular speed) must necessarily decrease as $\mathcal{J}_0$ increases as a result of global $U(1)$ charge conservation, provided $\rho \sim \rho_i \sim \rho_c$ is maintained in the bulk of the co- and counter-rotating regions.
Additionally, each of these counter-rotating regions is isolated, in stark contrast to the case of small $\mathcal{J}_0$, where approximately half of the domain is counter-rotating (see e.g., Fig.~\ref{fig:merger_conter_rot}).
Occasional mergers of such regions do however also occur in this regime at late times, as can be seen in the last panel of Fig.~\ref{fig:phidotlargec0}.

The counter-rotating regions begin forming at different locations at different times, as seen in the second panel of Fig.~\ref{fig:phidotlargec0}; see discussion in Sec.~\ref{sec:LargeJ0discussion}.
Over a rough time-span of $\approx 50/\mu$, all counter-rotating regions form. 
The domain boundary of each such small region is a site of vortex pair production and annihilation. 
As before, locally around each isolated counter-rotating region, this occurs periodically on timescales $\tau=\pi/|\langle\dot{\theta}|\rangle$.
However, since each counter-rotating region was formed at a different time, the global phase coherence of all such regions that was present in the case of small $\mathcal{J}_0$ is now lost.
This is possibly due to the fact that the counter-rotating regions are much smaller than the differentially rotating regions within which they emerge. 
This emergence requires, compared to the small-$\mathcal{J}_0$ case, additional nonlinear dynamics that moves energies from the fastest growing $k$-modes (due to parametric resonance) to the larger-$k$ modes that correspond to the inverse length scale of the counter-rotating regions. 
The higher-$k$ modes are created from a combination of many lower-$k$ modes with a large variety of initial $k$ and phases, and would likely not be phase coherent across the whole simulation box.
Consequently, complete annihilation of all vortices globally at any given time is unlikely.
Instead, vortex--anti-vortex pairs are produced and annihilated almost continually, with the times of production on any given domain boundary randomly offset from the times of production on any other given domain boundary (i.e., local periodicity of the vortex density is washed out on scales larger than a few times the wavelength of the most unstable Fourier mode).
Note also, in this case $\rho_{\min}\gg v$ implies efficient confinement of all vortices to the domain boundaries separating the islands of counter-rotating angular mode from the co-rotating environment.

\begin{figure}[t]
    \centering
    \includegraphics[width=1\linewidth]{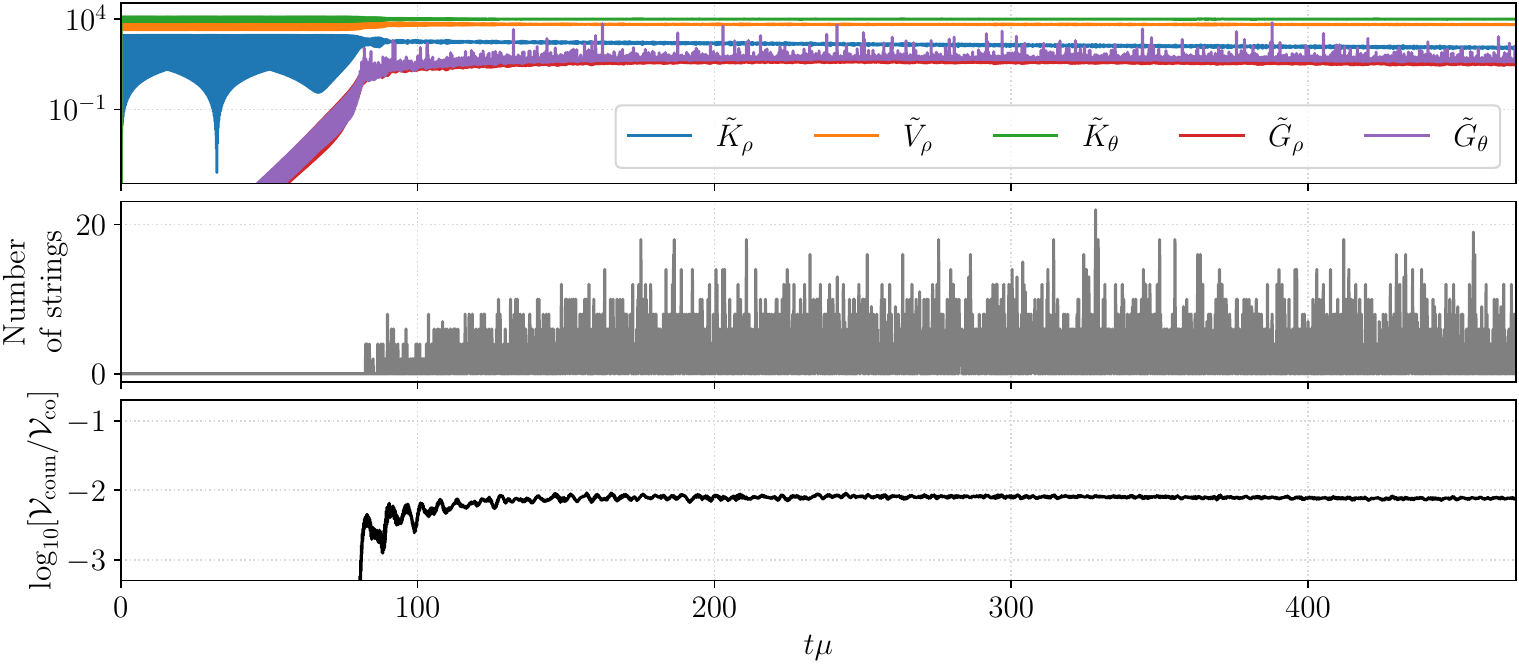}
    \caption{Quantitative measures of the state of the system shown in  Fig.~\ref{fig:phidotlargec0}.
    \underline{\textsc{Top panel:}}~%
    the evolution of the spatially averaged energy densities $\tilde{K}_\rho=K_\rho v^{-2}\mu^{-2}$ etc., from the onset of the parametric resonance at early times through vortex formation towards late times (note: this is an extension to later times of Fig.~\ref{fig:2dexpgrowthLargec0}). 
    \underline{\textsc{Center panel:}}~%
    The number of strings in the simulation domain of total volume $(20/\mu)^2$ at any given time. 
    \underline{\textsc{Bottom panel:}}~%
    The ratio of the total volumes of co- and counter-rotating regions throughout the evolution of the system. \ZenodoComment}
    \label{fig:largec0energies}
\end{figure}

Lastly, in Fig.~\ref{fig:largec0energies} we quantify the vortex density as well as ratio of co- to counter-rotating regions (in the two-dimensional setting). 
Once the unstable modes reach the nonlinear regime and counter-regions are formed, vortex production commences. 
As noted above, over roughly $\approx 50/\mu$ all counter-rotating regions appear and both the vortex density and the total counter-rotating volume increase towards a saturated quasi-equilibrium state. 
On a flat background spacetime, this state is long-lived and changes likely only on the timescales of the merging of the counter-rotating regions. 
Our numerical simulations are however of insufficient total duration to allow us to estimate this timescale. 
In the presence of cosmological expansion, this behavior would be altered, as we discuss in detail in the following section.

\section{Evolution in an Expanding Universe}
\label{sec:FLRW}

With the intuition built in the preceding sections, we are now ready to lift the assumption of a flat background spacetime (i.e., Minkowski spacetime) and place the system instead on a radiation-dominated%
\footnote{%
    We choose to solve the system in radiation domination because a long period of radiation domination is expected in the early Universe; however, the lessons learned from the numerical results presented here can be generalized to universes with different background cosmologies.} %
expanding background FLRW cosmology.
In this analysis, we assume that $\Phi$ is a subdominant energy component of the Universe as a whole, such that it does not drive or alter the cosmological expansion.

In the following, we describe the results of a numerical analysis designed to address two qualitatively non-trivial questions. 
First, do (and how do) the counter-rotating regions develop and behave as the Universe expands? 
In particular, how do $\rho$ and $\dot\theta$ inside the counter-rotating regions, and $\partial_i\theta$ on the domain boundaries, evolve as the Universe expands and the charge density $j^0$ decreases? 
Second, how does the vortex number (density) in two dimensions and string length (density) in three dimensions change as a function of time? 
In particular, does the system transit smoothly from the initial phase of periodic string formation (when the typical $\rho/v\gg1$, as described around Fig.~\ref{fig:largecirculation_energy}) to the late-time phase of total string annihilation (when the typical $\rho/v\simeq 1$, as described around Fig.~\ref{fig:string_energy_small_circulation})? 

The line element of the FLRW metric in $D$ spatial dimensions is $ds^2 = dt^2 - [a(t)]^2 \delta_{ij} dx^i dx^j$, with scale factor $a$,  where $dx^i$ is an infinitesimal increment of the $i$-th spatial direction in comoving coordinates. In this context, the equations of motion \eqref{eq:EoM} are modified:
\begin{align}
    \ddot{\Phi} + DH \dot{\Phi} - \frac{1}{a^2}\nabla^2 \Phi - \mu^2 \Phi + \lambda |\Phi|^2 \Phi &= 0 \:, \label{eq:FLRWeom}
\end{align}
along with its complex conjugate; here $H(t) \equiv \dot{a}/a$ is the Hubble parameter, $\nabla^2 \equiv \delta^{ij}\partial_i\partial_j$ is the flat-space Laplacian in terms of comoving coordinates, and $\dot{} \equiv \partial_t{}$.
In the radiation-dominated setting, the Hubble parameter behaves as $H \propto a^{-(D+1)/2}$, implying a scale factor $a(t) \sim t^{2/(D+1)}$.

In principle, if one considered inflationary initial conditions, then as the Universe cools to the point when $\sqrt{V''(\rho_i)} \simeq H$, the oscillation of the zero-mode of $\Phi$ would begin, and perturbations would grow due to parametric resonance, directly analogous to what we described in Sec.~\ref{sec:growth}. 
As the Universe continues to evolve, the Hubble parameter then decreases faster than the frequency of oscillations in the potential $V$, and the massive degree of freedom then lies well within the horizon. 
At this point, much of the qualitative phenomenology discussed in Secs.~\ref{sec:growth},~\ref{sec:formation}, and~\ref{sec:dynamics} would be expected to carry over to the evolving background cosmology. 

However, we set slightly different initial conditions for this cosmological simulation from those that would follow from the general setting described above.
First, we choose the initial $\theta$ to be constant in the simulation box~\cite{Tkachev:1998dc}, across many Hubble patches, so that we focus on strings that are produced well within the Hubble radius from our mechanism, without the contamination from the strings formed due to the standard Kibble--Zurek mechanism~\cite{Kibble:1976sj,Zurek:1985qw}. 
Second, for numerical convenience, we start the simulation with $\rho_i/v \in \{3,5,6\}$, when the fastest growing modes are already well inside the horizon; i.e., $\sqrt{V''(\rho_i)} \gg H$.
We also set white-noise perturbations in the initial conditions with $\sigma_{\rho_i}=10^{-2}\rho_i$ (see Sec.~\ref{sec:pertfullnonlinear}); these are much larger than the perturbations that would be seeded during inflation.
This choice either mocks up%
\footnote{%
    Of course, the Fourier spectrum of this noise is different from that which would obtain from taking white-noise conditions earlier and evolving them to grow, since we know that different $k$ modes have different growth timescales.
    } %
the perturbation growth that would otherwise have taken place if we were to have started the simulation when $\sqrt{V''(\rho)} \lesssim H$, or they could also emerge directly in scenarios motivated either by reheating after (hybrid) inflation~\cite{Kofman:1997yn,Tkachev:1998dc} or via the axion kinetic-misalignment mechanism~\cite{Co:2017mop}. 
The qualitative understanding we gain from our simulations from these initial conditions will thus give us insight to these cases, while allowing us to focus our attention numerically on the period when strings form and annihilate periodically, and then slowly disappear completely as the typical $\rho$ inside the simulation box approaches $v$.

\subsection{Preliminary Dependence on the Scale Factor}
\label{sec:FLRWanalysis}

Before we turn to numerical results, we begin our analysis analytically by considering the cosmological evolution of the zero-mode. 
To that end, we return to the simplified treatment of the problem introduced in Sec.~\ref{sec:growth}. 
In $D$-dimensional FLRW spacetime, the radial and angular degrees of freedom of the coherent component, $\rho_0$ and $\theta_0$, respectively, follow the equations of motion given by [cf.~\eqref{eq:FLRWeom}]
\begin{align}
    \ddot{\theta}_0+D H\dot{\theta}_0&=-2\dot{\theta}_0\frac{\dot{\rho}_0}{\rho_0}\:, & \ddot{\rho}_0+D H\dot{\rho}_0&=-\lambda\rho_0(\rho_0^2-v^2)+\rho_0\dot{\theta}_0^2\:; 
    \label{eq:backgroundhubble}\\
    \Rightarrow \frac{d}{dt}\left[ a^{D} \mathcal{J}_0 \right] &= 0 \:, & \Rightarrow \ddot{\rho}_0+D H\dot{\rho}_0&=-\lambda\rho_0(\rho_0^2-v^2)+\frac{\mathcal{J}_0^2}{\rho_0^3}\:,\label{eq:backgroundhubble2}
\end{align}
where, as before, $\mathcal{J}_0\equiv \rho_0^2\dot{\theta}_0$, which is \emph{covariantly} conserved: i.e., $\mathcal{J}_0 = (\text{const.}) \times a^{-D}$.
In addition to the $\rho_i$ initial conditions discussed above, we fix the rest of the initial conditions to be $\dot{\rho}_0(0)$, $\theta_0(0)=0$, and $\dot{\theta}_0(0)=\mathcal{J}_0/\rho_i^2$, depending on $\rho_0(0)=\rho_i$ (see discussion in footnote~\ref{ftnt:FLRWtime} below regarding the time $t=0$ in this context). 

Per the discussion above, because we fix our initial conditions when $\mu,\sqrt{V''(\rho_i)} \gg H$, the Hubble friction term in \eqref{eq:backgroundhubble} does not suppress the dynamics of the background field, but it cause alterations; moreover, the perturbations around this background with the shortest growth timescales through the parametric resonance are characterized by $k\lesssim \mu$ in the parameter space of interest, and so lie inside the Hubble horizon. 
Therefore, these modes immediately begin growing exponentially, driven by the zero-mode/background-field dynamics.

During the initial growth phase of the parametrically unstable perturbations (i.e., while these modes remain sub-dominant), the primary coherent/background dynamics of the system can be understood using~\eqref{eq:backgroundhubble}; see the detailed derivations of the following claimed scalings in App.~\ref{app:scalings}~\cite{Turner:1983he}. 
In the regime of large initial radial amplitudes $\rho_i/v\gg 1$, but independently of $\mathcal{J}_0$ as long as $\sqrt{V_{\rm eff}''(\rho)}\gg H$ still holds, we have $\rho_0 \propto a^{-D/3}$ initially and the relevant energy densities initially scale as $k_\rho, k_\theta, V_\rho\sim a^{-4D/3}$ for $D=2,3$.
The case where $\mathcal{J}_0\gg v^2\mu$ and $\rho_i\approx \rho_\text{min}$ exhibits the same scalings.%
\footnote{%
    At $\mathcal{J}_0 \gg \mu v^2$, we note that setting $\rho_i(a_i) = \rho_{\text{min}}(a_i)$ (i.e., the $\Phi$ field initial conditions are set such that the field instantaneously moves in a circular trajectory along the instantaneously circular degenerate potential minimum) does not necessarily result in having a non-oscillatory evolution of the $\rho$ field about the $a$-dependent minimum value $\rho_{\text{min}} \sim \mathcal{J}_0^{1/3} \propto a^{-D/3}$. 
    This is because $\rho_{\text{min}}$ does not necessarily change adiabatically on the timescale of $1/\sqrt{V_{\rm eff}''(\rho_{\text{min}})}$, which in general prevents $\rho$ from tracking $\rho_{\text{min}}$ adiabatically.
    Instead, imposing our initial conditions even with $\rho_i(a_i) = \rho_{\text{min}}(a_i)$ will generally result in an oscillation of $\rho$ similar to that which would obtain if one took instead $\rho_i(a_i) \neq \rho_{\text{min}}(a_i)$, but still $\rho_i(a_i) \sim \rho_{\text{min}}(a_i)$.
    We do not comment further on this case.
} %
These scalings of the zero-mode would change were the radial-mode amplitude closer to the vacuum expectation value; that is $\rho_i\approx v$. 
In that case, and assuming $\mathcal{J}_0\lesssim v^2\mu$, we have $\rho_0 -v \propto a^{-D/2}$ and the energy densities initially scale as $k_\rho, V_\rho\sim a^{-D}$ and $k_\theta\sim a^{-2D}$.

Once the parametric resonance drives $k$-modes into the nonlinear regime, the simplified picture of \eqref{eq:backgroundhubble} no longer applies and the scalings of various energy components become determined by the late-time evolution of the system. 
Qualitatively, during this phase, the system reaches a quasi-stationary period when strings periodically form on the boundaries of the counter-rotating regions as they expand. 
During this phase, we observe that the field within the counter-rotating regions rotates approximately circularly in field space, with constant velocity $\dot{\theta}_{\textsc{crr}}(a)$ at a fixed displaced $\rho_{\textsc{crr}}(a)$, with both quantities decreasing adiabatically as the scale factor $a$ increases. 
In the simulations, we also find that the total $j^0$ charge integrated over each counter-rotating region is approximately conserved, reflecting that the counter-rotating regions are of approximately constant comoving size, so that their physical volume grows as $a^D$.
As the Universe expands, the $j^0$ charge density inside each CRR scales as [see also \eqref{eq:noethercurrent}]
\begin{align}
    j^0 = \rho^2 \dot{\theta} \propto a^{-D}
\label{eq:J0scaling}
\end{align}
in $D$-dimensions. 
For fields that are rotating in circular motion (i.e., $\dot\rho,\ddot\rho\sim 0$) in the $\rho_{\textsc{crr}}/v \gg 1$ limit, the displacement $\rho_{\textsc{crr}}$ obeys an equation cognate to setting the RHS of the $\rho_0$ EoM in~\eqref{eq:backgroundhubble2} approximately to zero (we ignore spatial gradients here, as we are approximating the fields inside each CRR to be roughly constant throughout that region):
\begin{align}
    j_0^2/\rho_{\textsc{crr}}^3 \approx \lambda \rho_{\textsc{crr}}^3 \Rightarrow \rho_{\textsc{crr}} \propto j_0^{1/3} \propto a^{-D/3} \:. 
\label{eq:rhoscaling}
\end{align}
Combining~\eqref{eq:J0scaling} and~\eqref{eq:rhoscaling}, we find that both $\dot{\theta}_{\textsc{crr}}(a)$ and $\rho_{\textsc{crr}}(a)$ scale as $a^{-D/3}$, so long as $\rho_{\textsc{crr}}/v$ is sufficiently large that the $\mu^2\rho^2$ term in $V_{\text{eff}}$ can be neglected. 
The energy density in the system, which is dominated by the rotational energy density during this phase of the evolution, $k_{\theta} =\rho^2 \dot{\theta}^2$, would therefore scale as $a^{-4D/3}$: i.e., as $a^{-8/3}$ for $D=2$, and $a^{-4}$ for $D=3$. 
As a result, during both the initial parametrically unstable phase and the subsequent string-formation phase, the system's energy densities scale the same way with $a$, at least in the $\rho \gg v$ limit.

\subsection{Numerical Results}
\label{sec:FLRWnumerics}

Let us now turn to our numerical results. 
To be slightly more precise about the FLRW cosmology we assume and the initial conditions that we set, in both two and three spatial dimensions we start our simulations at $t=0$ from the initial conditions as described in Sec.~\ref{sec:pertfullnonlinear}, while setting%
\footnote{\label{ftnt:FLRWtime}%
    We can put this into a more familiar form by defining $t \equiv t'-t_0'$, where $\mu t_0' = 25$.
    Then $a(t') = (t'/t_0')^{2/(D+1)}$ and $H=2/((D+1)t')$, so that $H_0 = H(t=0) = H(t'=t_0') = 2/((D+1)t_0')$.
    } %
$a(t)=[tH_0(D+1)/2+1]^{2/(D+1)}$, such that $a(0)=1$ and $H_0/\mu\equiv H(0)/\mu=2/(25(D+1))$. 
Therefore, and as noted above, the most unstable modes of the parametric resonance, characterized by $k\lesssim \mu$ in the parameter space of interest, lie inside the Hubble horizon at $t=0$, and immediately begin growing exponentially. 
Further details on the numerical implementation and accuracy can be found in App.~\ref{app:simdetails}.
Note that the spatially averaged energy densities are now defined as $K_\theta= (2aL)^{-D}\int_{V_b} d^D\!x \ a^D k_\theta$, etc., with comoving spatial volume $V_b=(2L)^D$. 

\subsubsection{Small Initial \texorpdfstring{$\mathcal{J}_0$}{J0}}
\label{sec:FLRWsmallC0}

We begin by considering the scalar field dynamics for small $\mathcal{J}_0$ on an expanding background cosmology in both two and three spatial dimensions.

\paragraph{Two Spatial Dimensions}
\label{sec:FLRW_2D_smallJ0}

\begin{figure}[t]
    \centering
    \includegraphics[width=0.99\linewidth]{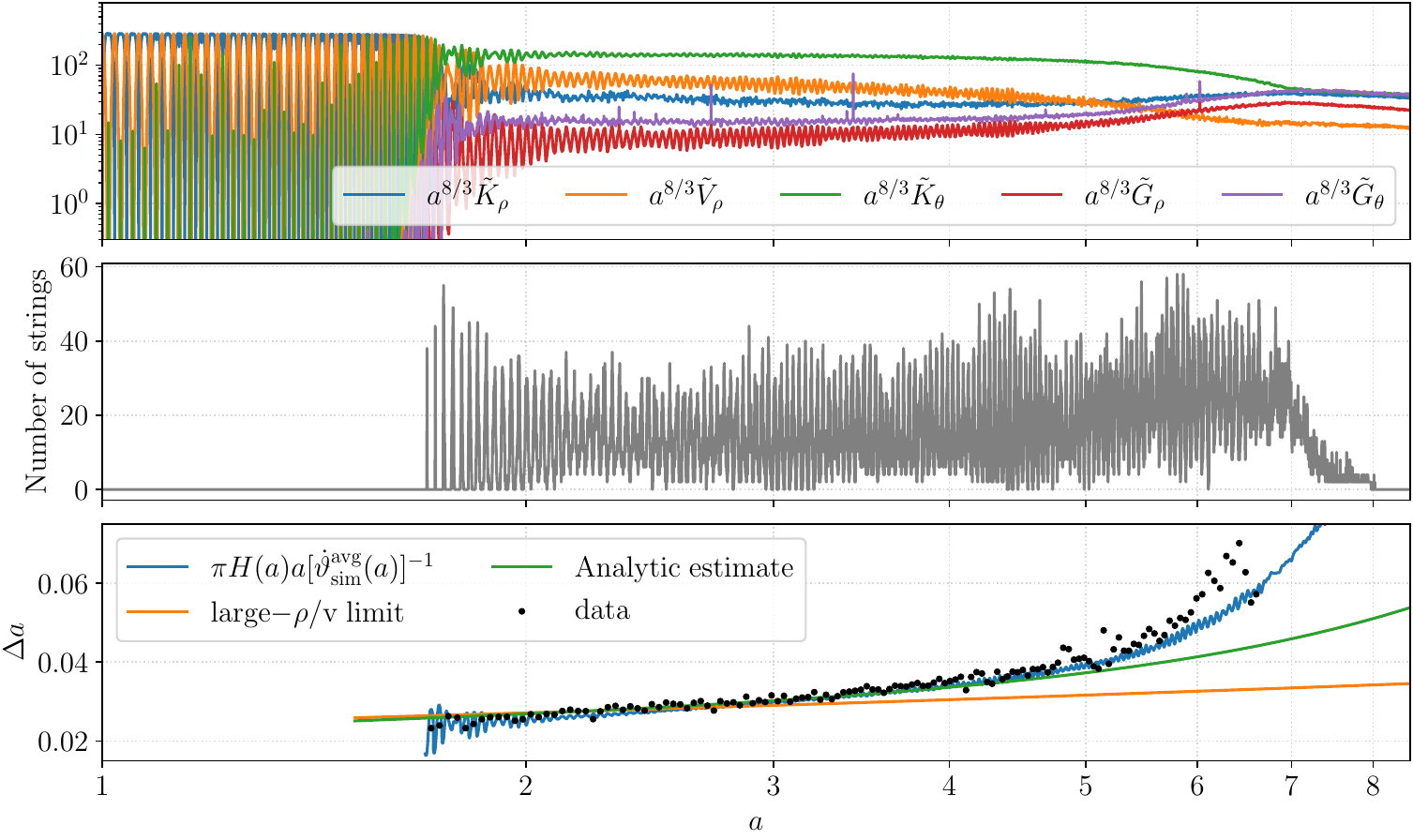}
    \caption{
    The evolution of the case with $\mathcal{J}_0=0.1v^2\mu$, $\rho_i/v=5$, and $\sigma_{\rho_i}=10^{-2}\rho_i$ in a $D=2$ radiation-dominated FLRW spacetime. 
    \underline{\textsc{Top panel:}}~%
    The energy densities $\tilde{K}_\rho\equiv K_\rho/(v\mu)^2$, etc.\ through the evolution of the parametric instability.
    We rescale the energy densities plotted by the scalings with $a$ that are expected for both the zero-mode evolution and the fields inside the counter-rotating regions, as discussed in Sec.~\ref{sec:FLRWanalysis}.
    \underline{\textsc{Center panel:}}~%
    The total number of strings in the comoving simulation volume $(10/\mu)^2$ [i.e., proper volume $(10a/\mu)^2$].
    \underline{\textsc{Bottom panel:}}~%
    The difference in scale factor $\Delta a$ between two peaks of the string number density as measured from the center panel (labelled ``data''). 
    Also shown are a variety of theoretical predictions for this quantity computed using \eqref{eq:dtVortex} [or, equivalently, \eqref{eq:PeriodDeltaa}]. 
    The blue curve shows the period computed by setting $|\dot{\theta}(a)|$ in \eqref{eq:PeriodDeltaa} equal to the moving time-average over roughly one oscillation period ($\Delta a = 0.03$) of $\langle |\dot\theta (a)|\rangle$, as obtained from the simulation; we denote this moving average as $\dot{\vartheta}^{\textrm{avg}}_{\text{sim}}(a)$. 
    The solid orange line is the analytical prediction of \eqref{eq:periodicityDdim}, while the solid green line is the improved theoretical prediction described in App.~\ref{app:burstp}. 
    }
    \label{fig:cosmo2d_rho3}
\end{figure}

In Fig.~\ref{fig:cosmo2d_rho3}, we show the evolution of the energy densities and number of strings in the simulation volume from the onset of the parametric resonance through the late-time endstate of the system. 
As before, the unstable modes grow exponentially until vortex pair production becomes efficient and the modes reach the nonlinear regime. 
The system then goes through cycles of efficient periodic vortex production (when $|\partial_i\theta|$ is super-critical) and complete annihilation (while $|\partial_i\theta|\approx 0$ on the domain boundaries), similar to the Minkowski case.

However, the periodicity of these cycles is now different.
In the large $\rho_i$ limit ($\rho_i \gg v$), we can compute it precisely (which we do here for general $D$ spatial dimensions).
Recall that, in this regime, once counter-rotating regions have formed, both $\dot{\theta}_{\textsc{crr}}(a)$ and $\rho_{\textsc{crr}}(a)$ scale as $a^{-D/3}$; cf.~\eqref{eq:rhoscaling}.
In the Minkowski case, we found that the spacing in time between bursts $i$ and $i+1$ (for $i=1,2,\ldots$) of vortex formation across a domain boundary satisfied 
\begin{align}
    \int_{t_i}^{t_{i+1}}dt \,|\dot{\theta}_{\textsc{crr}}(t)| &= \pi\:, \label{eq:dtVortex}
\end{align}
where $\dot{\theta}_{\textsc{crr}}(t)$ is the value of $\dot{\theta}_{\textsc{crr}}$ in a one of the co- or counter-rotating regions and we have used that the differential $|\dot{\theta}_{\textsc{crr}}|$ across a domain boundary is still approximately twice the value of $|\dot{\theta}_{\textsc{crr}}|$ that obtains in the two counter-rotating regions on either side of the boundary.
We hypothesize that this condition still holds locally in FLRW provided that the bursts are sufficiently close in time.
We can then adapt the condition \eqref{eq:dtVortex} to find the scale factors at bursts $i$ and $i+1$ in the FLRW case by using the definition of the Hubble constant and the field scalings noted above:
\begin{align}
     \int_{a_i}^{a_{i+1}} \frac{da }{H(a) a }  \times \left[|\dot{\theta}_{\textsc{crr}}(a_1)|\left(\frac{a_1}{a}\right)^{D/3} \right] &= \pi\:. \label{eq:daVortex}
\end{align}
Because the integrand in \eqref{eq:daVortex} is simply a power of $a$ once the scaling $H(a) \propto a^{-(D+1)/2}$ is inserted, it is then trivial to show that the solution for the scale factor at the $i$-th burst, $a_i$, is
\begin{align}
    \left(\frac{a_{i}}{a_1}\right)^{D/6+1/2} = 1+ (i-1) \frac{D+3}{6} \frac{\pi H(a_1)}{|\dot{\theta}_{\textsc{crr}}(a_1)|}\:.
    \label{eq:periodicityDdim}
\end{align}

Our numerical results in $D=2$ match well with \eqref{eq:periodicityDdim} when $\rho \gg v$; see Fig.~\ref{fig:cosmo2d_rho3}.
As the Universe expands and $\rho_{\textsc{crr}}(a)\rightarrow v$, $\dot\theta_{\textsc{crr}}$ redshifts much steeper as a function of the scale factor $a$ (i.e., $\dot{\theta}_{\textsc{crr}} \propto a^{-2D}$ in this regime, as discussed above), and the burst separation in $a$ also increases.
In this case, an improved prediction for $\theta_{\textsc{crr}}(a)$ can be obtained by numerically solving \eqref{eq:backgroundhubble2} with $\dot{\rho},\ \ddot{\rho} =0$; see App.~\ref{app:burstp} for further details.
The result of this refined prediction is also shown in Fig.~\ref{fig:cosmo2d_rho3} for $D=2$, along with an even more refined semi-analytical prediction that combines \eqref{eq:dtVortex} with numerical simulation data%
\footnote{%
    Numerically, $\dot{\theta}(a)$ and $\rho(a)$ inside the counter-rotating regions are not exactly constant, and it is somewhat arbitrary to assign which regions in the simulation box belong to the counter-rotating regions and which belong to the domain boundaries. 
    However, since the counter-rotating regions have volume that is much larger than the domain boundaries, we use the straight volume average of $|\dot{\theta}(a)|$ over the entire simulation box in the comparisons.} %
about $\dot{\theta}(a)$.

Besides this redshifting, the periodicity of the production and annihilation cycles is unaffected by the evolving background cosmology. 
On timescales larger than $\tau$, counter-rotating regions begin to merge.
Overall, even after vortex production begins, the spatially-averaged energy densities of the angular and radial degrees of freedom scale roughly as predicted by \eqref{eq:backgroundhubble}, together with $G_\rho,G_\theta\sim a^{-8/3}$. 

Crucially, as the (spatially averaged) potential energy density $V_\rho$ approaches $V(0)$, the system begins to transition away from a state of merging counter-rotating regions with cyclic vortex production and annihilation, and toward the radiation state found in Sec.~\ref{sec:evolution_small_displacement}, which is devoid of coherent field space rotation and vortices. 
This transition can be seen in Fig.~\ref{fig:cosmo2d_rho3} around $a\approx 7$. 
This is despite starting with large initial radial amplitude $\rho_i$; cf.~the Minkowski case shown in Fig.~\ref{fig:largecirculation_energy}. 
Just as in the Minkowski case, this radiation state is characterized by $\tilde{V}_\rho\ll 1$ together with the equipartition of $\tilde{K}_\rho, \tilde{K}_\theta$, and $\tilde{G}_\theta$, as well as the relation $\tilde{K}_\rho\gtrsim \tilde{G}_\rho$; therefore, the radiation is semi-relativistic with frequency $k_\rho,k_\theta\sim\mu$.

\begin{figure}[t]
    \centering
    \includegraphics[width=1\linewidth]{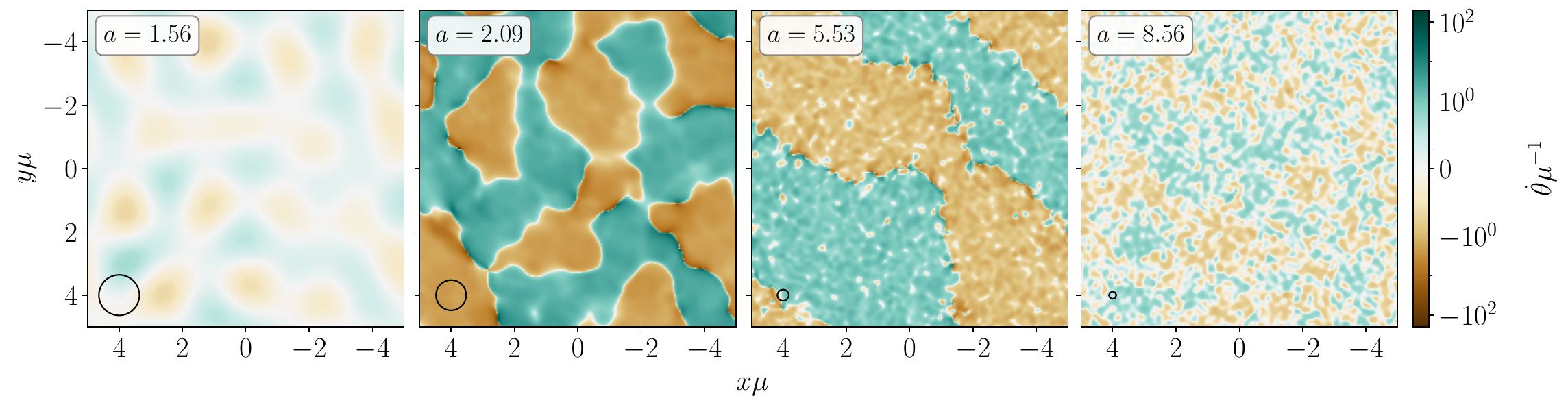}
    \caption{Snapshots of the evolution of counter-rotating regions in the two-dimensional cosmological simulation shown in Fig.~\ref{fig:cosmo2d_rho3}. 
    Each panel corresponds to a different scale factor $a(t)$; $x$ and $y$ are comoving co-ordinates. 
    For visual aid, the black circle in the lower left corner of each panel indicates a region of fixed \emph{physical} radius $\mu^{-1}$; its comoving radius is $(a\mu)^{-1}$.
    }
    \label{fig:cosmo_counter}
\end{figure}

Overall, the dynamics of the system from vortex formation through the transition to the radiation state can be seen in the spatial domain in Fig.~\ref{fig:cosmo_counter}. 
Unlike the flat-spacetime simulations, here there is less time for the counter-rotating regions to merge before expanding spacetime dynamics dominate and the system transitions to the radiation state.
Hence, the central panels in Fig.~\ref{fig:cosmo_counter} contain only marginally larger (in comoving coordinates) counter-rotating regions as compared to the left panel, together with small-scale perturbations marking the onset of the transition to the radiation state. 
At late times (right panel), the state of the system qualitatively resembles the end state discussed in Sec.~\ref{sec:evolution_small_displacement}.

Before we close this subsection, it is important that we comment on the synchronization of the string production and annihilation across the simulation box. 
The simulation box is much larger than the length scale of the initially growing perturbation and thus the size of the counter-rotating regions. 
The synchronization is observed in the presence of a relatively large initial perturbation of $\sigma_{\rho_i}=10^{-2}\rho_i$, and is preserved throughout the expansion of the Universe and the dilution of the energy densities. 
Our interpretation of this is based on the intuition from the separate-universe approach~\cite{Wands:2000dp}, which suggests that this $10^{-2}$ difference in $\rho_i$, and therefore the local energy density, would lead to an $\mathcal{O}(10^{-2})$ time offset in the growth of the short length-scale perturbations, and a $\mathcal{O}(10^{-2})$ difference in the terminal differential $\dot \theta$ between the co- and counter-rotating regions at saturation across the simulation box. 
Such a difference will result in an $\mathcal{O}(1)$ phase difference after roughly $\mathcal{O}(10^{2})$ periods of string burst production, which might be beyond the 137 periods we seen in this simulation.
In our Universe, where the initial perturbation seeded from inflation is $\mathcal{O}(10^{-5})$ [and $D=3$; see below], this would suggest that the string bursts could stay time-synchronized across many Hubble patches over, potentially, as many as $10^5$ periods. 
Note that this time synchronization depends only on whether this differential $\dot\theta$ stays a constant, and should not depend on whether the field $\theta$ stays the same across Hubble patches, though it remains to be understood if the strings coming from the Kibble mechanism, albeit dilute, can disrupt this time synchronization across Hubble patches.

\begin{figure}[t]
    \centering
    \includegraphics[width=0.99\linewidth]{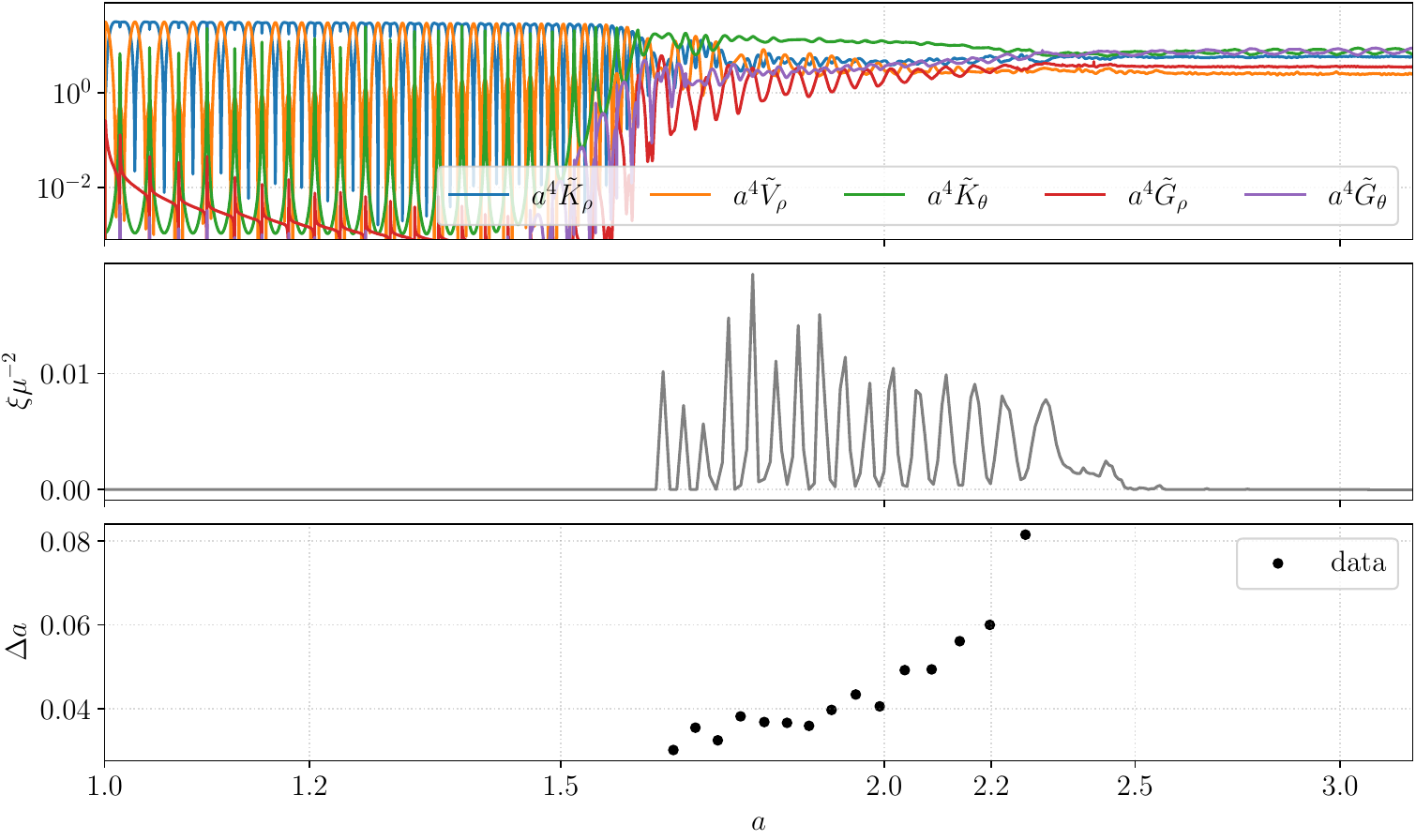}
    \caption{The evolution of the case with $\mathcal{J}_0=0.1v^2\mu$, $\rho_i/v=3$, and $\sigma_{\rho_i}=10^{-2}\rho_i$ in a $D=3$ radiation-dominated FLRW spacetime.
    \underline{\textsc{Top panel:}}~%
    The energy densities $\tilde{K}_\rho\equiv K_\rho v^{-2}\mu^{-2}$, etc.
    The primary $a$-scalings are different than in Fig.~\ref{fig:cosmo2d_rho3} but as predicted below \eqref{eq:backgroundhubble2} for $D=3$.
    \underline{\textsc{Center panel:}}~%
    The comoving string length density $\xi=\ell/V^{\text{comov.}}_b$, where $\ell$ is the total comoving length of strings in the comoving simulation volume $V^{\text{comov.}}_b = (10/\mu)^3$, whose proper (i.e., physical) volume is $V_b=a^3 V^{\text{comov.}}_b$.
    \underline{\textsc{Bottom panel:}}~%
    The difference in scale factor $\Delta a$ between two peaks of the string number density as measured from the center panel.    
     }
    \label{fig:cosmo3d_rho3}
\end{figure}

\paragraph{Three Spatial Dimensions}
Turning now to the evolution in $D=3$ spatial dimensions, we focus entirely on the case with $\rho_i/v=3$, highlighting only the differences from the $D=2$ case considered in the previous section, because the evolution proceeds qualitatively similarly.
After the exponentially growing modes disrupt the coherent oscillation of the fields, string loops form on domain boundaries separating counter-rotating regions.
As before, these efficiently annihilate cyclically and the system enters a phase of merging counter-rotating regions.
In Fig.~\ref{fig:cosmo3d_rho3}, we show the evolution of the energy densities and string length density with $\rho_i/v=3$ in three dimensions. 
The primary difference between this three-dimensional setting and the prior two-dimensional scenario is that the energy densities now scale as $\sim a^{-4}$, as expected.

Although we also show in Fig.~\ref{fig:cosmo3d_rho3} the spacing $\Delta a$ between sequential bursts of string production and annihilation, we do not attempt a comparison to our (semi-)analytical prediction for this quantity (see App.~\ref{app:burstp}) as we did in the lower panel of Fig.~\ref{fig:cosmo2d_rho3}.
This is because, for computational reasons, we are only able to simulate a $D=3$ case where $\rho \sim \mathcal{O}(v)$ at the time of the first burst of string production. 
This $\rho$ at first burst in 3D is comparable to the $\rho$ at $a=6$ in the 2D simulation, where our analytical estimate for $\Delta a$ starts to deviate from numerical simulation results.

\subsubsection{Large Initial \texorpdfstring{$\mathcal{J}_0$}{J0}}
\label{sec:FLRWlargeC0}

\begin{figure}[t]
    \centering
    \includegraphics[width=1\linewidth]{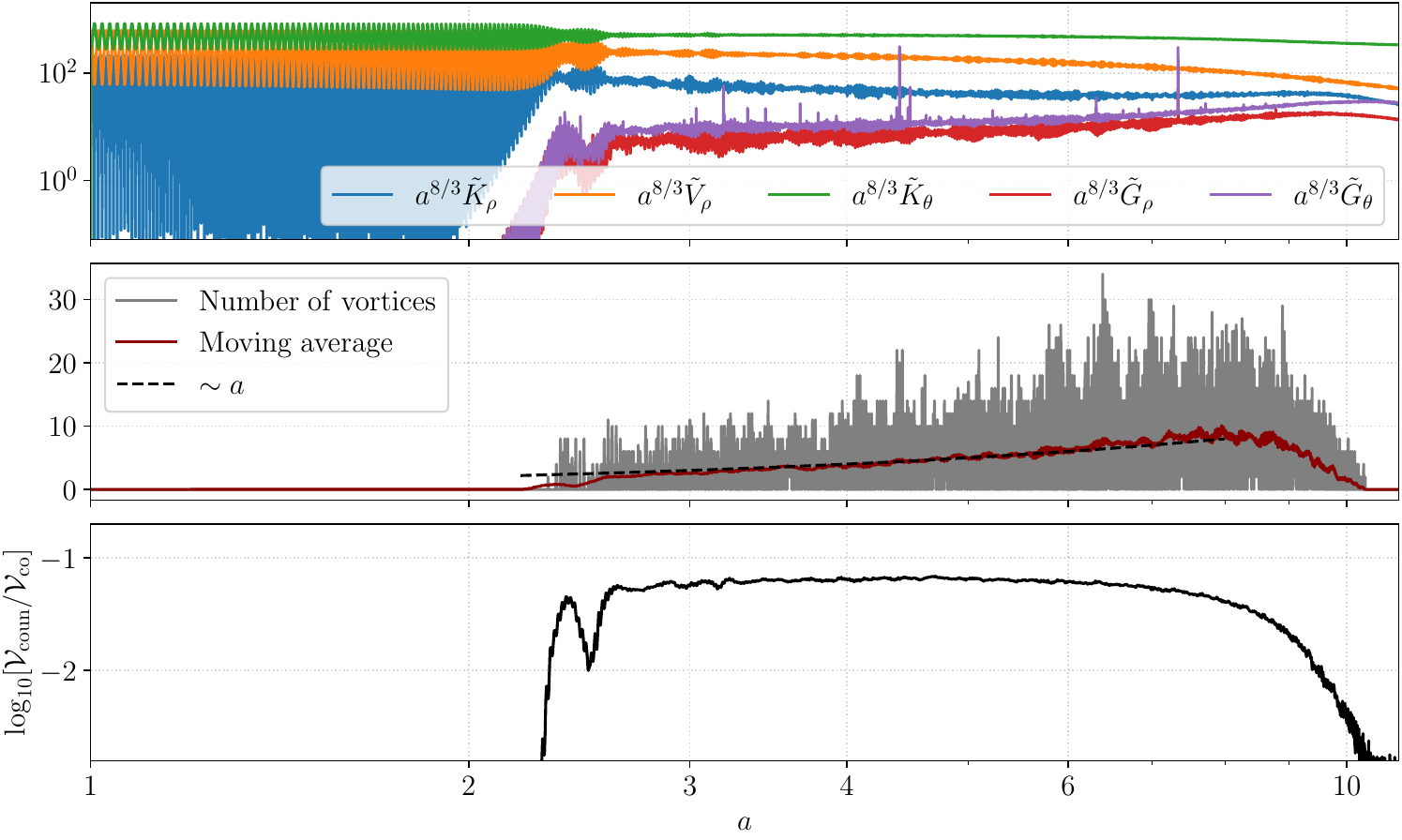}
    \caption{
    The evolution of the system with parameters $\mathcal{J}_0=10^2v^2\mu$, $\rho_i/v=6$, $\sigma_{\rho_i}=10^{-5}\rho_i$, and $D=2$, on an expanding background cosmology. 
    \underline{\textsc{Top panel:}}~%
    The spatially averaged energy densities $\tilde{K}_\rho\equiv K_\rho v^{-2}\mu^{-2}$, etc.
    \underline{\textsc{Center panel:}}~%
    The total number of vortices in the simulation domain with comoving volume $(10/\mu)^2$ [i.e., physical volume $(10a/\mu)^2$], as well as its moving average compared with a linear-in-scale factor scaling.
    \underline{\textsc{Bottom panel:}}~%
    The ratio of volume of co- to counter-rotating regions in the domain.}
    \label{fig:cosmo_largec0}
\end{figure}

Finally, we close this section by considering the cosmological evolution of systems in the large-$\mathcal{J}_0$ regime. 
As an example, we focus entirely on the case with $D=2$, $\rho_i/v=6>\rho_{\min}$ and $\mathcal{J}_0 = 10^3\mu v^2$, which undergoes the parametric resonance instability and saturates through efficient vortex pair production. 
In Sec.~\ref{sec:endstatelargec0}, we showed that in this regime of the parameter space, the ratio of counter- to co-rotating regions $\mathcal{V}_{\rm coun}/\mathcal{V}_{\rm co}$ is large, individual counter-rotating regions act primarily independently (except for occasion mergers), formation of such regions proceeds out of phase even for neighboring regions, and hence, the phase-coherence of vortex pair production found in systems with $\mathcal{J}_0\ll v^2\mu$ is lost. 
This behavior carries over to system on an expanding cosmological background, as can be seen in Fig.~\ref{fig:cosmo_largec0}. 
In fact, the ratio $\mathcal{V}_{\rm coun}/\mathcal{V}_{\rm co}$ remains largely constant as the radial and angular modes decrease as $\langle\rho\rangle,\langle|\dot{\theta}|\rangle\sim a^{-2/3}$ after vortex production commences. 
Since mergers of individual counter-rotating regions are unlikely and the ratio $\mathcal{V}_{\rm coun}/\mathcal{V}_{\rm co}$ remains largely constant during this phase, the proper length of the domain boundaries increases roughly linearly in the scale factor. 
This enables an approximately linear-in-$a$ increase of the comoving vortex number density, which can be seen in the center panel of Fig.~\ref{fig:cosmo_largec0}. 
During this period, the various energy densities shown in the top panel of Fig.~\ref{fig:cosmo_largec0} [except for $V(\rho)$] do not redshift faster than $K_{\theta}$.
This secular growth in the comoving vortex number density is halted when $\langle\rho\rangle\gtrsim v$, and it then begins to decrease around $a\approx9$. 
This eventually leads to the total disappearance of all vortices around $a\approx 10$, at which point $\rho \simeq 1.3 v$ and $j_0 \lesssim \mu v^2$, both comparable to when vortices disappear in the small-$\mathcal{J}_0$ case (see discussions at the end of App.~\ref{app:burstp}).

The comoving vortex number density begins to decrease when the domain boundaries on which they form, and the counter-rotating regions that sustain the domain boundaries, also begin to disappear. 
This is most evident from the bottom panel of Fig.~\ref{fig:cosmo_largec0}, where $\mathcal{V}_{\rm coun}/\mathcal{V}_{\rm co}$ also decreases dramatically around $a\approx 10$.
Due to the remaining $\langle j_0\rangle \simeq \mu v^2$ (resulting from the large $\mathcal{J}_0$), the system then enters a phase of stable circular motion dominated by large $K_\theta$,
together with a non-relativistic radiation component $K_\rho\gtrsim G_\rho$. 
However, there is no parametric difference between the size of $K_\theta$ vs.~$K_\rho$, $G_\rho$, and $G_\theta$ at this moment, due to continuous production of radiation through vortex dynamics.
Since $K_\theta$ would now redshift as $a^{-2D}$, while $K_\rho$ would redshift like matter ($a^{-D}$) and $G_\theta$ would redshift like radiation ($a^{-(D+1)}$), any late period in which the dominant energy-density component of the scalar field scales as it would in a kination-dominated epoch (i.e., $K_\theta\propto a^{-2D}$) would, if it exists, last for only a short duration. 
Note also that, even though $K_\theta$ is the largest component of the energy density during much of the earlier evolution shown in Fig.~\ref{fig:cosmo_largec0} during which vortices are present (spanning around a decade of growth in $a$), the scaling $K_\theta \propto a^{-8/3}$ (i.e., $a^{-4D/3}$ for $D=2$) is not a period of kination-like scaling~(i.e., $a^{-2D}$).%
\footnote{%
    In fact, in $D=3$, $a^{-4D/3}$ would be $a^{-4}$, the same scaling as radiation.
    } %

\section{Phenomenological Implications: Periodic Gravitational-Wave Bursts}
\label{sec:stringburst}

Having discussed how the dynamics of the scalar field that exists on an FLRW spacetime that is governed by \eqref{eq:lagrangianintro} with very coherent initial conditions naturally leads to a duration of periodic-in-$a$ vortex formation followed at late times by a state best characterized as scalar radiation, we will in this section make some preliminary comments on phenomenological implications.
Specifically, we focus on the possibility of periodic gravitational-wave production.

The periodic bursts of string formation and annihilation that we have found in the small-$\mathcal{J}_0$ case would be expected to lead to periodic bursts of gravitational-wave production.
However, while the string-production periodicity in the large-$\rho_i$ limit ($\rho_i \gg v$) was easily derived in Sec.~\ref{sec:FLRW_2D_smallJ0} (and can also be straightforwardly extended to the small-$\rho$ regime; see App.~\ref{app:burstp}), the frequency spectrum and amplitude of these gravitational waves, which can depend on $v$, $\mu$, and $\rho_i/v$, can span a much wider range.
The field dynamics we have described can happen as early as the inflation/reheating epoch, similar to~\cite{Tkachev:1998dc} (if $V''(\rho_i) \simeq H_I^2$); around the time of the would-be Peccei--Quinn (PQ) phase transition~\cite{Co:2017mop}; and as late as today, similar to~\cite{Boyle:2001du} (if $V''(\rho_i) \simeq H_0^2$). 
The corresponding gravitational-wave signal can, as a result, have frequencies ranging from the region probed by CMB B-modes to the region probed by proposed high-frequency gravitational wave detectors of $\mathcal{O}(\rm GHz)$~\cite{Aggarwal:2020olq}. 
The amplitude of the gravitational waves emitted in a string annihilation event is of order $h\sim Gv^2$ at emission, or equivalently, the total gravitational-wave energy density $\Omega_{\rm tot} (f) \sim (Gv^2)^2$, where $v$ can in principle be close to the Planck scale for small enough $\lambda$. 
This suggests that the amplitude can also have a wide range, with the upper bound coming from the current searches of stochastic gravitational waves~\cite{LIGOScientific:2019vic,NANOGrav:2023gor} as well as the $N_{\rm eff}$ constraints~\cite{Planck:2018vyg}.
However, although the peak frequency and the amplitude of the signal can vary, the periodic nature of the string formation and annihilation, and the corresponding gravitational-wave bursts, can imprint novel structure in the gravitational-wave spectrum. 

\paragraph{Frequency Content:} 
The observable signal today depends on both the amplitude and frequency of the individual burst, described by a certain GW spectrum $\Omega_i(f)$; the relative amplitudes of the different bursts; and how they sum up, accounting for both the redshifting of frequencies and the dilution of the energy density (the latter of which decreases the gravitational-wave amplitude). 
A precise spectrum $\Omega_i(f)$ can only be obtained through detailed numerical simulations; in the following, we discuss several novel qualitative observable features that may appear, in order to motivate further studies.

The frequency of the observable gravitational-wave signal depends on the frequency content of the individual bursts, as well as their (relative) redshifting. 
The frequency content of the individual burst may depend on as many as four distinct length scales: the Hubble size, as well as the typical length, separation, and core size of the strings.
In terms of the model parameters, in the large $\rho_i/v$ limit, they can depend on energy scales $H$ and $\sqrt{V''(\rho)}$, the latter of which is approximately $\lambda^{1/2}{\rho}_{\textsc{crr}}$ in the counter-rotating regions where $\rho \sim \rho_{\textsc{crr}} \gg v$, and approximately $\mu$ around the domain-boundary regions ($\rho\leq v$) where the strings form and annihilate. 
The gravitational-wave signal can, therefore, have emission frequency $f_{\Omega, {\rm emit}}$ (as measured at the epoch of emission) peaked at any of these energies, although one of the three is likely dominant. 
However, without a dedicated numerical analysis, we will not be able to predict which of these three frequencies dominates; as a result, we comment on all three possibilities.  

The main difference between these three possibilities arises from how the different energy scales, and hence $f_{\Omega, {\rm emit}}$, depend on the scale factor $a_{\text{emit}}$ at emission. 
We remind the reader that, in radiation domination, $H\propto a^{-2}$, $\lambda^{1/2}{\rho}_{\textsc{crr}} \propto a^{-1}$, and $\mu \propto a^0$.
This is to be compared with how the frequency of the emitted gravitational wave redshifts from emission to observation: $f_{\Omega, {\rm obs}} = f_{\Omega, {\rm emit}} \times ( a_{\text{emit}} / a_{\text{obs}} )$.
Let us examine each of the three cases.
(\emph{a}) If the energy scale that dominates the gravitational wave emission is $\lambda^{1/2}{\rho}_{\textsc{crr}}$, then $f_{\Omega, {\rm emit}}$ scales as $a_{\text{emit}}^{-1}$; in this case, all of the gravitational-wave bursts will be peaked at the same observed frequency today: i.e., $f_{\Omega, {\rm obs}} \propto a_{\text{emit}}^0$ is a constant (apart from the last few cycles when $\rho \sim v$, and the large $\rho/v$ approximation fails for the scaling of $\rho_{\textsc{crr}}$). 
(\emph{b}) If, on the other hand, the energy scale more relevant at the string locations $\mu$ sets the peak gravitational-wave frequency, then the different $\Omega_i(f)$ will be peaked at the same frequency $f_{\Omega, {\rm emit}}$ at emission, and the $f_{\Omega, {\rm obs}}^i \propto a_{\text{emit}}$ differ owing only to redshifting effects. 
In this case, we should thus expect a forest of peaks of GW radiation, with each peak frequency $f_{\Omega, {\rm obs}}^i \propto a_i$ satisfying [cf.~\eqref{eq:periodicityDdim} for $D=3$]
\begin{align}\label{eq:freqforest}
  \frac{f^{i+1}_{\Omega, {\rm obs}}}{f^1_{\Omega, {\rm obs}}} = 1+ i \frac{\pi H(a_1)}{|\dot{\theta}(a_1)|} \qquad \Rightarrow \qquad \frac{\Delta f^{i}_{\Omega, {\rm obs}}}{f^1_{\Omega, {\rm obs}}} \equiv \frac{f^{i+1}_{\Omega, {\rm obs}} - f^{i}_{\Omega, {\rm obs}}}{f^1_{\Omega, {\rm obs}}} = \frac{\pi H(a_1)}{|\dot{\theta}(a_1)|} \equiv \frac{\Delta f_{\Omega, {\rm obs}}}{f^1_{\Omega, {\rm obs}}}\:,
\end{align}
where the last $\equiv$ defines the constant ${\Delta f_{\Omega, {\rm obs}}}$.
Note that these peaks will be \emph{equally spaced in frequency} (again, apart from the last few bursts). 
If the spectrum $\Omega_i(f)$ of the individual bursts is a broken power-law around the peak frequency, then the fractional half-width of the spectrum at emission $\sigma^f_{\Omega, {\rm emit}}/f_{\Omega, {\rm emit}}$, and hence of the spectrum at observation $\sigma^{f,i}_{\Omega, {\rm obs}}/f^i_{\Omega, {\rm obs}}$, is a constant. 
As a result, the ratio of the inter-peak spacing to the peak half-width, $\Delta f_{\Omega, {\rm obs}} / \sigma^{f,i}_{\Omega, {\rm obs}} \propto (f^i_{\Omega, {\rm obs}})^{-1}$, decreases as the frequency grows. 
This suggests that it will be easier to resolve the individual peaks at low frequencies; while, at higher frequencies, the peaks would sum up to a much smoother spectrum.
(\emph{c}) There is also the possibility that some other scale, or combination of scales, set the peak frequency of the gravitational waves at emission. 
These scales could depend on $a$ as $a^{-P}$; e.g., $H\propto a^{-2}$. 
Correspondingly, \eqref{eq:freqforest} generalizes, and the forest of peaks would be related by
\begin{align}\label{eq:freqK}
    \frac{f^{i+1}_{\Omega, {\rm obs}}}{f^1_{\Omega, {\rm obs}}}  = \left(1+ i \frac{\pi H(a_1)}{|\dot{\theta}(a_1)|}\right)^{1-P} \:;
\end{align}
in this case, the inter-peak frequency spacing is no longer constant, and a more complicated spectrum arises.
A dedicated study beyond the scope of this work would be needed to know the total number of bursts that are obtained for a given set of initial conditions, as well as the corresponding index $P$.

\paragraph{Amplitude:} 
The amplitude of the observable gravitational-wave bursts at the peak frequency $f^i_{\Omega, {\rm obs}}$ depends also on the amplitude at emission, as well as redshifting effects. 
The gravitational-wave burst, once emitted, redshifts like radiation $h^i \propto (a_i/a)$. 
The peak amplitude of the individual burst, just like its frequency $f^i_{\Omega, {\rm emit}}$, can depend on the scale factor at emission $a^i$. 
Such a dependence stems from how the various (energy) densities and energy scales depend on $a$. 
As shown in Fig.~\ref{fig:cosmo3d_rho3}, the total energy density in the system, as well as the gradient energy density in the domain boundary regions, both redshift as $a^{-4}$. 
On the other hand, the string number density redshifts as%
\footnote{%
    This statement follows from the peak values of $\xi$ shown in the center panel of Fig.~\ref{fig:cosmo3d_rho3} being approximately constant with $a$; because $\xi$ is a ratio of a comoving length to a comoving volume, it follows that the physical number density redshifts as $\sim \text{const.}\times a/a^3\propto a^{-2}$.
} %
$a^{-2}$ (ignoring the burst oscillations).
This suggests different relations between the peak amplitudes of the forest of peaks.
Whereas a dedicated numerical study will be needed to understand which sets the relative size of the peak amplitude, some qualitative understanding can be obtained. 
Let us focus on the case where the peak frequencies are related by \eqref{eq:freqforest} and assume that the spectrum of each burst $\Omega_i(f)$ scales as $a_i^{-Q}$ at the peak frequency.
In this case, the total signal $\Omega_{{\rm obs},i}(f)$ will scale as $a_i^{4-Q} \propto (f^{i}_{\Omega, {\rm obs}})^{4-Q}$. 
Correspondingly, in the region where the individual peaks significantly overlap, the different gravitational-wave bursts sum up to a power-law that scales at high frequencies as
\begin{align}\label{eq:gwpowerlaw}
   \frac{{\rm d} \Omega_{\rm tot}(f) }{{\rm d} \log f} \propto f^{5-Q}\:.
\end{align}
As was discussed previously, this power-law behavior is correct in the region of large $\rho_{\textsc{crr}}/v$, and becomes more complicated for the last few bursts at the highest frequencies.

\paragraph{Summary:} 
To summarize, the periodic bursts of string formation and annihilation can lead to periodic gravitational-wave bursts.
In turn, these periodic bursts can lead to observable gravitational-wave signals with frequency peaks that are equally spaced over a wide frequency range [or, more generally, satisfy \eqref{eq:freqK}]. 
At low frequencies, these peaks might be widely spaced and either partially or fully resolvable, which could be a ``smoking gun'' signal.
On the other hand, they will most likely sum up to a featureless power law at high frequencies; cf.~\eqref{eq:gwpowerlaw}.
The shape of the individual bursts, as well as the two indices $P$ and $Q$ that we have used to roughly parametrize them above, would need to be determined numerically.

\paragraph{Further comment:} 
The preceding discussion is based on the dynamics of the vortex production and annihilation because these periodic dynamics are associated with the evolution of regions of high energy density in the field; these are expected to have associated GW production.
However, it is unclear whether these dynamics dominate the GW spectrum, and/or whether there are other GW components that are sourced merely by the overall evolution of the inhomogeneous scalar field at late times, such that the overall spectrum might have both periodic-in-frequency and broad features. 
Recall also that even unambiguously delineating the energy density that is associated with ``vortices'' vs.~that in the ``background field'' is a conceptual and operational challenge in the context of an excited field with nonlinear interactions; in turn, it may be difficult to unambiguously associate GW emission with ``vortices'' vs.~``the rest of the field''.
Ultimately, the GW spectrum arising from the full field evolution must be studied numerically; this is a task we defer to future work.

\section{Conclusion and Remarks}
\label{sec:conclusion}

In this paper, we studied the dynamics of a complex scalar field in a theory with a spontaneously broken global $U(1)$ symmetry. 
If the scalar field is initially displaced radially from the minimum, then, independent of the initial $U(1)$ charge, field perturbations grow due to parametric resonance after the field starts oscillating when the Hubble friction drops below its mass. 
We utilize numerical time-domain simulations to explore the behavior of the system once these unstable modes reach the nonlinear regime.
In this regime, counter-rotating regions (i.e., $U(1)$-charge separated regions) emerge and global strings are formed on domain boundaries separating these counter-rotating regions.
Importantly, in simulations conducted in both two and three spatial dimensions, we found that these global strings are confined to the domain boundaries; this enables the complete annihilation of the string network after formation.
Moreover, these dynamics (creation and subsequent annihilation of the string network) occur \emph{periodically}, potentially for many cycles. 
For small initial $U(1)$-charge density [more concretely, small $\mathcal{J}_0/(\lambda^{1/2} \rho_i^3$)], the periodic formation and annihilation of strings is synchronized on all domain boundaries over large spatial scales, and for a large number of periods. 

These synchronized periodic dynamics likely lead to periodic gravitational-wave bursts. 
Generically, it is known that string-annihilation events that take place in a more quiescent background typically produce gravitational waves with a frequency content determined by the radial mode mass $\mu$, while the amplitude of the wave is determined by the string tension $T\sim v^2$ to be of order $G T \sim G v^2$~\cite{Vachaspati:1984gt,Gorghetto:2021fsn}. 
Importantly, the scale $\mu$ can be comparable or larger than the Hubble scale at the time of the burst, while the scale $v$ can be at the GUT scale or even close to the Planck scale. 
In our case, owing to the more excited field configurations and the confinement of the strings to domain boundaries, the spectrum of the gravitational-wave signal from each burst of string production and annihilation, together with the dependence of both its frequency and amplitude on $\rho_i/v$, can only be determined with more dedicated numerical simulations, which we leave to future work.
Observationally, however, we note that these periodic gravitational-wave bursts could in certain cases lead to a forest of resolvable peaks in the observed gravitational-wave spectrum.
Were such a forest of peaks to be observed, it would strongly hint toward some manner of periodic dynamics occurring in the early Universe, and could even be a smoking-gun signal for the type of periodic string dynamics we have found in this paper.

The results presented in this paper may also be useful for understanding preheating after inflation~\cite{Dolgov:1989us,Traschen:1990sw,Kofman:1994rk,Kofman:1997yn,Tkachev:1998dc}.%
\footnote{%
    We thank Sergey Sibiryakov for discussions on this point and pointing out~\cite{Tkachev:1998dc}, in which string formation was first observed in the study of preheating after inflation.
} %
Separately, an interesting question for future work is whether the dynamics we discovered in this paper survive if the $U(1)$ is weakly gauged, or if the $U(1)$ is extended to a larger symmetry group (e.g., the case of the Standard Model electroweak sector)~\cite{Dufaux:2010cf,Rajantie:2000nj,Amin:2014eta}.
Would defects form periodically if the Standard Model Higgs were displaced in the radial direction, and would those dynamics remain synchronized across Hubble patches?

Our result may also have implications for mechanisms that rely on a coherent rotating field motion in theories with an (approximate) $U(1)$ symmetry, including various realizations of Affleck--Dine baryogenesis, as well as the kinetic-misalignment axion-production mechanism. 
Even though we started our simulations with highly coherent initial conditions inspired by those assumed in kination (i.e., a coherently rotating scalar field, as in our large $\mathcal{J}_0$ case), our simulations show that parametric resonance effects drive the growth of spatial perturbations deep into the nonlinear regime, resulting in the appearance of small counter-rotating regions and the subsequent production of topological defects (i.e., strings or vortices). 
This indicates that the role of spatial inhomogeneities likely cannot be disregarded in models that attempt to set up a coherent rotating field motion, and simulations in \emph{position} space, instead of semi-analytical analysis in momentum space, are required to uncover these dynamics.
Moreover, we find in our simulations that, once string production has commenced, the system enters a radiation-dominated scaling of its dominant energy components \emph{before} the cessation of string production at late times.
Because kination assumes a non-radiation-dominated field configuration without topological defects, our simulations suggest that, even with highly coherent initial conditions, field dynamics drive the system to a configuration that would be inconsistent with kination-like conditions at later times, at least within the realms of validity of the simulations we have run.
That said, we do not make a definitive conclusion regarding the implications for kination in this work. 
We do however suggest that the role of unstable inhomogeneous modes in kination-like scenarios may need to be revisited to more fully understand whether such mechanisms are robust to these effects.

Extending prior work~\cite{East:2022ppo,East:2022rsi} that showed similar conclusions for the gauged $U(1)$ case, in this paper we showed that in a theory with a $U(1)$-global symmetry, the formation and evolution of topological defects is important in determining the dynamics of the theory in the nonlinear regime, as well as understanding the fate of the nonlinear system. 
These dynamics leave intriguing and novel gravitational-wave signals, which hope will be uncovered in future work.

\phantomsection\addcontentsline{toc}{section}{Acknowledgments}
\section*{Acknowledgments}

We thank Will East, Sung-Sik Lee, Sergey Sibiryakov, and Zach Weiner for helpful discussions, and Anson Hook for comments on the draft. 
This work used \texttt{anvil} at Purdue University through allocation PHY240191 from the Advanced Cyberinfrastructure Coordination Ecosystem: Services \& Support (ACCESS) program \cite{access}, which is supported by National Science Foundation (NSF) Grants Nos.~OAC-2138259, -2138286, -2138307, -2137603, and -2138296. 
J.H.~would like to thank KITP for hospitality during the completion of this work.
This research was supported in part by NSF Grant No.~PHY-2309135 to the Kavli Institute for Theoretical Physics (KITP). 
M.A.F.~thanks the Aspen Center for Physics for hospitality during early stages of this work, supported by NSF Grant No.~PHY-2210452.
Research at Perimeter Institute is supported in part by the Government of Canada through the Department of Innovation, Science and Economic Development and by the Province of Ontario through the Ministry of Colleges and Universities.

\appendix

\section{Concise Review of Vortices in Condensed Matter Systems}\label{app:cmtvortex}

In this appendix, we provide a concise review about vortices in condensed matter systems, in the hope of providing intuition and motivation for the investigations in the main text. 
These include seminal results in the study of the Ginzburg--Landau model of superfluidity in three dimensions and the classical rotor model (classical XY Model, which describes a two-dimensional superfluid film) in two dimensions~\cite{ANDERSON:1966bps,Ao:1993zz,Thouless:1996mt,1985PhRvL..55.2887H,Sonin_1997,1993PhyA..200...42T,Volovik:2003fe,RevModPhys.59.1001}. 
However, since $\dot\theta$ is nonzero and large in our system, this analogy to the condensed matter systems is not exact, as we have already pointed out in several places in the main text. 
Therefore, unlike in the case of gauged vortices in~\cite{East:2022rsi}, where direct comparisons are possible, here we will only point out some qualitative similarities. 

The Ginzburg--Landau free energy can be written as~\cite{Ginzburg:1950sr,Mahan}
\begin{align}
    \mathcal{F} = \alpha(T) |\Psi|^2 +\frac{\beta(T)}{2}|\Psi|^4 +\frac{1}{2 m_*} \left|\vec{p}\Psi\right|^2 \:, 
\end{align}
where $\Psi$ is the wavefunction and $\vec{p}$ its momentum, $\alpha(T)$ and $\beta(T)$ are parameters of the theory that can in principle depend on the temperature, and $m_*$ is the mass of the field $\Psi$, which is, e.g., the mass of the helium atom in superfluid helium.  
This free energy has the same $U(1)$ global symmetry as the Lagrangian we studied in this paper [cf.~\ref{eq:lagrangianintro}], and can be considered as its non-relativistic limit.%
\footnote{%
    Note that the Ginzburg--Landau model has three free parameters while the non-relativistic limit of~\eqref{eq:lagrangianintro} would have two. 
    The difference stems from the breaking of Lorentz symmetry in a superfluid.
    } %
At zero temperature, the $U(1)$ symmetry is spontaneously broken
\begin{align}
    |\Psi|^2 = -\frac{\alpha(0)}{\beta(0)} \equiv n_s\: ,
\end{align}
where $n_s$ is the number density of the superfluid, and the wavefunction 
\begin{align}
    \Psi = \sqrt{n_s} \exp [ i \theta(x)] \:.
\end{align}
The current corresponding to the $U(1)$ symmetry is
\begin{align}
    \vec{J} = - i \frac{\Psi^{\dagger} \partial_{i}\Psi-\Psi \partial_{i}\Psi^{\dagger} }{2 m_*}= \frac{n_s}{m_*} \nabla \theta(x) \equiv  n_s \vec{v}_s \: ,
\end{align}
where $\vec{v}_s = \frac{\nabla \theta(x)}{ m_*}$ is the velocity of the fluid~\cite{Mahan}. 

Vorticity in viscous fluids is produced at a critical velocity of the fluid when the flow becomes turbulent, characterized by the Reynolds number. 
The production of vortices in superfluid was first considered by Feynman when he identify the condition for vortex formation in a thin, narrow orifice with radius $a$, via a Kelvin--Helmholtz-like instability~\cite{feynman1955chapter,ANDERSON:1966bps}. 
The Feynman critical velocity is
\begin{align}
    v_{\rm Feynman} = \frac{2}{m_* a} \: ,
\end{align}
corresponding to a critical $|\partial_i \theta| = 2/a$, depending only on the system geometry.
Later, theoretical and experimental studies identified logarithmic corrections to this critical velocity, stemming from the fact that global vortices have logarithmically divergent energy~\cite{ANDERSON:1966bps}. 
In our study, an artificial orifice-like configuration is created by the counter-rotating regions with finite extent, where the field $|\partial_i \theta|$ is substantial only on the domain boundary with a finite length; cf.~Fig.~\ref{fig:flow}. 
Although the field profile is not quite the same, this motivates us to find a similar critical $|\partial_i \theta|$ above which vortices form.

In superfluids, the forces that vortices experience, and hence their movement, have been studied extensively, especially in two dimensions (superfluid film).
The action
\begin{align}
    S = -\sum_j n_s m_* K_j \int X_j d Y_j +\sum_{i\neq j}\frac{n_s m_*}{2 \pi} K_i K_j \int \ln \left|\vec{R}_i-\vec{R}_j\right|dt \:,
\end{align}
first discussed in~\cite{1985PhRvL..55.2887H,1993PhyA..200...42T}, is an effective description of vortex interactions in the Ginzburg--Landau model in 2D (classical XY model).
Here, $\vec{R}_i = \left\{X_i,Y_i\right\}$ is the location of the vortex, while the circulation $K = \pm 2\pi/m_*$ for vortices ($+$) and anti-vortices ($-$). 
It was noted in \cite{1993PhyA..200...42T} that the first term in the action is a Berry phase and that this action resembles the action describing the motion of a charged particle (in the XY-plane) in an external magnetic field pointing in the Z-direction (see~\cite{Tong:2016kpv} for a detailed discussion about the connection with the quantum Hall effect). 
The resulting equation of motion of a vortex $i$ in the background of all the other vortices can be found to be
\begin{align}
    \dot{Y}_i =\sum_{j\neq i} \frac{K_j}{2\pi} \frac{X_i -X_j}{\left|\vec{R}_i-\vec{R}_j\right|^2} \:,\quad  
    \dot{X}_i =-\sum_{j\neq i} \frac{K_j}{2\pi} \frac{Y_i -Y_j}{\left|\vec{R}_i-\vec{R}_j\right|^2} \: .
\end{align}
The force can then be found to be~\cite{Thouless:1996mt}
\begin{align}
    \vec{F}_i = n_s K_i \hat{z}\times (\vec{v}_V - \vec{v}_s) \: ,
\end{align}
where $\vec{v}_V \equiv \dot{\vec{R}}_i$ is the velocity of the vortex $i$. 
This force, which is known as the ``Magnus force'', has been shown to determine the motion of vortices in a superfluid~\cite{Thouless:1996mt,Varoquaux_2015}. 
Whereas we have mainly described the case of a superfluid (with a global $U(1)$ symmetry), these discussions also apply to superconductors (with a gauged $U(1)$ symmetry), in which case the vortices are Pearl or Abrikosov vortices~\cite{1964ApPhL...5...65P,Abrikosov:1956sx}, and the critical velocity becomes a critical bias current~\cite{PhysRevB.84.174510}.

Motivated by these seminal papers, we analyzed our simulation and found that the vortices move in directions perpendicular to the direction of $\partial_i \theta$, and vortex--anti-vortex pairs separate once produced, which is in agreement with findings of Pearl vortex in superconductor with a large background bias current~\cite{Varoquaux_2015,PhysRevB.84.174510}.
On the other hand, we also found that the vortices in our simulation are confined to domain boundaries with large $|\partial_i \theta|$, and do not move into regions where $|\dot\theta|$ is large. 
A large $|\dot\theta|$ does not exist in the superfluid context; however, if we simply boost to the vortex rest frame, $\partial_i \theta$ becomes non-zero in that frame, and the vortex would be pushed out of the region with large $|\dot\theta|$. 
This might not be the main effect that confines the vortices on the domain boundaries; for example, the vortices would prefer to reside in locations with smaller $\rho$. 
However, it does give a plausible explanation for the confinement, and calls for an extended framework to include the effect of $\dot\theta$.

Particle--vortex duality sheds light on this problem~\cite{Peskin:1977kp,Dasgupta:1981zz,Karch:2016sxi,Senthil:2018cru,beekman2011electrodynamics}. 
In two dimensions, particle--vortex duality can be established between the Abelian Higgs model and the XY model. 
Specifically, the Coulomb phase of the Abelian Higgs model is dual to the Higgsed phase of the XY model.
The charged Higgs in the Abelian Higgs model is dual to the vortex in the XY model, and the gauge field is dual to the velocity field through~\cite{Senthil:2018cru} (see footnote~\ref{ftnt:LeviCivita2Ddefn} for the definition of the Levi-Civita tensor in 2+1-dimensions)
\begin{align}\label{eq:duality}
   \frac{1}{2\pi} \epsilon_{\mu\nu\rho}F^{\mu\nu} = \partial_\rho \theta \:;
\end{align}
that is, $\dot\theta$ is dual to $F^{ij}$, the magnetic field, while $\partial_i\theta$ is dual to  $F^{0j}$, the electric field. 
In the dual description, the Magnus force on the vortices in superfluid is a Coulomb force, while the force the vortices experience when they enter the regions with non-zero $\dot\theta$ could be interpreted as the corresponding Lorentz force.

The duality is only exact in the IR fixed point. 
However, it does offer some intuitive understanding of vortex pair production at finite field strength. 
The critical velocity at which vortices pair produce would correspond to a critical electric field beyond which charged particles pair produce. 
In the simulation, it is observed that pair production occurs if $\partial_\mu \theta \partial^{\mu} \theta$ is large and space-like, also in agreement with the fact that Schwinger pair production depends on the Lorentz scalar $F^{\mu\nu} F_{\mu\nu}$~\cite{Cohen:2008wz,Kim:2003qp}.
These intriguing correspondences suggest that this duality might offer valuable insights about processes that occur at finite field strengths. 
We leave further theoretical investigation of these points to future work.

\section{Simulation Details}
\label{app:simdetails} 

\paragraph{Numerical Implementation:}
Let us elaborate on the numerical methods employed to solve the set of partial differential equations \eqref{eq:EoM}, and their FLRW analogues \eqref{eq:FLRWeom}, fully nonlinearly. 
To that end, we utilize the framework introduced in \cite{Pretorius:2004jg,Siemonsen:2020hcg}, restricted to a fixed background spacetime. 
In both Minkowski and expanding FLRW spacetimes, we use Cartesian coordinates in each spacelike slice, such that the metric takes the form: $ds^2=dt^2-a^2(t)d\bm{x}^2$ [with $a(t)\equiv 1$ for Minkowski spacetime]. 
In order to capture vortex formation, we simulate $\Phi_R\equiv\rm{Re}\ \Phi$ and $\Phi_I\equiv\rm{Im}\ \Phi$, which are well-defined at vortex locations (as opposed to $\rho$ and $\theta$), ensuring numerical stability throughout the simulation. 
The spatial and temporal derivatives of the angular and radial modes are computed using
\begin{align}
    (\partial_\mu\theta)(t,\bm{x})&\equiv \rho(t,\bm{x})^{-2}[\Phi_R(t,\bm{x})\partial_\mu\Phi_I(t,\bm{x})-\Phi_I(t,\bm{x})\partial_\mu\Phi_R(t,\bm{x})]\:; \\
    (\partial_\mu\rho)(t,\bm{x})&\equiv \rho(t,\bm{x})^{-1}[\Phi_R(t,\bm{x})\partial_\mu \Phi_R(t,\bm{x})+\Phi_I(t,\bm{x})\partial_\mu \Phi_I(t,\bm{x})]\:, 
\end{align}
respectively. 
The gradients of the former, just like $\theta$, are ill-defined at vortex locations, leading to numerical artifacts also in the gradient energy density $g_\theta$. 
The equations are discretized using fourth-order accurate finite-difference stencils in conjunction with fourth-order accurate Runge-Kutta time stepping. 
The grid spacing is $\Delta x$ and related to the timestep size $\Delta t$ by the Courant factor: $\Delta t=\Delta x/2$. 
Periodic boundary conditions are imposed in both two- and three-dimensional settings, where the box with volume $(2L)^D$ has linear size $L=10\mu^{-1}$ and $L=5\mu^{-1}$ for all two-dimensional simulations on a Minkowski and expanding FLRW background spacetime, respectively. 
Kreiss--Oliger numerical dissipation is used for numerical stability. 
This implies that modes of wavelength comparable to the grid spacing, $\lambda\sim\mathcal{O}(1)\Delta x$, are damped efficiently and removed from the evolution. 
This technique also automatically damps out unresolved (i.e., sub-grid/high-momentum) modes in the initial conditions after some transient; we find these transients to be irrelevant to our conclusions (see also comments at footnote \ref{ftnt:KOdissipation}).
The scale-invariant Gaussian random field, entering the initial conditions described in detail in Sec.~\ref{sec:pertfullnonlinear}, is constructed using the Box--Muller approach. 
Lastly, vortices are located using the algorithm introduced in \cite{Fleury:2015aca}. 
In two spatial dimensions ($D=2$), this unambiguously identifies the vortices piercing a cell. 
In three spatial dimensions ($D=3$), we search for strings through only a single face of a given cell (say, the face orthogonal to the $+x$-direction) to find the following lower bound on the string length density $\xi$: this density $\xi=\Delta x/V_b (N_S/N_T)$ is obtained by comparing the total number of cells, $N_T$, to the total number of cells with a face pierced by a string, $N_S$, where the total simulation volume is $V_b$. 
This generally underestimates the length density, since the length of a string's segment associated with a cell is at least%
\footnote{\label{ftnt:lengthInCell}%
    Because the string cores are well-resolved, the string cannot be highly curved within a single cell. 
    While there may be a small, $\mathcal{O}(1)$ inaccuracy in these per-cell length bounds owing to string curvature, this is likely a negligible effect.
    } %
$\Delta x$.
In principle, the length of this segment could be as long as the simulation box's linear size $2L$ if the string is aligned with the faces. 
However, in practice, the string network lacks order on scales comparable to the box size. 
Therefore, the density $\xi$ is likely at most a factor of a few below the true string length density. 
In those cases with complete string annihilation, the absence of strings piercing all six faces of each cell in the simulation domain was checked.

\begin{figure}
    \centering
    \includegraphics[width=0.48\linewidth]{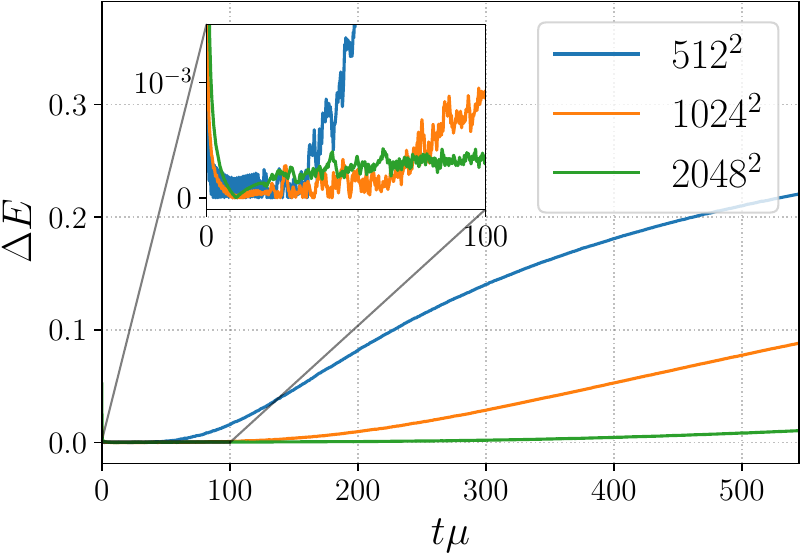}
    \hfill
    \includegraphics[width=0.48\linewidth]{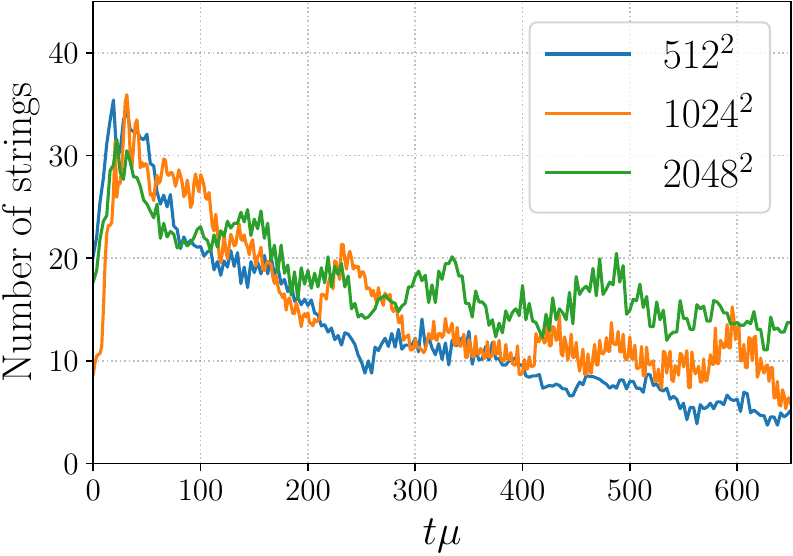}
    \caption{The convergence behavior of the solution with parameters $D=2$, $\mathcal{J}_0=0.1v^2\mu$, $\rho_i/v=5$, and $\sigma_{\rho_i}=10^{-2}\rho_i$ on a Minkowski background spacetime. 
    \underline{\textsc{Left panel:}}~%
    The relative difference $\Delta E=|E_{\rm T}(t)-E_{\rm ref}|E^{-1}_{\rm ref}$ throughout the simulation up to late times with different resolutions, where the reference energy is defined as $E_{\rm ref}=E_{\rm T}(t=10\mu^{-1})$.
    \underline{\textsc{Right panel:}}~%
    The moving time-average of the total number of strings in the domain. 
    For convenience, we sampled the number of strings at coarser time steps here, as compared to the results shown in Fig.~\ref{fig:largecirculation_energy} in the main text.}
    \label{fig:2dconv}
\end{figure}

\paragraph{Convergence in Flat Spacetime:}
To probe the validity of our results in the continuum limit, we perform a sequence of convergence studies. 
In Fig.~\ref{fig:2dconv}, we show the convergence behavior of an example solution discussed throughout this work, obtained using varying resolutions.
This is representative of other two- and three-dimensional simulations performed. 
Recall, the parametric resonance reaches the nonlinear regime and induces vortex pair production at around $t\mu\approx 25$. 
First, numerical dissipation efficiently damps the high-frequency content of the scalar field at early times, $t\mu\lesssim 10$, which leads to the non-conservation of the total energy as shown in the inset of the left panel of Fig.~\ref{fig:2dconv}. 
This is expected and has little effect on the bulk dynamics, as these are dictated by well-resolved wavenumbers driving the parametric resonance. 
Therefore, the relative error in the total energy throughout the simulation, $\Delta E$, as defined at the reference time $t\mu=10$, by which the high-frequency content was removed and the total energy $E_{\rm T}(t)$ stabilized.
The relative difference $\Delta E$ converges towards zero with increasing resolution; at $t\mu=100$ this convergence is consistent with $\sim (\Delta x)^3$. 
Once vortices are forming, we find that the rolling time-averaged total number of vortices in the simulation domain is largely independent of resolution of up to $t\mu\approx 300$. 
Beyond this, the vortex number increases slightly with resolution. 
Vortex pairs are produced by small-scale perturbations as argued in Sec.~\ref{sec:vortexformationdynamics}; high-resolution simulations resolve smaller scales, allowing for higher vortex densities. 
However, a direct comparison of this between resolutions is challenging, since we obtain a new noise realization for each simulation. 
The default resolution in two dimension is $1024^2$, whereas it is $512^3$ in three dimensions (this leads to the same spatial resolution, as we are using $L=10\mu^{-1}$ and $L=5\mu^{-1}$ for the box sizes, respectively).

\paragraph{Critical Field for Vortex Formation:}
Subtleties arise when determining the critical radial displacements $\rho_c$ as a function of $\mathcal{J}_0$ shown in Fig.~\ref{fig:veffmin}. 
Not only could the threshold depend on resolution as outlined in the above discussion, but due to the small number density of vortices for $\rho_i\gtrsim\rho_c$, the critical threshold $\rho_c$ may depend on the noise realization $\mathcal{N}(\bm{x})$. 
Precisely identifying this threshold requires considering an average over a series of noise realizations, as well as varying $\sigma_{\rho_i}$. 
In this work, we simply obtain approximate thresholds as follows. 
At fixed $\mathcal{J}_0$, we perform a suite of simulations spanning a range of initial displacements $\rho_i$ from just below to just above the threshold $\rho_c$. 
For each such displacement, we consider only a single noise realization and fix the variance to $\sigma_{\rho_i}=10^{-6}\rho_i$.
We then estimate the threshold as the mean $\rho_c\approx (\rho_i^<+\rho_i^>)/2$, between the initial displacement $\rho_i=\rho_i^<$, which marginally produced vortex pairs, and $\rho_i=\rho^>_i$, which lead to no vortices throughout its evolution. 
The errorbars shown in Fig.~\ref{fig:veffmin} are then simply $\pm 2(\rho_i^>-\rho_i^<)$. 
Importantly, we do not strictly rule out that vortices may form for $\rho_{\text{min}} < \rho_i<\rho_i^<$, due to the subtleties listed above.

\paragraph{Growth Rates for Perturbations:}
For the comparison of the growth rates of the various spatial Fourier modes, shown in the right panel of Fig.~\ref{fig:2dexpgrowthLargec0} for $D=2$, we proceeded as follows.
At each time step throughout the evolution, we compute the spatial Fourier transform $\tilde{\rho}(t,\bm{k})$ of $\rho(t,\bm{x})-\langle\rho(t,\bm{x})\rangle$ and transform to cylindrical coordinates $\bm{k}=(k_x,k_y)\rightarrow (k_r,k_\varphi)$. 
Since the spatial dependence of $\rho(t,\bm{x})-\langle\rho(t,\bm{x})\rangle$ is initialized to be a scale-invariant Gaussian random field, $\tilde{\rho}(t,\bm{k})$ exhibits a high degree of spherical symmetry in momentum space. 
We define the angular-averaged Fourier transform $\tilde{\rho}_\varphi(t,k_r)=(2\pi)^{-1}\int_0^{2\pi} d k_\varphi\, k_r \tilde{\rho}(t,k_r,k_\varphi)$. 
Finally, for a given $k_r$-mode, we fit an exponential $\sim \exp(\Gamma t)$ to $\tilde{\rho}_\varphi(t,k_r)$ to determine $\Gamma(k_r)$. 
Notice, the value $\Gamma(k_r)$ depends on the fitting time window as follows. 
At early times, $\tilde{\rho}_\varphi(t\mu \ll 1,k_r)$ is constant in $k_r$ (up to fluctuations varying between noise realizations), sourcing all unstable modes at roughly the same amplitude. 
After a few e-folding times of the most unstable mode, the amplitudes of these modes differ by orders of magnitude. 
This implies that the numerical discretization error associated with the most unstable mode mixes into those unstable modes with smaller $\Gamma(k_r)$. 
Therefore, at late times (before the modes reach the nonlinear regime) most $k_r$-modes grow with similar growth rates dictated by the most unstable mode.
This is a numerical artifact and can be remedied to a degree by increasing $L$ and spatial resolution. 
To obtain $\Gamma(k_r)$ for the case presented in the right panel of Fig.~\ref{fig:2dexpgrowthLargec0}, we restrict to a fitting window early on in the exponential growth of $\tilde{\rho}_\varphi(t,k_r)$, where the growth rate of $k_r$-modes with smaller growth rates are not significantly affected by the truncation error of the most unstable mode. 
The errorbars shown in Fig.~\ref{fig:2dexpgrowthLargec0} are determined taking into account different fitting windows, a simulation with $\times 2$ the resolution, and a simulation with $\times 4$ the box volume. 
Notice, this uncertainty is larger for those $k_r$-modes with smaller $\Gamma(k_r)$. 
Finally, extracting $\Gamma(k_r)$ from $\theta(t,\bm{x})-\langle\theta(t,\bm{x})\rangle$ yields results consistent with those presented in the right panel of Fig.~\ref{fig:2dexpgrowthLargec0}.

\begin{figure}[t]
    \centering
    \includegraphics[width=0.49\linewidth]{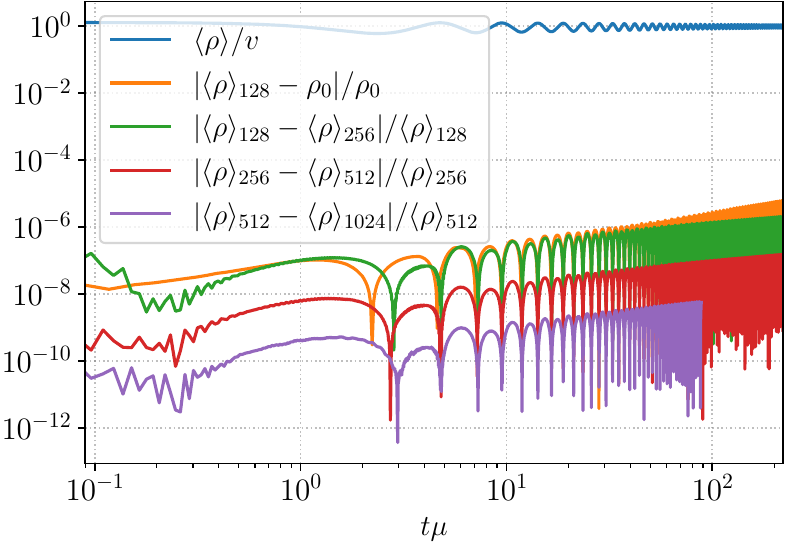}
    \hfill
    \includegraphics[width=0.485\linewidth]{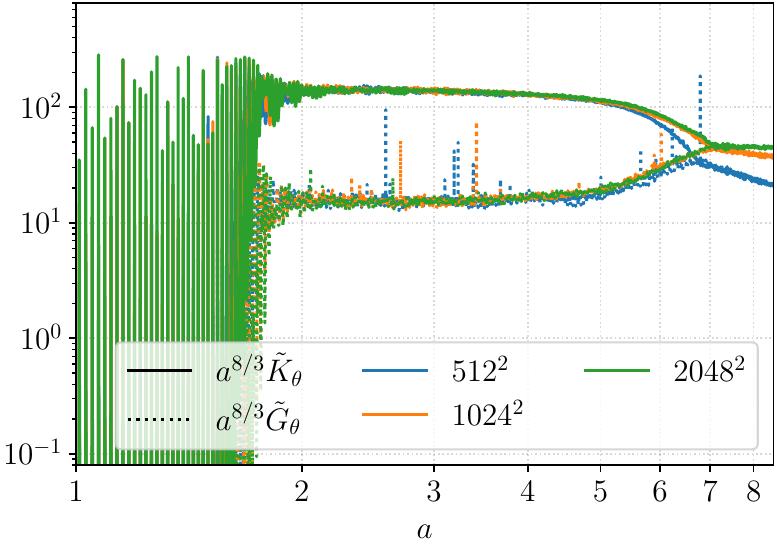}
    \includegraphics[width=0.99\linewidth]{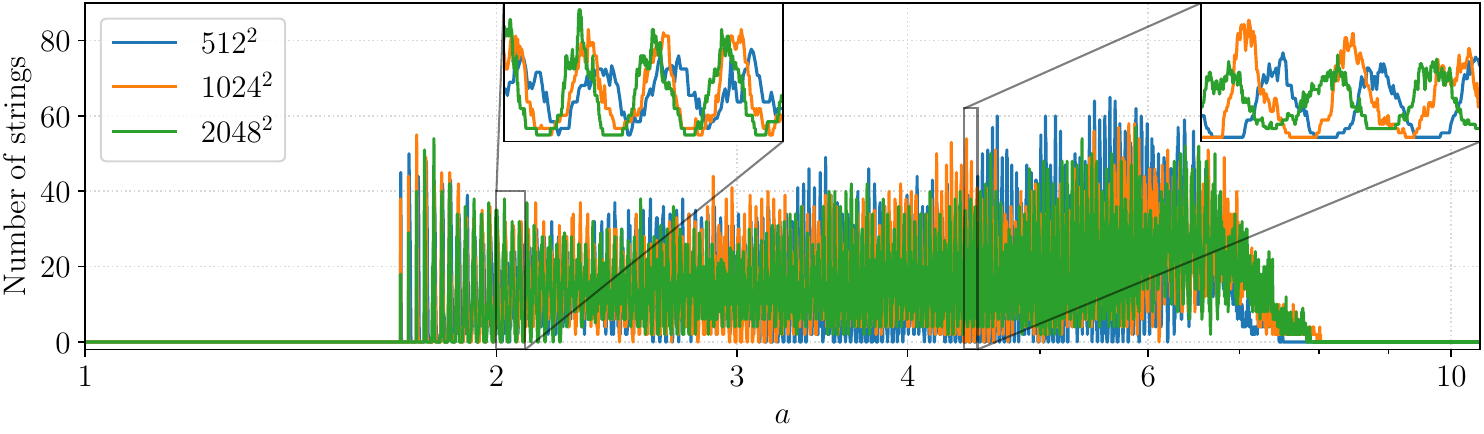}
    \caption{The convergence behavior of the solutions on an expanding FLRW background. 
    \underline{\textsc{Top left panel:}}~%
    Focusing on the case with $D=2$, $\rho_i/v=3$, and $\mathcal{J}_0=0.1v^2\mu$. 
    The solution for the radial mode, $\rho_0(t)$, obtained from numerically integrating \eqref{eq:backgroundhubble}, as well as relative differences between this solution and the spatial average of $\rho(t,\bm{x})$ obtained using our nonlinear numerical methods solving the full scalar field equations on an expanding FLRW background with spatial resolutions between $128^2$ and $1024^2$ grid points, shown as different colored lines, as indicated by the subscripts in the legend.
    \underline{\textsc{Top right panel:}}~%
    Turning to the case with associated with $D=2$, $\rho_i/v=5$, and $\mathcal{J}_0=0.1v^2\mu$. 
    The spatially averaged kinetic and gradient energy densities, $\tilde{K}_\theta$ (solid lines) and $\tilde{G}_\theta$ (dotted lines), through the transition to the late-time radiation state for three different spatial resolutions, as denoted by the different colored lines shown in the legend.
    \underline{\textsc{Bottom panel:}}~%
    The total number of vortices in the simulation box of size $(10/\mu)^2$ with parameters $D=2$, $\rho_i/v=5$, and $\mathcal{J}_0=0.1v^2\mu$ for the three different resolutions. }
    \label{fig:cosmoconv}
\end{figure}

\paragraph{Convergence of Cosmological Implementation:}
To test our cosmological implementation, we solve for the evolution of the zero-mode $\rho_0$ governed by \eqref{eq:EoM0re}, augmented by the appropriate Hubble friction in two spatial dimensions, as shown in \eqref{eq:backgroundhubble}. 
This simpler set of ordinary differential equations can be readily solved using standard numerical integration techniques. 
The solution can be compared with the early-time (i.e., before unstable modes acquire a significant amplitude) solution of a cosmological simulation solving the full complex scalar field equations on a radiation-dominated FLRW background spacetime. 
For the purposes of this comparison, we set $\sigma_{\rho_i}=10^{-14}\rho_i$ to ensure the unstable modes are sourced at sufficiently small amplitudes.
In Fig.~\ref{fig:cosmoconv}, we compare these two solutions. 
Clearly, at early times these two solutions are in excellent agreement. 
Furthermore, with increasing resolution of our fully nonlinear numerical methods, the difference converges at the expected fourth order. 
To convince ourselves that the vortex annihilation and transition to the radiation state at late times during the cosmological evolution is not due to insufficient resolution, we show the relevant energy densities in the left panel of Fig.~\ref{fig:cosmoconv}. 
Evidently, the densities are converged in the relevant regime, and the transition to radiation is independent of resolution. 
The curves only begin deviating from each other once $(a\mu)^{-1}$ approaches the grid scale. 
In the bottom panel of Fig.~\ref{fig:cosmoconv}, we show the change of the vortex density with resolution. 
The periodicity is a consistent feature across resolutions, while the overall density may vary from burst to burst between resolutions (for all the reasons outlined above, including---but not limited to---necessarily different random realizations of the initial conditions [i.e., phases and amplitudes] across the different-resolution simulations).
Lastly, the bursts of vortex production in different-resolution simulations experience a mild dephasing between early and lates times. 
This originates both from an initial phase-offset at the time of the first burst, as well as numerical drift accumulated throughout the simulations. 
However, the total dephasing of the bursts accumulates to $\sim \mathcal{O}(1)$ radians across $\sim\mathcal{O}(100)$ bursts.

\begin{figure}[t]
    \centering
    \includegraphics[width=0.5175\linewidth]{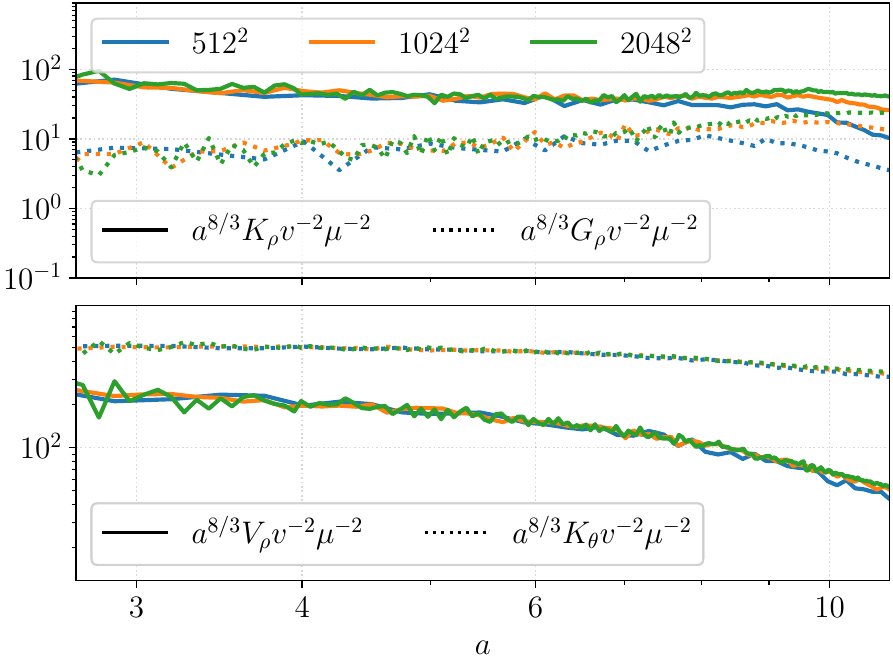}
    \hfill
    \includegraphics[width=0.4725\linewidth]{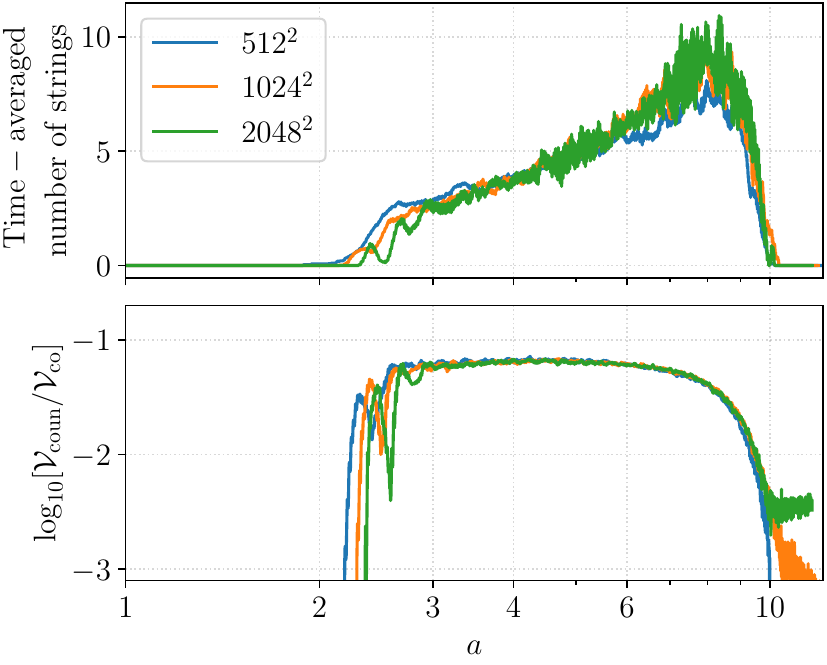}
    \caption{The convergence properties of the solution with parameters $\mathcal{J}_0=10^2v^2\mu$, $\rho_i/v=6$, and $\sigma_{\rho_i}=10^{-5}\rho_i$ for $D=2$, with increasing resolution from $512^2$ through $2048^2$ grid points (as denoted by the different color lines specified in the legends). 
    This set of parameter choices was discussed in the main text at Fig.~\ref{fig:cosmo_largec0}.
    \underline{\textsc{Left column:}}~%
    Various relevant spatially- and temporally-averaged energy densities, shown here only in the regime after saturation of the parametric resonance. 
    Different energy-density components are plotted as different line styles, as denoted in the legends for each panel.
    \underline{\textsc{Right column:}}~%
    The evolution of the rolling time-average of the total number of vortices in the simulation domain of size~$(10/\mu)^2$~[top panel], as well as the ratio of the area of co- to counter-rotating regions~[bottom panel].}
    \label{fig:cosmoconvmean6}
\end{figure}

In Fig.~\ref{fig:cosmoconvmean6}, we show the convergence of the solution on an expanding background spacetime associated with large initial $\mathcal{J}_0$ discussed in the main text; cf.~Fig.~\ref{fig:cosmo_largec0}.
The primary properties of the solution are converged and consistent across resolutions. 
In particular, the disappearance of vortices and reduction in area of counter-rotating regions around $a\approx 8$, as well as the densities $V_\rho$ and $K_\theta$ are well-resolved.
In contrast, the densities $K_\rho$ and $G_\rho$ vary significantly across resolutions towards late times.
This can be understood as follows: since the initial radial amplitude $\rho_i$ is large, the transition from a state with vortices to a state without occurs only around $a\approx 10$. 
At these times, physically relevant modes with co-moving wavelengths $\lesssim (a\mu)^{-1}$ approach the grid-scale and begin to be dissipated by numerical dissipation. 
This manifests itself as a spurious reduction of energy densities dominated by such modes (e.g., $K_\rho$ and $G_\rho$) for $a\gtrsim 7$. 
Lastly, with increasing resolution the radiation state (in addition to the residual large-scale coherent field rotation) for $a\gtrsim 10$ is better resolved, leading to an enhancement of $\mathcal{V}_{\rm coun}/\mathcal{V}_{\rm co}$ after vortices disappeared (see bottom right panel of Fig.~\ref{fig:cosmoconvmean6}).

\section{Perturbative Stability Analysis at \texorpdfstring{$\mathcal{J}_0 \ll \mu v^2$}{J0 << mu v-squared} and \texorpdfstring{$\bar{\zeta}_0 \ll 1$}{zeta0bar << 1}}
\label{app:mathieuAnalysis}

In this appendix, we undertake the linearized stability analysis of the perturbations $\delta \rho$ and $\delta\sigma$ that was referred to in Sec.~\ref{sec:SimplifiedAnalytics}.
We work here in the limits $\mathcal{J}_0 \ll \mu v^2$ and $\bar{\zeta}_0 \ll 1$.

A note of caution: we abuse notation in this appendix by otherwise silently redefining some quantities already defined otherwise in Sec.~\ref{sec:SimplifiedAnalytics}; any contrary definitions given in Sec.~\ref{sec:SimplifiedAnalytics} should be disregarded when reading this appendix.
In all cases however, once the limits $\mathcal{J}_0 \ll \mu v^2$ and $\zeta_0(t)\ll 1$ limits are taken in the definitions given in this appendix, the expressions given in Sec.~\ref{sec:SimplifiedAnalytics} for the relevant quantities are recovered.

First, we consider more carefully the linearization of \eqref{eq:EoM0re2} in the $\zeta_0(t)\ll 1$ limit.
To start, let us write $\rho_0(t) = \bar{\rho}_0 ( 1 + \zeta_0(t) )$ where $\langle \zeta_0 \rangle = 0$ (angle brackets denoting the time average) and $\bar{\rho}_0$ is defined below. 
We also define $\tilde{\omega}_i \equiv \omega_i/\mu$ and $\tilde{t} = \mu t$, and recall that $\lambda = \mu^2/v^2$ by definition.
Inserting $\rho_0(t) = \bar{\rho}_0 ( 1 + \zeta_0(t) )$ into \eqref{eq:EoM0re2}, expanding in $\zeta_0$, and keeping terms only up to $\mathcal{O}(\zeta_0)$, we have 
\begin{align}
    \partial_{\tilde{t}}^2 {\zeta}_0 + \left(  3\bar{\rho}_0^2 / v^2  + 3 \mathcal{J}^2_0/(\mu^2\bar{\rho}_0^4) - 1 \right) \zeta_0 + \left[ \bar{\rho}_0^2 / v^2  - \mathcal{J}^2_0/(\mu^2 \bar{\rho}_0^4) - 1 \right] &= 0\:. \label{eq:zetaEoMLinear}
\end{align}
The term in the $[\,\cdots]$ bracket must be set to zero in order to enforce the definition $\langle \zeta_0 \rangle = 0$, thereby implicitly defining $\bar{\rho}_0$ via $\bar{\rho}_0^2 / v^2 - \mathcal{J}_0^2/(\mu^2 \bar{\rho}_0^4)  - 1 = 0$; while this can actually be solved in closed form, the resulting expression is algebraically complicated so we omit it here.
However, let us define
\begin{align}
    \bar{\rho}_0^2 &\equiv v^2(1+\rho'_0) &&\Rightarrow& \rho'_0 - (\mathcal{J}_0/(\mu v^2))^2( 1 + \rho'_0)^{-2}  &= 0\:;  \label{eq:rhohat0}
\end{align}
note that $\rho'_0>0$ is required by its defining equation.
Re-arranging the defining equation for $\bar{\rho}_0$ gives $\mathcal{J}_0^2/(\mu^2 \bar{\rho}_0^4)= \bar{\rho}_0^2 / v^2 - 1 = \rho'_0$; therefore, if $\mathcal{J}_0 \ll \bar{\rho}_0^2\mu$, then $\rho'_0 \ll 1$.
In this appendix, we assume $\mathcal{J}_0\ll \mu\rho_0^2$, so we shall set $\rho'_0=0 \Rightarrow \bar{\rho}_0 = v$ in what follows.
Substituting into \eqref{eq:zetaEoMLinear} in this limit leads to
\begin{align}
    \partial_{\tilde{t}}^2 {\zeta}_0 + 2 \zeta_0 &= 0 \:,
\end{align}
so that
\begin{align}
    \zeta_0(\tilde{t}) &= \bar{\zeta}_0 \cos ( \sqrt{2} \cdot \tilde{t} + \varphi_{\zeta} )\:. \label{eq:zetaEoMLinearSoln}
\end{align}

Now consider that we showed in Sec.~\ref{sec:pert} that the two perturbations $\delta \sigma$ and $\delta \rho$ are governed in the $\mathcal{J}_0 \ll \mu v^2$ limit by \eqref{eq:XiEqn}, where $\tilde{\omega}_i$ is defined via \eqref{eq:omegaTildeDefn} to be 
\begin{align}
    \tilde{\omega}^2_i(\tilde{t}) = \tilde{k}^2 - 1 + c_i [ \rho_0(t) / v ]^2 \: ,
\end{align}
where $\tilde{k} \equiv k/\mu$, and the $c_i$ are defined at \eqref{eq:ciDefn}.
Defining $\bar{c}_i \equiv (\bar{\rho}_0/v)^2 c_i$ and substituting for $\rho_0(t)$, we have
\begin{align}
    \tilde{\omega}^2_i(\tilde{t}) &= \tilde{k}^2  - 1 + \bar{c}_i \left[ 1 + \zeta_0(\tilde{t}) \right]^2 \\ 
        &= ( \tilde{k}^2 + \bar{c}_i - 1 ) + 2 \bar{c}_i \zeta_0(\tilde{t}) + \bar{c}_i \zeta_0(\tilde{t})^2 \label{eq:omegaExpr}\:,
\end{align}
Now, taking the $\zeta_0(\tilde{t})\ll 1$ limit in order to focus only on small-amplitude field excursions, and also assuming the $\mathcal{J}_0\ll \mu\rho_0^2$ limit in which $\bar{c}_i \approx c_i$, we have
\begin{align}
    \partial_{\tilde{t}}^2 \mathbb{X}_i + \tilde{\omega}_i^2 \mathbb{X}_i&=0\:; &
    \tilde{\omega}_i^2 &\equiv \frac{ \omega_i^2}{\mu^2} \approx ( \tilde{k}^2 + c_i - 1 )  + 2 c_i \zeta_0(\tilde{t}) \:. \label{eq:omegaSmall2}
\end{align}

As noted in the main text in the paragraph below \eqref{eq:ciDefn}, this casts the approximate EoM into the form of the Mathieu equation $\partial_{\tilde{t}}^2 f + \tilde{\omega}_i^2(\tilde{t})\cdot f= 0 $ where $\tilde{\omega}_i^2(\tilde{t}) \equiv \omega_{0,i}^2 + 2\epsilon_i \cos( \sqrt{2} (\tilde{t} - \tilde{t}_0) )$, which exhibits both narrow and broad parametric resonance phenomena.
For $\epsilon_i \ll \omega_{0,i}^2/2$, narrow resonance occurs~\cite{Kovacic:2018weg}: the primary narrow resonance exists for $1/2-\epsilon_i \lesssim \omega_{0,i}^2 \lesssim 1/2+ \epsilon_i$, while a secondary narrow resonance occurs when $2 - \epsilon_i^2/6 \lesssim \omega_{0,i}^2 \lesssim 2 + 5\epsilon_i^2/6$; in fact, resonances exist for all $\omega_{0,i}^2 \sim n^2/2$ for $n\in\mathbb{Z}, n>0$ (the centers of these bands can also be offset by amounts $\propto \epsilon_i^j$ for some $j\in\mathbb{Z}, j>0$), with widths $\Delta(\omega_0^2) \propto \epsilon_i^n$.
However, the higher-$n$ resonances are increasingly inefficient at driving growth: for instance, at small $\epsilon_i$, it can be shown that the largest Floquet exponents%
\footnote{%
    The growing solution amplitude increases by $\text{exp}[\tilde{\Gamma} \tilde{T}]$ in each period $\Delta \tilde{t} = \tilde{T}=\pi\sqrt{2}$ of the $\tilde{\omega}_{i}^2$ oscillation.
} %
$\tilde{\Gamma}$ for the primary band ($n=1$) are $\tilde{\Gamma} \sim \epsilon_i/\sqrt{2}$ for $\omega_{0,i}^2=1/2$ (i.e., at the band center), while those in the secondary band ($n=2$) are $\tilde{\Gamma} \sim \epsilon_i^2/(4\sqrt{2})$ for $\omega_{0,i}^2=2+\epsilon_i^2/3$ (i.e., at the band center); see, e.g., \cite{MagnusWinkler,Kovacic:2018weg} for brief introductions to Floquet analysis.
For $\epsilon_i \gtrsim \omega_{0,i}^2/2$, the resonance is instead broad: the field $f$ exhibits durations where $\tilde{\omega}^2_i(\tilde{t})<0$, yielding explosive field growth~\cite{Kofman:1997yn} (except in certain narrow bands of parameter space).\\

We now undertake a perturbative stability analysis.
Consider first the $\delta \sigma$ perturbation, for which $c_\sigma = 1$.
Mapping this onto the Mathieu equation $\omega_0$ and $\epsilon$ variables defined above, we have
\begin{align}
    \omega_{0,\sigma}^2 &= \tilde{k}^2 \: ; & \epsilon_{\sigma} &= \bar{\zeta}_0  \:.
\end{align}
It is straightforward to then show that the primary narrow resonance is accessed when 
\begin{align}
   \tfrac{1}{2} - \bar{\zeta}_0  &\leq \tilde{k}^2 \leq  \tfrac{1}{2} + \bar{\zeta}_0 & \Rightarrow &&
    \sqrt{\max\left[ 0 ,  \tfrac{1}{2} - \bar{\zeta}_0 \right]} &\leq \tilde{k} \leq \sqrt{  \tfrac{1}{2} + \bar{\zeta}_0  }\:. \label{eq:lowerBandEdge1}
\end{align}
This analysis is perturbatively consistent and in the narrow resonance regime ($2\epsilon_{\sigma} < \omega_{0,\sigma}^2$) when $\bar{\zeta}_0 \ll \tilde{k}^2/2$.
If we assume that $\bar{\zeta}_0<1/2$ (which will be seen to be consistent \emph{a posteriori}) then the most stringent version of this narrow-resonance consistency constraint is at the lower band edge, from which we must have 
\begin{align}
    \bar{\zeta}_0 \leq \frac{\tfrac{1}{2} - \bar{\zeta}_0 }{2} \quad \Rightarrow \quad \bar{\zeta}_0 \leq \frac{1}{6} \:.
\end{align}
On the other hand, the narrow-resonance consistency constraint is slightly looser at the upper band edge:
\begin{align}
    \bar{\zeta}_0 \leq \frac{\tfrac{1}{2} + \bar{\zeta}_0 }{2} \quad \Rightarrow \quad \bar{\zeta}_0 \leq \frac{1}{2} \:.
\end{align}
For consistency, it therefore follows that unstable band is [cf.~\eqref{eq:sigmaUnstable}]
\begin{align}
    \sqrt{ \tfrac{1}{2} - \bar{\zeta}_0 } \leq \tilde{k} \leq \sqrt{ \tfrac{1}{2} + \bar{\zeta}_0  } \qquad [\delta \sigma \text{ unstable}]\:.
\end{align}
At the largest allowed value of $\bar{\zeta}_0$, we see that the unstable narrow-resonance band stretches up to $\tilde{k} \sim 1$.
Moreover, the largest Floquet exponent for this band is $\tilde{\Gamma} = \epsilon_\sigma/\sqrt{2} = \bar{\zeta}_0/\sqrt{2}$, and occurs for $\omega_{0,\sigma}^2 = 1/2 \Rightarrow \tilde{k} = 1/\sqrt{2}$.
For larger values of $\bar{\zeta}_0$, this perturbation component may exhibit broad resonance.
It is also possible for $\delta \sigma$ to access the higher narrow resonances for larger values of $\tilde{k}$.

Now consider the $\delta \rho$ perturbation component. 
The mapping onto the Mathieu equation $\omega_0$ and $\epsilon$ variables is now
\begin{align}
    \omega_{0,\rho}^2 &= \tilde{k}^2 + 2  \: ; & \epsilon_{\rho} &= 3 \bar{\zeta}_0  \:.
\end{align}
The condition for the primary resonance to be accessed is now $0 \lesssim \tilde{k} \lesssim \sqrt{3(\bar{\zeta}_0-1/2)}$, which gives a window that is open only for $\bar{\zeta}_0>1/2$.
However, consistency of being in the narrow-resonance regime demands $\bar{\zeta}_0 \leq (\tilde{k}^2 + 2 )/6$.
It is easy to show that this cannot be satisfied in that band of $\tilde{k}$ values for $\bar{\zeta}_0>1/2$. 
The narrow primary resonance is thus not available to this mode.

On the other hand, the second narrow resonance band is available when
\begin{align}
    2 - \frac{1}{6} \left( 3 \bar{\zeta}_0 \right)^2 \leq  \tilde{k}^2 + 2  \leq 2 + \frac{5}{6} \left( 3  \bar{\zeta}_0 \right)^2\:,
\end{align}
leading to [cf.~\eqref{eq:rhoUnstable}]
\begin{align}
    \Rightarrow 0 &\leq \tilde{k} \leq \sqrt{\frac{15}{2}} \cdot \bar{\zeta}_0\qquad [\delta\rho \text{ unstable}]\:.
\end{align}
This implies that $\omega_{0,\rho}^2 \sim 2 + \mathcal{O}(\bar{\zeta}^2_0)$, so this is in the narrow resonance regime so long as $\bar{\zeta}_0 \lesssim 1/3$.
Note also that, at the largest consistent value of $\bar{\zeta}_0$, the upper edge of the band is at $\tilde{k} \sim \sqrt{5/6} \sim \mathcal{O}(1)$.
Moreover, the largest Floquet exponent for this band is $\tilde{\Gamma} = \epsilon_\rho^2/(4\sqrt{2}) = (3\bar{\zeta}_0/2)^2/\sqrt{2}$, and occurs for $\omega_{0,\rho}^2 = 2 + \epsilon_{\rho}^2/3 \Rightarrow \tilde{k} = \sqrt{3}\cdot \bar{\zeta}_0$.
Again, access to the higher narrow resonances is possible.

\section{Derivations of Some Parametric Scalings}
\label{app:scalings}

In Sec.~\ref{sec:FLRW}, we noted certain scalings observed in the numerics of the fields $\rho, \theta$ and their energy-density components with the scale factor $a$, in certain limits.
In this appendix, we show how these scalings arise analytically from \eqref{eq:backgroundhubble2}.

\subsection{Limiting Case of \texorpdfstring{$\rho \gg v$}{rho-i >> v}}
\label{app:scalings1}

In the limit $\mathcal{J}_0 \ll \mu v^2$ and $\rho \gg v$, from \eqref{eq:backgroundhubble2} we have $\ddot{\rho}_0 + DH \dot{\rho}_0 + \rho_0^3 \approx 0$.
Rescaling $t \equiv x^{3(D+1)/(D+3)}$ and $\rho_0 \equiv a^{-D/3} f$, we find $ f'' + (3(D+1)/(D+3))^2 f^3 - 2D(D-3)/(D+3)^2 f/x^2 = 0$, where ${}^\prime \equiv d/dx$. 
But for $0<D\leq 3$, this is clearly an \emph{undamped} anharmonic oscillator equation with positive (and, for $D\neq3$, time-dependent) frequency squared, so $f$ just oscillates with roughly fixed amplitude: $f\propto a^0 \times g$, where $g$ is an oscillatory function (which we will drop below); this is strictly true for $D=3$, and true at late time for $D=2$.

These facts confirm that, in amplitude, $\rho_0 \propto a^{-D/3}$.
For large $\rho \gg v$, $V_\rho \propto a^{-4D/3}$ follows immediately.
From $\mathcal{J}_0 \sim \rho_0^2 \dot{\theta}_0 \propto a^{-D}$, we then also have $\dot\theta_0 \propto a^{-D/3}$, from which it follows that $k_\theta \propto a^{-4D/3}$.
Finally, in $D=3$ and in $D=2$ at late times, the EoM for $f$ takes the form $f'' + c f^3 \approx 0$ for some constant $c$.
Multiplying through by $f'$ and integrating once, we have $(f')^2/2 + c f^4/4 \sim \text{const.}$
Since $f \propto a^0$, we must then have $f' \propto a^0$.
Moreover, it is fairly straightforward to show that $\dot{\rho} \propto x^{-4D/(D+3)}f'$ at late times, from which it follows that $\dot{\rho} \propto a^{-2D/3}$ and therefore $k_\rho \propto a^{-4D/3}$.

We also note that, with $\mathcal{J}_0$ large,  $\rho_0^3 \propto a^{-D}$ and also $\mathcal{J}_0^2/\rho_0^3 \propto a^{-2D}/(a^{-D}) \propto a^{-D}$.
As a result, one can see from \eqref{eq:backgroundhubble2} that the scalings we have derived here will also hold at non-zero $\mathcal{J}_0$.
In particular, it can be shown that the same scalings of $V,k_\rho,k_\theta\propto a^{-4D/3}$ obtain even when $\rho \sim \rho_{\text{min}} \propto \mathcal{J}_0^{1/3} \propto a^{-D/3}$ in this case.

\subsection{Limiting Case of \texorpdfstring{$\rho \sim v$ (for $\mathcal{J}_0 \ll \mu v^2$)}{rho-i ~ v (for J0 << mu v-squared)}}

In the limit $\rho \simeq v$ with $\mathcal{J}_0 \ll \mu v^2$, different scalings obtain.
To see this, set $\mathcal{J}_0=0$ and $\rho = v + \delta$ in \eqref{eq:backgroundhubble2}, expand in powers of $\delta$, and keep only the lowest-order terms.
This gives $\ddot{\delta} + DH \dot{\delta} + 2\mu^2 \delta \approx 0$.
Then take $\delta \equiv a^{-D/2} f$, which yields $\ddot{f} + \left[ 2\mu^2 + D/(D+1)^2/t^2 \right] f \approx 0$, which is again an oscillator equation with a time-dependent frequency, which goes to a simple harmonic oscillator equation at late times.
As a result, we expect a roughly fixed-amplitude oscillation: $f \propto a^0 \times h$, where $h$ is an oscillatory function (which we omit in what follows).
Moreover, at late times, we have (by a similar argument as in App.~\ref{app:scalings1}) that $\dot{f}^2/2+\mu^2f^2 \approx \text{const.}$, which implies that we must have $f' \propto a^0$ in amplitude too.
But $\mathcal{J}_0 \approx v^2 \dot\theta \propto a^{-D} \Rightarrow \dot{\theta}\propto a^{-D}$.
Putting this all together, it follows that $V \propto \mu^2 \delta^2 \propto a^{-D}$, $k_{\rho} \propto \dot{\rho}^2 \propto a^{-D} \dot{f}^2 \propto a^{-D}$ (at late times), and $k_\theta \propto v^2 \dot{\theta}^2 \propto a^{-2D}$.

\section{String Burst Periodicity for General \texorpdfstring{$\rho/v$}{rho/v}}
\label{app:burstp}

In this appendix, we provide more details about the semi-analytical and numerical theory predictions shown in Fig.~\ref{fig:cosmo2d_rho3}. 
In Sec.~\ref{sec:FLRW}, we discussed the periodicity of the string burst cycles in the large $\rho/v$ limit in arbitrary dimensions, and found the analytical solution in \eqref{eq:periodicityDdim}. 
However, to better resolve the transition to radiation, we chose to simulate a smaller ratio, $\rho_i/v =5$, in two dimensions. 
In this two dimensional simulation, the vortices first form when $a = 1.7$, at which point the average $\rho/v$ in the simulation box is $\sim 2.5$. 
The average $\rho/v$ continues to decrease as the Universe expands, until the average $\rho/v \simeq 1$, when all string burst stops at $a \simeq 8$. 
The large $\rho/v$ approximation, on which \eqref{eq:periodicityDdim} is based, therefore does not apply very well to our simulated parameter choice; a more precise semi-analytical computation is needed to compare with the simulation results. 

In this improved semi-analytical computation, we continue to assume that, once the counter-rotating regions form, the $U(1)$ charge is covariantly conserved over a period of string formation and annihilation. 
As such, the $j^0$ charge density inside the counter-rotating region scales as $j^0 \propto a^{-D}$ [cf.~\eqref{eq:J0scaling}], and the equation of motion for the radial mode (which we treat as approximately spatially constant within each region) is [cf.~\eqref{eq:backgroundhubble2}]%
\footnote{%
    In this appendix, $\rho$ and $\dot\theta$ correspond to $\rho_\textsc{crr}$ and $\dot\theta_\textsc{crr}$ in the main text, respectively; we have dropped the subscript for simplicity of notation.} %
\begin{align}
    \ddot{\rho} + D H \dot{\rho} -  \mu^2 \rho- {j}_0^2 / \rho^3 + \lambda \rho^3 =0\:. \label{eq:appRhoEqn}
\end{align}

During the initial phase of the evolution, the field within each counter-rotating region evolves on nearly circular trajectories in field space, so the first two terms in \eqref{eq:appRhoEqn}, $\ddot{\rho}$ and $D H \dot{\rho}$, can be neglected.
As such, we have
\begin{align}\label{eq:rhoevolve}
   -\mu^2 \rho+ \lambda \rho^3 ={j}_0^2 / \rho^3 \rightarrow -\mu^2\rho^4 +\lambda \rho^6 ={j}_0^2 \propto a^{-2D}\:. 
\end{align}
Note that, in the large $\rho/v$ limit, the term $-\mu^2\rho^4$ can also be dropped, which led to the $\rho \propto a^{-D/3}$ scaling we found in the main text at \eqref{eq:rhoscaling}.
However, as $\rho/v \rightarrow 1$, the $-\mu^2\rho^4$ term cannot be neglected, and the solution for $\rho(a)$ must be found from \eqref{eq:rhoevolve}.%
\footnote{%
    Given the structure of \eqref{eq:rhoevolve}, which is a cubic equation for $\rho^2$, this can in principle be done in closed form, but the expression is algebraically complicated, so we omit it here.} %
Correspondingly, $\dot \theta(a)$ inside the counter-rotating regions can then be obtained with $\dot\theta (a) = j_0/[\rho(a)]^2$.
The spacing $\Delta a$ between sequential string bursts (``inter-burst spacing'') can then be determined as a function of $a$ directly from \eqref{eq:dtVortex}, yielding 
\begin{align}\label{eq:PeriodDeltaa}
    \Delta a = \frac{\pi H(a) a}{|\dot\theta(a)|}\:,
\end{align}
where we have assumed that $\Delta a/a \ll 1$.

A comparison with simulation data is shown in Fig.~\ref{fig:cosmo2d_rho3}.
Similar to the first few periods in Fig.~\ref{fig:largecirculation_energy}, the period $\Delta a$ in the simulation is slightly below this theory prediction (blue line) because, before the saturation of the growth of perturbation, the orbit in field space within most of the simulation volume is not yet fully circularized. 
This theory prediction and the simulation results agree well for scale $2\lesssim a \lesssim 5$, and are much improved compared to the analytical prediction \eqref{eq:periodicityDdim} that assumed the large $\rho/v$ limit (orange line).  
For $a\gtrsim 5$, the charge density in the counter-rotating regions has fallen to $|j_0| \lesssim \mu v^2$, while the average radial mode displacement $\rho \approx 1.2 v$. 
At this point, the charge that can be exchanged between counter-rotating regions over one period of string production and annihilation cannot be neglected,%
\footnote{\label{ftnt:J0change}%
    We can understand this at a parametric level as follows. 
    On the domain boundaries, $|\partial_i \theta| \sim \mu$, leading to $|j^i| \sim v^2 \mu$ when $\rho \sim v$.
    On the other hand, we have at these late times that $|j^0| \lesssim \mu v^2$ in the counter-rotating regions.
    Applying the conservation law $\partial_\mu j^\mu = 0$ in integrated form over a counter-rotating region with characteristic physical spatial extent $\sim \lambda$, we find that $\lambda^2 \Delta j^0 / \Delta t \sim v^2 \mu \lambda$, or $\Delta j^0 \sim v^2 \mu \Delta t /\lambda$.
    Since we expect $\Delta t \sim \lambda$ (i.e., the spacing between sequential string production--annihilation cycles is comparable to the light-crossing time of the region), we have $\Delta j^0 \sim v^2 \mu \sim j^0$.
    This indicates that $j^0$ can change significantly within each counter-rotating region over one cycle of string formation and annihilation in this limit.
} %
and the counter-rotating regions are beginning to break up or dissipate; cf.~Fig.~\ref{fig:stringsmallcirtulation}. 
As expected, the number of strings in the box then decreases (similar to what we observe in Fig.~\ref{fig:string_energy_small_circulation}), and the inter-burst period also significantly deviates from the theory prediction we have developed in this appendix.
Nevertheless, even at these late times, if we take $\dot \theta(a)$ as measured from the full numerical simulation and compute the inter-burst spacing with \eqref{eq:PeriodDeltaa} directly (red line), we still find good agreement with the inter-burst spacing obtained directly from simulation.
Because \eqref{eq:PeriodDeltaa} is simply a rewriting of \eqref{eq:dtVortex} in the $\Delta a/ a\ll1$ limit, this validates that we do have the correct vortex-production criterion, even at late times.

\def\bibsection{\phantomsection\addcontentsline{toc}{section}{References}{\normalfont\large\bfseries References}}
\bibliographystyle{JHEP}
\bibliography{bib.bib}

\end{document}